\documentclass[%
 reprint,
 superscriptaddress,
 nofootinbib,
 amsmath,amssymb,
 aps,prd
]{revtex4-2}

\usepackage{graphicx}
\usepackage{dcolumn}
\usepackage{bm}
\usepackage{enumitem}
\usepackage[caption=false]{subfig}
\usepackage{verbatim}

\usepackage{graphicx}
\usepackage{color}
\usepackage{natbib}
\usepackage{epsfig}
\usepackage{float}
\usepackage{xcolor}

\usepackage{amsmath,amssymb,amsfonts}
\usepackage{subfig}
\usepackage{comment}

\begin{document}

\title{Frequency deviations in universal relations of isolated neutron stars and postmerger remnants}
\author{Georgios Lioutas}\email{g.lioutas@gsi.de}
\affiliation{GSI  Helmholtzzentrum  f\"ur  Schwerionenforschung,  Planckstra{\ss}e  1,  64291  Darmstadt,  Germany}
\affiliation{Department of Physics and Astronomy, Ruprecht-Karls-Universit{\"a}t Heidelberg, Im Neuenheimer Feld 226, 69120 Heidelberg, Germany}

\author{Andreas Bauswein}
\affiliation{GSI  Helmholtzzentrum  f\"ur  Schwerionenforschung,  Planckstra{\ss}e  1,  64291  Darmstadt,  Germany}
\affiliation{Helmholtz Research Academy Hesse for FAIR (HFHF), GSI Helmholtz Center for Heavy Ion Research, Campus Darmstadt,  Germany}

\author{Nikolaos  Stergioulas}
\affiliation{Department of Physics, Aristotle University of Thessaloniki, 54124 Thessaloniki, Greece}

\date{\today}

\begin{abstract}
We relate the fundamental quadrupolar fluid mode of isolated non-rotating NSs and the dominant oscillation frequency of neutron star merger remnants. Both frequencies individually are known to correlate with certain stellar parameters like radii or the tidal deformability, which we further investigate by constructing fit formulae and quantifying the scatter of the data points from those relations. Furthermore, we compare how {\it individual} data points deviate from the corresponding fit to all data points. Considering this point-to-point scatter we uncover a striking similarity between the frequency deviations of perturbative data for isolated NSs and of oscillation frequencies of rapidly rotating, hot, massive merger remnants. The correspondence of frequency deviations in these very different stellar systems points to an underlying mechanism and EoS information being encoded in the frequency deviation. We trace the frequency scatter back to deviations of the tidal Love number from an average tidal Love number for a given stellar compactness. Our results thus indicate a possibility to break the degeneracy between NS radii, tidal deformability and tidal Love number. We also relate frequency deviations to the derivative of the tidal deformability with respect to mass. Our findings generally highlight a possibility to improve GW asteroseismology relations where the systematic behavior of frequency deviations is employed to reduce the scatter in such relationships and consequently increase the measurement accuracy. In addition, we relate the $f-$mode frequency of static stars and the dominant GW frequency of merger remnants. We find an analytic mapping to connect the masses of both stellar systems, which yields particularly accurate mass-independent relations between both frequencies and between the postmerger frequency and the tidal deformability.
\end{abstract}

\maketitle

\section{Introduction}
Fluid oscillation frequencies are one of the most fundamental properties of a stellar system. In the case of isolated neutron stars (NSs), the fundamental $f-$mode is one of the main characteristics of the system and is particularly important because it leads to strong emission of gravitational waves (GWs) (see \cite{1999LRR.....2....2K}). Similarly, binary neutron star (BNS) mergers which do not lead to a prompt collapse to a black hole, produce remnants in which fluid modes are excited. At a frequency $f_{\mathrm{peak}}$ in the range of a few kHz, the dominant fluid oscillation of the merger remnant is an efficient emitter of GWs, see e.g. \cite{1994PhRvD..50.6247X, PhysRevLett.94.201101, 2005PhRvD..71h4021S, 2007PhRvL..99l1102O, Stergioulas2011, 2012PhRvL.108a1101B, 2012PhRvD..86f3001B, 2013PhRvD..88d4026H, 2015PhRvD..91f4001T, 2015PhRvL.115i1101B,2016CQGra..33h5003C} as well as the reviews \cite{Bauswein2016,2019JPhG...46k3002B, 2019PrPNP.10903714B, 2020IJMPD..2941015F,2020GReGr..52..108B,2020arXiv200402527D,2020arXiv201208172S} and references therein. Hence, it is an important target of current GW searches.
The sensitivity of the LIGO and Virgo GW detectors was not sufficient to detect the post-merger phase of the BNS merger GW170817  \cite{2017PhRvL.119p1101A,2017ApJ...851L..16A}, or the likely BNS merger GW190425 \cite{2020ApJ...892L...3A}. However, a post-merger detection is expected to be achieved in the near future, either with upgraded or next-generation detectors \cite{2014PhRvD..90f2004C, 2016CQGra..33h5003C, 2017PhRvD..96l4035C, 2018PhRvL.120c1102B, 2018PhRvD..97b4049Y, 2019PhRvD..99d4014T, 2019PhRvD..99j2004M, 2019MNRAS.485..843O, 2019PhRvD.100d3005E, 2019PhRvD.100d4047T, 2019PhRvD.100j4029B, 2019CQGra..36v5002H, 2020PhRvD.102d3011E, 2020PASA...37...47A, 2020PhRvL.125z1101H, 2020arXiv200708766P, 2020arXiv201112414A, 2021PhRvD.103b2002G}.

The frequency of fluid oscillations depends on the stellar structure. However, the equation of state (EoS) of NS matter is only incompletely known e.g. \cite{2007PhR...442..109L, 2012ARNPS..62..485L, 2016ARA&A..54..401O, 2017RvMP...89a5007O, 2019EPJA...55...10L, 2021PhR...890....1H}. In order to decipher the high-density EoS, many different relations have been proposed between the frequency of fluid oscillations and stellar parameters which are uniquely linked to the EoS. These relations are the basis of GW asteroseismology. There exists a variety of relations with different independent variables exhibiting a different degree of accuracy for both the $f-$mode in isolated NSs \cite{1998MNRAS.299.1059A, 2005MNRAS.357.1029T, 2010ApJ...714.1234L, 2014PhRvD..90l4023C}, as well as for the dominant fluid oscillation in BNS mergers e.g. \cite{2012PhRvL.108a1101B, 2012PhRvD..86f3001B,2015PhRvD..91f4001T, 2015PhRvL.115i1101B, 2020PhRvD.101h4039V, 2020PhRvD.102l3023B}. In practice, these relations are obtained by fitting frequencies as function of some chosen stellar parameter for a sizable number of EoS models.

In this paper we present a systematic comparison between previously proposed relations based on a consistent data set. We investigate the accuracy of these relations for isolated stars and merger remnants by quantifying the scatter in these relations. As a figure of merit, we use the mean and maximum deviation of data points from the corresponding relations in Hz. By determining absolute values of frequency deviations for all relations, we can quantitatively compare different relationships based on the set of EoSs considered here. Work along this direction for the case of isolated NSs was carried out in the past, but for a significantly smaller set of EoSs and a subset of the relations considered here \cite{2015PhRvD..91d4034C}.

Furthermore, we focus on the exact distribution of points with respect to these relations. This aspect is largely unexplored and has not yet been addressed before. Specifically, we investigate the point-to-point scatter comparing individual models, i.e.\ where a specific model is located with respect to a fit to the complete set of models. As the main result of this study, we point out that the individual models follow a systematic behavior.

In particular, we uncover a striking similarity between the frequency deviations w.r.t.\ a fit to the full sample of models in isolated stars and merger remnants described by the same EoSs. The agreement in how individual points scatter is surprising, since the frequencies refer to two very different systems and are obtained independently using different approaches and numerical codes. As a side note, this result further supports that the dominant fluid oscillation in BNS merger remnants is produced by the $f-$mode \cite{Stergioulas2011, Bauswein2015, Bauswein2016, 2020MNRAS.497.5480C}. We further investigate the underlying mechanism for the frequency deviations in both systems. We find that it is directly related to the tidal Love number $k_2$ \cite{2008ApJ...677.1216H, 2010PhRvD..81l3016H, 2010PhRvD..81h4016D} which indicates future applications for improved EoS constraints based on an understanding of the frequency deviations. Also, frequency deviations can be related to the derivative of the tidal deformability with respect to mass. Finally, with a deeper understanding of the frequency deviations we explore direct relations between the $f-$mode frequency of static stars and the dominant postmerger frequency of BNS merger remnants.

The paper is organized as follows: In Sec.\ \ref{Sec_SetupandData} we describe our data sets for both isolated NSs and merger remnants, as well as the set of EoSs we employ in this study. In Sec.\ \ref{Sec_systematicComp} we systematically investigate the accuracy of proposed relations between stellar pulsation frequencies and stellar parameters. Initially we focus on relations for the $f-$mode frequency in isolated NSs and then investigate also relations involving the dominant fluid oscillation frequency for BNS mergers. In Sec.\ \ref{Sec_fpertfpeakCon} we point out the similarity in how individual models distribute with respect to the respective relation for isolated NSs and for BNS mergers. We further investigate the source of these deviations and highlight future applications. In Sec.\ \ref{SecTheorRel} we introduce direct relations between the $f-$mode and postmerger frequencies. Finally, in the last section we provide a summary and conclusions. Throughout the whole work we set $c=G=1$, unless otherwise specified.

\section{Perturbative setup and merger data}\label{Sec_SetupandData}

In this study we consider two different sets of data for stellar pulsations. We discuss frequencies of static isolated stars, which we determine based on perturbative calculations. Moreover, we describe the oscillation frequencies of NS merger remnants. These data are based on relativistic hydrodynamics calculations, where we extract the frequency from the GW spectrum. In this section we provide more details on the data.

\subsection{Linear perturbations}\label{SubSecPertdata}
Neutron star pulsations can lead to GW emission. Since GWs carry away energy, they act as a damping mechanism. In a perturbative approach the pulsations are treated as damped linear oscillations, which are analyzed in terms of quasi-normal modes (QNMs). This ansatz assumes a $e^{i\omega t}$ time dependency, where $\omega$ is the complex eigenfrequency of the QNM. The complex nature of the eigenfrequency accounts for the damping. It reads
\begin{equation}
 \omega=2\pi f_{\mathrm{pert}} + \frac{i}{\tau_{\mathrm{damp}}},
\end{equation}
where $f_{\mathrm{pert}}$ is the pulsation frequency and $\tau_{\mathrm{damp}}$ the damping time of the oscillation. Extensive reviews on the formulation of linear oscillations can be found in \cite{1999LRR.....2....2K, 2013rrs..book.....F}.

We focus on the fundamental ($f-$)mode. We obtain the frequencies using the code presented in \cite{2018GReGr..50...12L}. We compute perturbative frequencies for stellar models in the range from $1.1~M_\odot$ to $1.9~M_\odot$ with a spacing of $0.05~M_\odot$ for different EoSs. We do not include the most compact stellar models for a given EoS, because our main purpose is to compare with binary neutron star mergers, which for the binary mass range under consideration do not reach such high densities/compactness. The most massive merger system we consider has a total mass of $3~M_\odot$ (see Section \ref{SubSecMergerdata}). The central rest-mass densities $\rho_c$ in these systems are comparable to the rest-mass densities of static stars up to about $1.9~M_\odot$.

\subsection{BNS mergers data sets}\label{SubSecMergerdata}
We simulate binary neutron star mergers with a 3D smoothed particle hydrodynamics (SPH) code employing the results from \cite{2002PhRvD..65j3005O, 2007A&A...467..395O, 2010PhRvD..81b4012B, 2012PhRvD..86f3001B}. The code adopts the conformal flatness condition \cite{1980grg1.conf...23I, 1996PhRvD..54.1317W} to solve Einstein's field equations. We choose a resolution of about $300,000$ SPH particles. For more details on the code and simulations we refer to \cite{2002PhRvD..65j3005O, 2007A&A...467..395O, 2010PhRvD..81b4012B, 2012PhRvD..86f3001B}. For $6$ EoS models which do not provide the full temperature dependence, we add a thermal pressure component with an ideal-gas index $\Gamma_{\mathrm{th}}=1.75$ (see \cite{2010PhRvD..82h4043B} for a detailed discussion). 

We extract the dominant postmerger GW frequencies (hereafter $f_{\mathrm{peak}}$) for a total of $57$ equal-mass binary systems. Among them $16$ are $1.2+1.2~M_\odot$ systems, $19$ are $1.35+1.35~M_\odot$ systems, $16$ refer to $1.4+1.4~M_\odot$ systems and finally $6$ correspond to $1.5+1.5~M_\odot$ systems. There are only a few models with $M_{\mathrm{tot}}=3~M_\odot$ because for most EoSs these binary systems lead to a prompt collapse of the merger remnant \cite{2020arXiv201004461B}. A detailed overview of which EoSs are simulated for the different binary systems can be found in Table \ref{TabSPHsystems}. Section \ref{SubSecEoS} provides more information on the different EoS models. We set up irrotational binaries, i.e. stars without intrinsic spin, and choose an initial orbital separation such that the system completes about three orbits before merging. For a subset of binary configurations we test that the dominant postmerger GW frequency is largely insensitive to the initial orbital separation and the resolution. We further comment on this aspect below, and also refer to e.g.~\cite{2012PhRvL.108a1101B, 2012PhRvD..86f3001B} for additional tests and information. We run all simulations for about 20~milliseconds after merging until the GW amplitude sufficiently decayed and the determination of $f_\mathrm{peak}$ is not affected by the simulation time.

\begin{table}[h!]
\caption{\label{TabSPHsystems} EoSs simulated for each binary system. The first column displays the masses of the binary system, while the second column lists all EoSs simulated for these particular masses. }
\begin{ruledtabular}
\begin{tabular}{cc}
\textrm{System masses [$M_\odot$]}&
\textrm{Simulated EoSs}\\
\colrule \vspace{0.25cm}
$1.2+1.2$    & \begin{tabular}{@{}c@{}} APR, BHBLP, BSK20, BSK21, \\ DD2, DD2F, DD2Y, WFF2, \\ LS220, LS375, GS2, SFHO,\\ SFHOY, SFHX, SLY4, TMA \end{tabular} \\ \vspace{0.25cm}
$1.35+1.35$  & \begin{tabular}{@{}c@{}}  ALF2, APR, BHBLP, BSK20,\\ BSK21, DD2, DD2F, DD2Y,\\ WFF2, LS220, LS375, GS2,\\ NL3, SFHO, SFHOY, SFHX,\\ SLY4, TM1, TMA \end{tabular} \\ \vspace{0.25cm}
$1.4+1.4$    & \begin{tabular}{@{}c@{}} APR, BHBLP, BSK20, BSK21,\\ DD2, DD2F, DD2Y, WFF2,\\ LS220, LS375, GS2, SFHO,\\ SFHOY, SFHX, SLY4, TMA \end{tabular} \\
$1.5+1.5$    & \begin{tabular}{@{}c@{}} BHBLP, BSK21, DD2, LS375,\\  GS2, TMA \end{tabular} \\
\end{tabular}
\end{ruledtabular}
\end{table}

Finally, we refer to \cite{Stergioulas2011, Bauswein2015} for evidence that the remnant's oscillation at $f_{\mathrm{peak}}$ is indeed produced by the $f-$mode, which motivates a comparison between the frequencies in static stars and in NS mergers.

\subsection{Equations of state}\label{SubSecEoS} 
 We consider a set of $20$ EoSs (ALF2 \cite{Alford2005,Read2009a}, APR \cite{Akmal1998}, BHBLP \cite{Banik2014}, BSK20 \cite{Goriely2010}, BSK21 \cite{Goriely2010}, DD2 \cite{Hempel2010,Typel2010}, DD2F \cite{Typel2005,Typel2010,Alvarez-Castillo2016}, DD2Y \cite{Fortin2018,Marques2017}, WFF2 \cite{Wiringa1988}, LS220 \cite{Lattimer1991}, LS375 \cite{Lattimer1991}, GS1 \cite{Shen2011}, GS2 \cite{Shen2011}, NL3 \cite{Hempel2010,Lalazissis1997a}, SFHO \cite{Steiner2013}, SFHOY \cite{Fortin2018,Marques2017}, SFHX \cite{Steiner2013}, SLY4 \cite{Douchin2001}, TM1 \cite{Sugahara1994a,Hempel2012}, TMA \cite{Toki1995,Hempel2012}) for which we calculate perturbative frequencies. Postmerger GW frequencies are computed for a slightly smaller subset of EoSs (see Section \ref{SubSecMergerdata}). 
 
 All of the EoSs in this study yield a maximum gravitational mass larger than $1.97~M_\odot$, which is in agreement with current observational constraints at the two sigma level \cite{2010Natur.467.1081D, 2013Sci...340..448A, 2018ApJS..235...37A, 2018ApJ...859...54L, 2020NatAs...4...72C}. Most of the EoS models are compatible with a tidal deformability of $1.37~M_\odot$ stars being smaller than $800$. Thus, they are in agreement with the less strict tidal deformability constraint from the analysis of the inspiral of GW170817 \cite{2017PhRvL.119p1101A, 2019PhRvX...9a1001A}. Six EoSs (LS375, GS1, GS2, NL3, TM1, TMA) yield $\Lambda_{1.37}>800$. We still include them in order to increase the available data set. This is useful, because it strengthens the reliability of our relations and allows us to verify our observations for a larger set of models. No quark or hybrid EoSs are taken into account.
 
 \begin{figure}[h!]
    \includegraphics{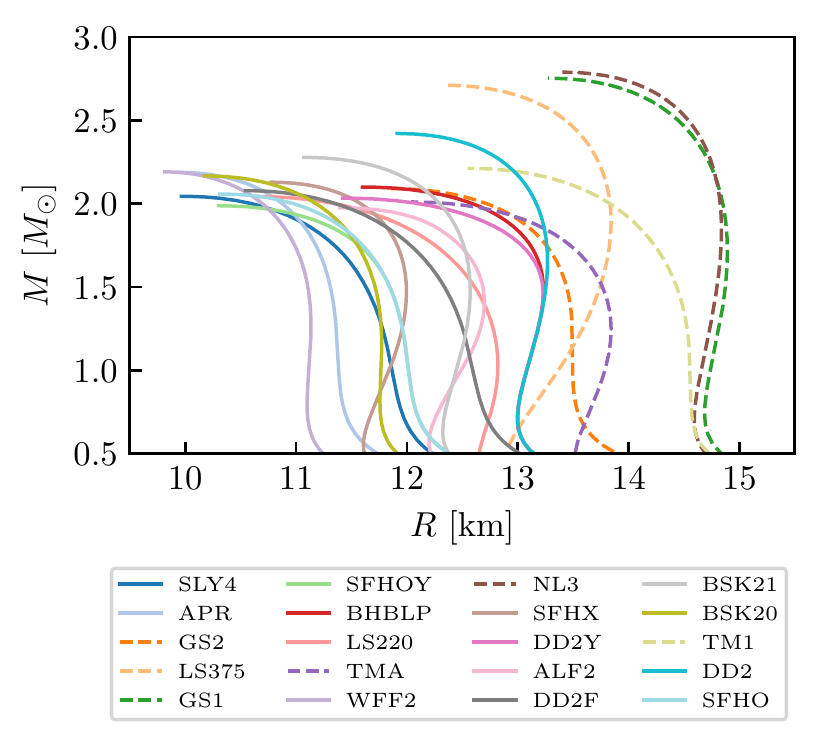}
    \caption{Gravitational mass $M$ versus radius $R$ for all EoSs considered in this work. Dashed curves refer to EoSs incompatible with current constraints on the tidal deformability.}
    \label{MRdiagram}
\end{figure}
 
 Figure \ref{MRdiagram} shows the gravitational mass $M$ versus radius $R$ relation for each EoS. EoSs which are excluded based on current GW measurements are depicted with dashed curves. Evidently, the EoS sample covers a broad range in the $M-R$ diagram.

\section{Frequency relations of the $f-$mode}\label{Sec_systematicComp}
In the case of isolated, static stars many relations have been proposed relating $f_{\mathrm{pert}}$ to stellar parameters such as the mass $M$, radius $R$, moment of inertia $I$ and tidal deformability $\Lambda=\frac{2}{3} k_2 (\frac{c^2R}{GM})^5$ (see e.g. \cite{1998MNRAS.299.1059A, 2005MNRAS.357.1029T, 2010ApJ...714.1234L, 2014PhRvD..90l4023C}). They exhibit a different level of accuracy. Similarly, in the case of binary systems, such relations involving $f_{\mathrm{peak}}$ have been proposed for systems of different masses and binary mass ratios (\cite{2012PhRvL.108a1101B, 2012PhRvD..86f3001B,2015PhRvD..91f4001T, 2015PhRvL.115i1101B, 2020PhRvD.101h4039V, 2020PhRvD.102l3023B}). These relations are sometimes presented for a broad range of masses, while in some cases they focus on a fixed mass.

Relations involving the $f-$mode frequency of either a static star or a merger remnant can be used to extract information about the stellar parameters from GW observations. Generally tighter relations allow for a better determination of the involved parameters. Hence relations with a smaller scatter are favored. In this section we focus on quantifying the scatter in different relations to enable an objective comparison between them.  

For static stars, we examine different relations proposed in the literature and newly introduced in this work. We discuss how the points scatter in each case. Furthermore, we consider new relations involving only properties referring to the innermost part of the star, containing $90\%$ of its mass. Our results are then applied to binary mergers as well.

As figure of merit to assess the tightness of a given relation, we consider the maximum and mean deviation between the data points and the respective fit to the data. The maximum deviation may be biased by the most extreme model and does not represent how most of the points scatter. However, it may provide a conservative measure of the accuracy of a relation and thus an upper estimate of the error if the relation is employed in GW measurements. Conversely, the mean deviation captures the point distribution, but it may not be fully representative of the error because the EoS sample is not a statistical ensemble. Hence, the mean deviation may be less suited to describe the error when using such relations to determine stellar parameters. We present both deviations, since a combination of the two provides a more complete picture. Generally, we find a consistent behavior of both measures.

In terms of notation, given a set of data $(X_i,Y_i)$ with $N$ points, we denote deviations by $\delta_X Y$. We define deviations between the data points and the corresponding fit as
\begin{equation}
 \delta_X Y_i = Y_i - Y_\mathrm{fit}(X_i),
\end{equation}
for which we express the maximum and average deviation as
\begin{align}
 \max{\left(\delta_X Y\right)} &= \max_i{\left( \left| \delta_X Y_i \right| \right)}, \\
 \overline{\delta_X Y} &= \frac{\sum_{i=1}^{N}{ \left| \delta_X Y_i \right| }}{N},
\end{align}
where $|\cdot|$ is the absolute value.

\subsection{Isolated stars}\label{Sec_stat_stars}
We start by discussing relations between $f_{\mathrm{pert}}$ and the stellar mass $M$ and radius $R$. All relations that we describe in the following are provided in Table \ref{Fit_table} based on our data. Table \ref{Fit_table} includes the deviations of the relations.

A very well known relation was proposed by Andersson and Kokkotas in \cite{1998MNRAS.299.1059A} between $f_{\mathrm{pert}}$ and the mean density of the star. Later, Tsui and Leung used a different scaling, which accurately describes both $f_{\mathrm{pert}}$ and $\tau_{\mathrm{damp}}$, as well as other families of modes \cite{2005MNRAS.357.1029T}. By this, the mass-scaled frequency $Mf_{\mathrm{pert}}$ is found to yield a tight correlation with the compactness $M/R$. 

\begin{figure}
    \subfloat[\label{AK_original}]{%
         \includegraphics[width=\columnwidth]{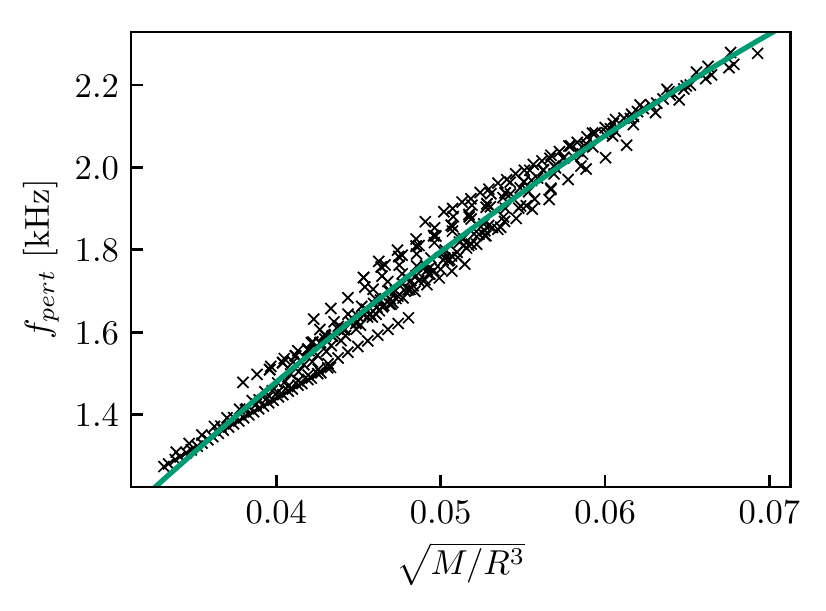}}

    \vspace*{8pt}%
    
    \subfloat[\label{TL_original}]{%
        \includegraphics[width=\columnwidth]{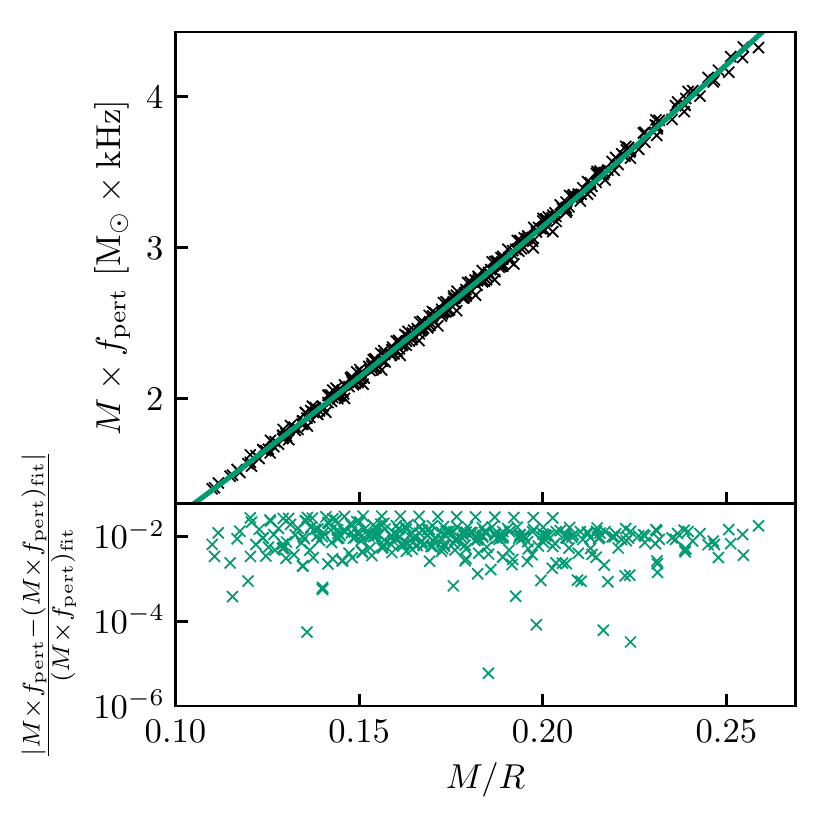}}
    
    \caption{Relations between $f_{\mathrm{pert}}$ and the mass and radius of a non-rotating neutron star. Panel \protect\subref{AK_original} displays the relation between the $f-$mode frequency and the mean density of the star as proposed in \cite{1998MNRAS.299.1059A}. In panel \protect\subref{TL_original} we plot the mass-scaled frequency versus the compactness as suggested in \cite{2005MNRAS.357.1029T}. In both panels the solid curve shows a second-order fit based on our data. }
    \label{fpert_Original}
\end{figure}

In Figs. \ref{AK_original}-\ref{TL_original} we plot our $f-$mode data for both relations. The solid curves display second-order fits to our data in both diagrams. The purpose of these fits is to accurately quantify how the data scatter around the respective function. Table~\ref{Fit_table} provides the fit parameters and the mean and maximum deviation of the data points from the fit. Based on the deviations it is evident that the second relation is more accurate, but still both fits exhibit some scatter\footnote{Note that for a fair comparison, throughout the whole paper, we compare absolute frequencies, also for relations with mass-scaled frequencies.}.

\begin{table*}
\caption{\label{Fit_table}Relations between $f-$mode frequencies of static stars and different stellar parameters. The tightness of the relations is quantified by the average and maximum deviation between fit and underlying data. First column provides a reference to the work where the relation of the respective form has been proposed. Frequencies are in kHz, masses in $M_\odot$, radii in units of $GM_\odot/c^2$ and moments of inertia in units of $G^2M^3_\odot/c^4$. The tidal deformability $\Lambda$ is dimensionless.}
\begin{ruledtabular}
\begin{tabular}{cclcc}
\textrm{Reference}&
\textrm{Position}&
\multicolumn{1}{c}{\textrm{Fit}}&
\textrm{Mean dev.}&
\textrm{Max dev.} \\
 & & & \textrm{[Hz]} & \textrm{[Hz]}\\ 
\colrule \vspace{0.2cm}
\cite{1998MNRAS.299.1059A} & Fig.\ (\ref{AK_original})        & $f_{\mathrm{pert}}  = -0.133 +47.23\sqrt{\frac{M}{R^3}}                             -173.2\frac{M}{R^3}$                     & $31$  & $102$  \\ \vspace{0.1cm}
This work                  & Fig.\ (\ref{AK_90percent})       & $f_{\mathrm{pert}}  = -0.2   +37.68\sqrt{\frac{M}{(R^{90\%})^3}}                    -92.14\frac{M}{(R^{90\%})^3}$            & $12$  & $34$ \\ \vspace{0.5cm}
This work              & Text                             & $f_{\mathrm{pert}}  = -0.106 +37.15\sqrt{\frac{M^{\mathrm{cc}}}{(R^{\mathrm{cc}})^3}} -57.56\frac{M^{\mathrm{cc}}}{(R^{\mathrm{cc}})^3}$            & $19$  & $54$ \\ \vspace{0.1cm}
\cite{2005MNRAS.357.1029T} & Fig.\ (\ref{TL_original})        & $Mf_{\mathrm{pert}} = -0.427 +14.95\frac{M}{R}                                              +14.43\left(\frac{M}{R}\right)^2$        & $19$  & $49$ \\ \vspace{0.1cm}
This work                  & Fig.\ (\ref{TL_90percent})       & $Mf_{\mathrm{pert}} = -0.626 +13.69\frac{M}{R^{90\%}}                                       +11.67\left(\frac{M}{R^{90\%}}\right)^2$ & $10$  & $34$ \\ \vspace{0.5cm}
This work              & Text                             & $M^{\mathrm{cc}}f_{\mathrm{pert}} = -0.586 +13.67\frac{M^{\mathrm{cc}}}{R^{\mathrm{cc}}}    +17.92\left(\frac{M^{\mathrm{cc}}}{R^{\mathrm{cc}}}\right)^2$ & $18$  & $64$ \\ \vspace{0.1cm}
\cite{2010ApJ...714.1234L} &  Text                           & $Mf_{\mathrm{pert}} = -0.117 +3.966\sqrt{\frac{M^3}{I}}               +18.97 \frac{M^3}{I}$                    & $0.9$ & $5$   \\ \vspace{0.5cm}
\cite{2010ApJ...714.1234L} & Text                            & $Mf_{\mathrm{pert}} = -0.117 +4.161\sqrt{\frac{M^3}{I}}               +16.93 \frac{M^3}{I}    
                                                                            +6.995\left(\frac{M^3}{I}\right)^{3/2}                       -7.855\left(\frac{M^3}{I}\right)^2$      & $0.8$ & $5$ \\ \vspace{0.1cm}
This work                  & Fig.\ (\ref{TidalDefFreq_Static})& $Mf_{\mathrm{pert}} = -0.656 +12.26\Lambda^{-1/5}                     -5.471 \Lambda^{-2/5}$                    & $3$   & $17$  \\ \vspace{0.5cm}
This work                  &  Text                           & $Mf_{\mathrm{pert}} = -0.24  +7.726\Lambda^{-1/5}                     +11.877 \Lambda^{-2/5}       
                                                                            -27.653\Lambda^{-3/5}                                    +15.387 \Lambda^{-4/5}$                   & $0.12$ & $0.7$  \\ \vspace{0.1cm}
\cite{2014PhRvD..90l4023C} &  Text                           & $Mf_{\mathrm{pert}} =  6.939  -9.294\times10^{-1}\ln{\Lambda}                       +3.267\times10^{-2}\left(\ln{\Lambda}\right)^2$      & $4$   & $19$  \\ \vspace{0.1cm}
\cite{2014PhRvD..90l4023C} &  Text                           & $Mf_{\mathrm{pert}} =  5.965 -0.2814\ln{\Lambda}                       -0.1214\left(\ln{\Lambda}\right)^2
                                                                            +1.555\times10^{-2}\left(\ln{\Lambda}\right)^3  -5.619\times10^{-4}\left(\ln{\Lambda}\right)^4$   & $0.14$ & $0.8$
\end{tabular}
\end{ruledtabular}
\end{table*}

Another relation was discussed by Lau et al.\ \cite{2010ApJ...714.1234L} involving the moment of inertia. They remarked that the previously suggested relations cannot describe quark stars, because neutron stars and quark stars result in different density profiles. They argued that the moment of inertia $I$ is sensitive to the matter distribution within the star. Thus defining an effective compactness through $I$ leads to a relation which successfully describes both types of stars. We include a second-order fit, which was also employed in \cite{2010ApJ...714.1234L}, as well as a fourth-order fit. The relation is very tight (see mean and maximum deviations in Table~\ref{Fit_table}). Its accuracy does not improve with the order of the fit.

Chan et al.\ \cite{2014PhRvD..90l4023C} suggested that there should exist a tight correlation between $f_{\mathrm{pert}}$ and $\Lambda$ (see also \cite{2019PhRvC..99d5806W}) based on the I-Love-Q relations \cite{2013Sci...341..365Y}, since $f_{\mathrm{pert}}$ tightly correlates with the moment of inertia $I$ \cite{2010ApJ...714.1234L}. Consequently, they consider a relation of the form $Mf_{\mathrm{pert}}(\ln{\Lambda})$. 

\begin{figure}
    \includegraphics{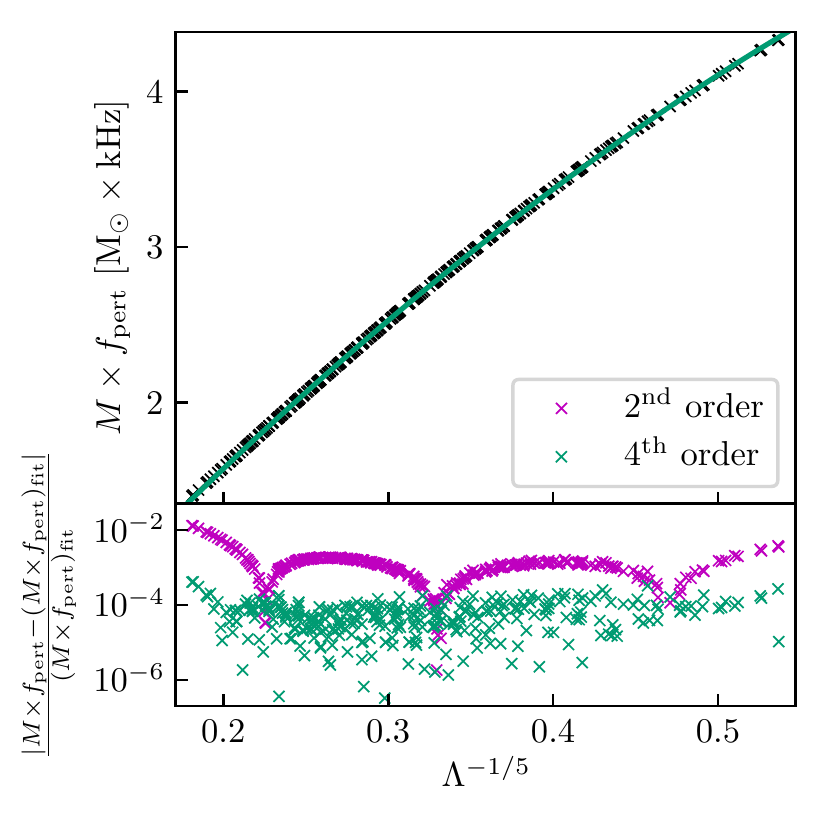}
    \caption{ Relation between mass-scaled $f-$mode frequency and $\Lambda^{-1/5}$. Solid curve displays a second-order fit based on our data. Bottom panel shows the fractional error for the second-order fit plotted in the upper panel, as well as the fourth-order fit discussed in the text. The legend explains the symbols' colors in the bottom panel.} 
    \label{TidalDefFreq_Static}
\end{figure}

Moreover, $\Lambda^{-1/5}$ is directly related to the compactness. This motivates a relation of the form $Mf_{\mathrm{pert}}(\Lambda^{-1/5})$, which we plot in Fig.\ \ref{TidalDefFreq_Static} and model by a second-order fit. According to the mean and maximum deviation of 3~Hz and 17~Hz, respectively, this relation is very tight (see Table \ref{Fit_table}). For comparison, the mean and maximum deviation for a second-order $Mf_{\mathrm{pert}}(\ln{\Lambda})$ relation are $4$~Hz and $19$~Hz respectively. Considering fourth-order fits, as proposed in \cite{2014PhRvD..90l4023C}, relations with $\Lambda^{-1/5}$ and those with $\ln{\Lambda}$ are identically accurate (see also bottom panel of Fig. \ref{TidalDefFreq_Static} for relations w.r.t.\ $\Lambda^{-1/5}$).

Interestingly, second-order fits involving $\Lambda^{-1/5}$ or $\ln{\Lambda}$ are less accurate than the relation involving the moment of inertia $I$ . However, increasing the order of the fits leads to tighter relations for $\Lambda$, while the accuracy of the moment-of-inertia relation remains practically the same regardless of the order. As a result, fourth-order relations with $\Lambda^{-1/5}$ or $\ln{\Lambda}$ are more accurate than the fourth-order relation with $I$.

Comparing entries in Table \ref{Fit_table}, it is clear that relations with $\Lambda$ are more accurate than those involving $M$ and $R$. One main difference between the tidal deformability and the compactness is that the first is less sensitive to the low-density parts of the star, in particular the crust \cite{2019PhRvC..99d5802P, 2020CQGra..37b5008G}. One may thus pose the question of whether the scatter in relations involving $M$ and $R$ can be solely attributed to the crust, since the radius $R$ is somewhat sensitive to the low-density EoS.

In order to investigate this point we define a new effective radius $R^{90\%}$ for static stars. It refers to the radius of a sphere containing $90\%$ of the gravitational mass of the configuration. By disregarding the outer shell containing $10\%$ of the mass, we obtain a radius which is insensitive to the crust and the low-density EoS. Using this definition, we excise the outermost $1.26-2.41$ km of the stellar configuration. An additional feature of this newly defined quantity is that, for all EoSs and models considered, the pressure at this radius corresponds to $3-5\%$ of the central pressure $p_c$\footnote{Whereas the pressure on the sphere containing 90\% of the mass is universally about 4\% of the central pressure, the density at this point in the star shows a larger scatter and corresponds to about $25-47\%$ of the central rest-mass density $\rho_c$.}. Thus, one could equivalently define a fixed pressure surface of e.g. $p_\ast=0.04 \times p_c$.

\begin{figure}
    \subfloat[\label{AK_90percent}]{%
         \includegraphics[width=\columnwidth]{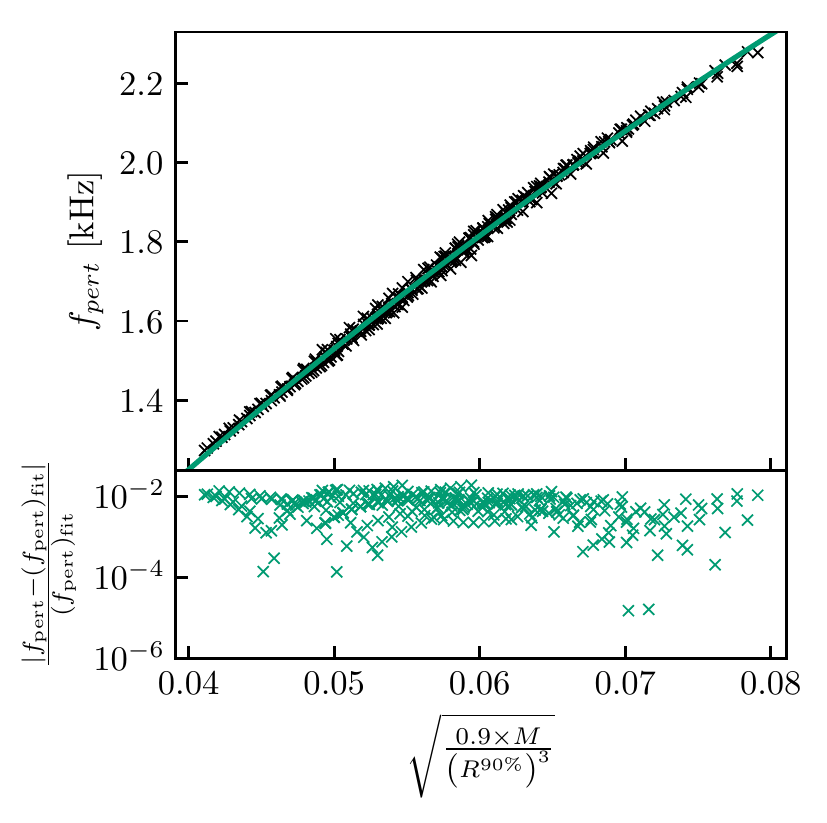}}

    \vspace*{8pt}%
    
    \subfloat[\label{TL_90percent}]{%
        \includegraphics[width=\columnwidth]{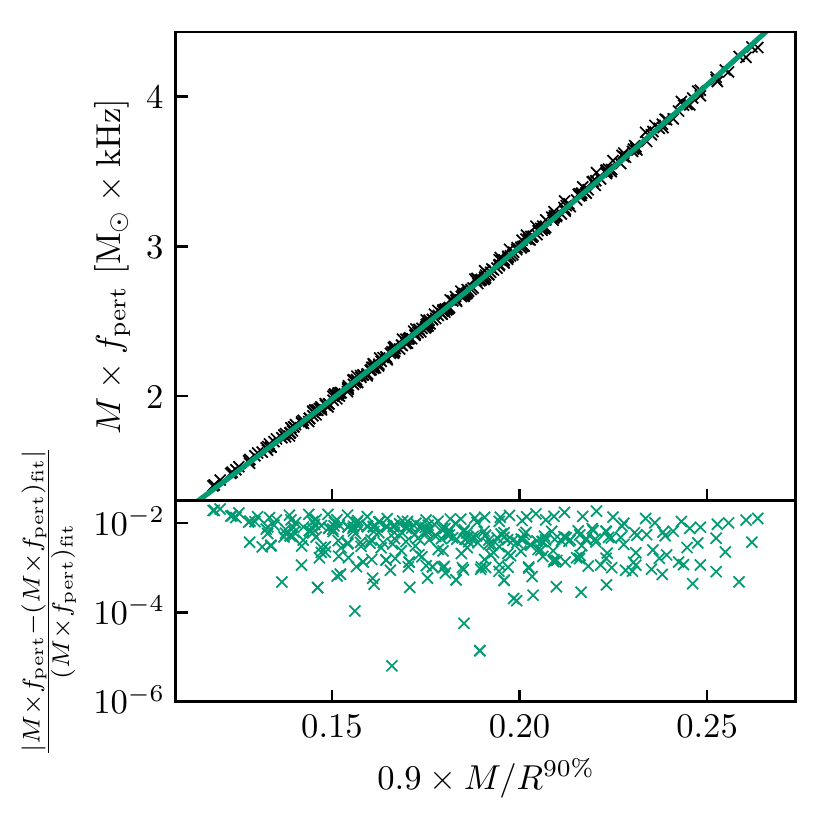}}
    
    \caption{Same as in Fig.\ \ref{fpert_Original}, but the stellar parameters refer to a sphere containing only $90\%$ of the mass of the corresponding configuration. Compared to Fig.\ \ref{fpert_Original} both relations get tighter.}
    \label{fpert_90pc}
\end{figure}

Based on the newly defined radius we introduce the mean density and compactness of the corresponding sphere. In Fig.\ \ref{fpert_90pc} we plot the relations shown in Fig.\ \ref{fpert_Original}, but employing the new quantities which omit the low-density material. Both relations become tighter. This is clearly shown in Table \ref{Fit_table}, where we explicitly give relations and characterize their quality by the corresponding mean and maximum deviation. In particular, for the relation involving the mean density the improvement is significant. Both relations involving the redefined mean density and compactness are practically identically accurate compared to each other. We conclude that $f_{\mathrm{pert}}$ is an excellent measure of the mean density of the star, when referring to the interior part comprising $90\%$ of its mass.

In addition, we employ the crust-core transition density $\rho_\mathrm{cc}$ to define a second excision procedure dismissing all material with a density below $\rho_\mathrm{cc}$. For most EoS models the exact density of the crust-core transition is not publicly available. Therefore, we estimate the crust-core transition density by extracting approximately the slope of the symmetry energy $L$ from the EoS table for neutrino-less beta equilibrium via the pressure at saturation density. We then employ a relation between $L$ and the dynamical crust-core transition density from \cite{2011PhRvC..83d5810D}. For each stellar model we identify the radius of the crust-core transition $R^{\mathrm{cc}}$, alongside the mass $M^{\mathrm{cc}}$ contained within this radius.

Analogously to $R^{90\%}$ we compute the mean density and compactness of the core region defined through $R^{\mathrm{cc}}$. In Table \ref{Fit_table} we provide relations of the same functional form as in Fig.\ \ref{fpert_Original}. The relation involving the mean density $\sqrt{(M^{\mathrm{cc}}/(R^{\mathrm{cc}})^3)}$ is almost twice as accurate compared to the case where the whole stellar configuration is considered. On the other hand, the accuracy of the relation considering the compactness does not change significantly. Both relations can potentially become tighter if one employs a more accurate definition of the crust-core transition density for each EoS considered.

We note that the relation involving $\Lambda^{-1/5}$ is still more accurate than the relations involving $R^{90\%}$ or $R^{\mathrm{cc}}$. Although low-density material has been removed, still some scatter is visible. We thus conclude that the scatter in these relations does not entirely result from the low-density description. This implies that the distribution of data points with respect to the fit may also contain additional information about high-density properties of the EoS, which affect the $f-$mode frequency.

\subsection{Merger remnants}
In the case of BNS mergers, tight relations have been found for systems with fixed total binary mass relating the dominant postmerger frequency to radii of static stars of a fixed mass \cite{2012PhRvL.108a1101B, 2012PhRvD..86f3001B,2014PhRvD..90b3002B}. Employing radii of static stars is a choice which is empirically found to yield tight relations bearing in mind that one cannot define the mass and radius of merger remnants in an unambiguous way. Mass-scaled relations are not as accurate \cite{2015PhRvL.115i1101B, Bauswein2016}. We consider relations between $f_{\mathrm{peak}}$ and $R^{90\%}$, $R^{\mathrm{cc}}$ or $\Lambda^{-1/5}$ to investigate whether these relations also become tighter as those for static stars.

\begin{figure}
    \subfloat{%
         \includegraphics[width=\columnwidth]{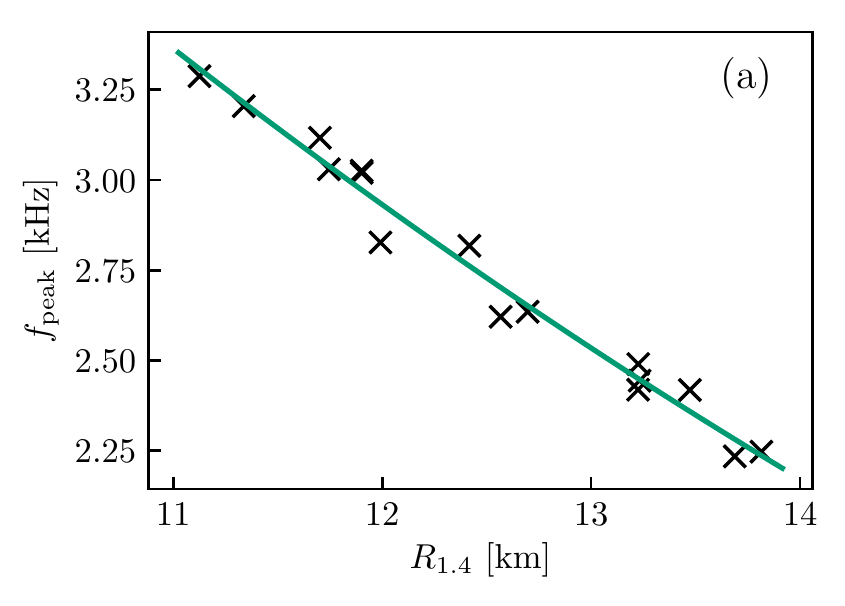}}

    \vspace*{8pt}%
    
    \subfloat{%
        \includegraphics[width=\columnwidth]{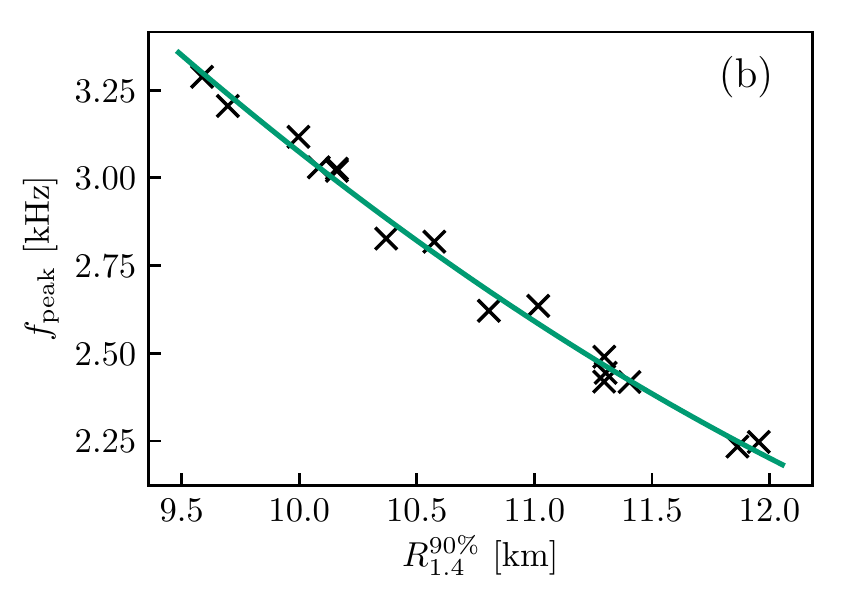}}
            
    \vspace*{8pt}%
    
    \subfloat{%
        \includegraphics[width=\columnwidth]{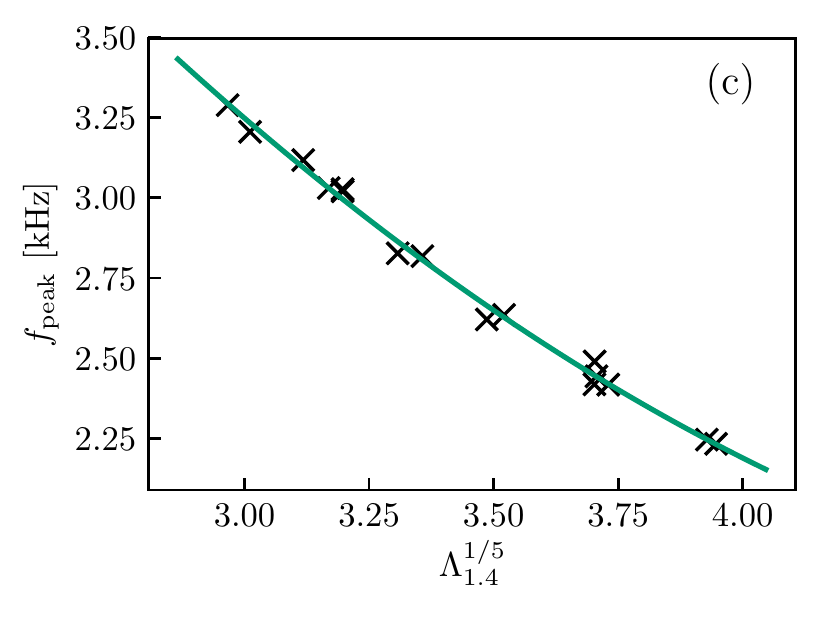}}

    \caption{Postmerger frequencies $f_{\mathrm{peak}}$ as function of various stellar parameters of static stars with different EoSs. $R_{1.4}$ (top panel) refers to the radius of a $1.4~M_\odot$ non-rotating NS, $R^{90\%}_{1.4}$ (middle panel) is the radius of a star with artificially excised low-density region (see Section \ref{Sec_stat_stars}) and $\Lambda^{1/5}_{1.4}$ (bottom panel) is the fifth-root of the tidal deformability. The frequencies refer to $1.2+1.2~M_\odot$ binary systems. }
    \label{AB_plots}
\end{figure}

Table \ref{Fit_table_Mergers} lists empirical relations of the form $f_{\mathrm{peak}}(R_x)$, $f_{\mathrm{peak}}(R^{90\%}_x)$, $f_{\mathrm{peak}}(R^{\mathrm{cc}}_x)$ and $f_{\mathrm{peak}}(\Lambda_x)$ for all binary systems considered in this study. Here $x$ stands for the mass of a static star. In order to choose an appropriate stellar mass for each binary system, we consider the maximum rest-mass densities in the first few milliseconds of the postmerger phase (for a more extensive discussion see Appendix \ref{Appendix_A}). Then, we determine the mass of static stellar configurations, which have roughly comparable central densities. For instance, we relate systems with a total mass of $2.4~M_\odot$ to static stars of $1.4~M_\odot$. Similarly, $1.6$, $1.7$ and $1.75~M_\odot$ static stars are chosen for binary systems with a total mass of $2.7$, $2.8$ and $3~M_\odot$, respectively. As an example, Fig.~\ref{AB_plots} displays the empirical relations for $1.2+1.2~M_\odot$ systems.

Quantifying deviations in terms of frequencies allows us to compare the quality of all relations to each other keeping in mind that the quantitative results to some extent depend on the chosen fiducial mass $x$. We find that using either $R^{90\%}$ or $\Lambda^{-1/5}$ leads to tighter empirical relations compared to $R$ or $R^{\mathrm{cc}}$ (see Fig.\ \ref{AB_plots} for an example with $1.2+1.2~M_\odot$ mergers). In the case of $R^{\mathrm{cc}}$ this is mostly an artifact of the fiducial mass chosen in Table \ref{Fit_table_Mergers}, which leads to more accurate relations for $R^{90\%}$ than for $R^{\mathrm{cc}}$ (see Table \ref{Optimal_fiducial_masses} for more details). $R^{90\%}$ relations are marginally less accurate than those with $\Lambda^{-1/5}$. This is in line with the findings for the relations of static stars considering a large mass range\footnote{We note that for $1.5+1.5~M_\odot$ systems the improvement is not as pronounced. This results from the fact that less systems are considered in this case, since many EoS models result in a prompt collapse and thus the data set is significantly smaller compared to the other binary masses.}. Furthermore, we point out that relations between $f_{\mathrm{peak}}$ and the tidal deformability of static stars are more accurate for fiducial masses higher than the mass of the inspiraling stars for all binary systems considered (see Table \ref{Fit_table_Mergers}). We in particular refer to the more thorough analysis of this aspect in Appendix~\ref{Appendix_A} (see also~\cite{2020PhRvD.101h4039V} for relations for a range of binary masses).

\begin{table*}
\caption{\label{Fit_table_Mergers}Fits (third column) to the data of postmerger frequencies for different total binary masses (first column) employing various independent variables given in the second column. Fourth and fifth column provide the average and maximum deviation between fit and data in absolute frequencies. Frequencies are in kHz, radii $R$, $R^{90\%}$ and $R^{\mathrm{cc}}$ in km and $\Lambda^{1/5}$ is dimensionless. Deviations for all relations are in Hz, so they can be directly compared to each other.}
\begin{ruledtabular}
\begin{tabular}{cclcccc}
\textrm{Binary masses}&
\textrm{Independent}&
\multicolumn{1}{c}{\textrm{Fit}}&
\textrm{Mean dev.}&
\textrm{Max dev.}\\
 \textrm{[$M_\odot$]} & \textrm{variable} & & \textrm{[Hz]} & \textrm{[Hz]}\\
\colrule \vspace{0.1cm}

$1.2+1.2$ & $R$                        &  $f_{\mathrm{peak}}  = 10.428  -8.347\times10^{-1}R_{1.4}        +1.749\times10^{-2}R_{1.4}^2$                                           & $41$ & $109$ \\ \vspace{0.1cm}
$1.2+1.2$ & $R^{90\%}$                 &  $f_{\mathrm{peak}}  = 12.963  -1.449R_{1.4}^{90\%}              +4.604\times10^{-2}\left(R_{1.4}^{90\%}\right)^2$                       & $31$ & $58$  \\ \vspace{0.1cm}
$1.2+1.2$ & $R^{\mathrm{cc}}$ &  $f_{\mathrm{peak}}  = 16.526 -1.9593R_{1.4}^{\mathrm{cc}}       +6.568\times10^{-2}\left(R_{1.4}^{\mathrm{cc}}\right)^2$                & $45$ & $112$  \\ \vspace{0.1cm}
$1.2+1.2$ & $\Lambda^{1/5}$            &  $f_{\mathrm{peak}}  =  9.74  -2.994\Lambda^{1/5}_{1.4}          +2.767\times10^{-1}\Lambda^{2/5}_{1.4}$                                 & $18$ & $44$  \\ \vspace{0.5cm}
$1.2+1.2$ & $\Lambda^{1/5}$            &  $f_{\mathrm{peak}}  =  9.74  -2.432\Lambda^{1/5}_{1.2}          +1.771\times10^{-1}\Lambda^{2/5}_{1.2}$                                 & $39$ & $72$  \\ \vspace{0.1cm}

$1.35+1.35$ & $R$                        &  $f_{\mathrm{peak}}  = 12.61 -1.134R_{1.6}                       +2.87\times10^{-2}R_{1.6}^2$                                            & $48$ & $84$  \\ \vspace{0.1cm}
$1.35+1.35$ & $R^{90\%}$                 &  $f_{\mathrm{peak}}  = 12.63 -1.306R_{1.6}^{90\%}                +3.79\times10^{-2}\left(R_{1.6}^{90\%}\right)^2$                        & $31$ & $60$  \\ \vspace{0.1cm}
$1.35+1.35$ & $R^{\mathrm{cc}}$ &  $f_{\mathrm{peak}}  = 14.653 -1.5432R_{1.4}^{\mathrm{cc}}       +4.597\times10^{-2}\left(R_{1.4}^{\mathrm{cc}}\right)^2$                & $47$ & $110$  \\ \vspace{0.1cm}
$1.35+1.35$ & $\Lambda^{1/5}$            &  $f_{\mathrm{peak}}  = 9.063 -2.912\Lambda^{1/5}_{1.6}           +0.276\Lambda^{2/5}_{1.6}$                                              & $26$ & $61$ \\ \vspace{0.5cm}
$1.35+1.35$ & $\Lambda^{1/5}$            &  $f_{\mathrm{peak}}  = 8.886 -2.147\Lambda^{1/5}_{1.35}          +1.397\times10^{-1} \Lambda^{2/5}_{1.35}$                               & $46$ & $88$ \\ \vspace{0.1cm}

$1.4+1.4$ & $R$                        &  $f_{\mathrm{peak}}  = 12.61 -1.085R_{1.7}                       +2.54\times10^{-2}R_{1.7}^2$                                            & $53$ & $151$ \\ \vspace{0.1cm}
$1.4+1.4$ & $R^{90\%}$                 &  $f_{\mathrm{peak}}  = 15.16 -1.716R_{1.7}^{90\%}                +5.51\times10^{-2}\left(R_{1.7}^{90\%}\right)^2$                        & $38$ & $130$ \\ \vspace{0.1cm}
$1.4+1.4$ & $R^{\mathrm{cc}}$ &  $f_{\mathrm{peak}}  = 16.389 -1.7824R_{1.4}^{\mathrm{cc}}       +5.467\times10^{-2}\left(R_{1.4}^{\mathrm{cc}}\right)^2$                & $53$ & $148$  \\ \vspace{0.1cm}
$1.4+1.4$ & $\Lambda^{1/5}$            &  $f_{\mathrm{peak}}  = 11.11 -4.584\Lambda^{1/5}_{1.7}           +5.821\times10^{-1}\Lambda^{2/5}_{1.7}$                                 & $36$ & $124$ \\ \vspace{0.5cm}
$1.4+1.4$ & $\Lambda^{1/5}$            &  $f_{\mathrm{peak}}  = 9.34 -2.342\Lambda^{1/5}_{1.4}           +1.533\times10^{-1}\Lambda^{2/5}_{1.4}$                                  & $65$ & $159$ \\ \vspace{0.1cm}

$1.5+1.5$ & $R$                        &  $f_{\mathrm{peak}} = -34.89 +6.19 R_{1.75}                      -2.519\times10^{-1}R_{1.75}^2$                                           & $29$ & $76$ \\ \vspace{0.1cm}
$1.5+1.5$ & $R^{90\%}$                 &  $f_{\mathrm{peak}} = -7.534 +2.275R_{1.75}^{90\%}               -1.189\times10^{-1}\left(R_{1.75}^{90\%}\right)^2$                       & $25$ & $64$ \\ \vspace{0.1cm}
$1.5+1.5$ & $R^{\mathrm{cc}}$ &  $f_{\mathrm{peak}}  = -7.829 +2.1565R_{1.4}^{\mathrm{cc}}       -1.0424\times10^{-1}\left(R_{1.4}^{\mathrm{cc}}\right)^2$                & $41$ & $88$  \\ \vspace{0.1cm}
$1.5+1.5$ & $\Lambda^{1/5}$            &  $f_{\mathrm{peak}} =   3.74 +0.755 \Lambda^{1/5}_{1.75}         -3.798\times10^{-1} \Lambda^{2/5}_{1.75}$                                & $30$ & $73$ \\ \vspace{0.1cm}
$1.5+1.5$ & $\Lambda^{1/5}$            &  $f_{\mathrm{peak}} =   -9.88 +8.624 \Lambda^{1/5}_{1.5}         -1.427 \Lambda^{2/5}_{1.5}$                                              & $52$ & $109$
\end{tabular}
\end{ruledtabular}
\end{table*}

The analysis shows that using a frequency measurement the determination of $R^{90\%}$ is up to twice as accurate as that of $R$ as the maximum deviation should be included as an error estimate. For all binary systems the mean deviation in $f_{\mathrm{peak}}(R^{90\%}_x)$ is about $70$m (based on the inverted relations $R^{90\%}_x (f_{\mathrm{peak}})$). Therefore, $R^{90\%}$ of a fixed mass static star can be determined with high accuracy from an observation of an equal-mass binary system. We emphasize that $R^{90\%}$ is as informative about the EoS as $R$. As $R$, the redefined radius $R^{90\%}$ is uniquely linked to the EoS and, moreover, is only sensitive to the high-density regime of the EoS.

\section{Connection between $f_{\mathrm{pert}}$ and $f_{\mathrm{peak}}$ frequencies}\label{Sec_fpertfpeakCon}
In this section we address how individual data points are distributed with respect to the corresponding relation, i.e.\ with respect to the fit to all points.

\subsection{Point scatter in $f_{\mathrm{peak}}$ relations and $f_{\mathrm{pert}}$ relations}\label{PointScatterObs}
In Fig.\ \ref{AB_plots}, we plot empirical relations between $f_{\mathrm{peak}}$ and $3$ different stellar parameters, namely $R$, $R^{90\%}$ and $\Lambda^{-1/5}$ for $1.2+1.2~M_\odot$ systems. Considering the exact location of individual points in the plots, the points deviate from the respective fit in a very similar way in all panels. EoSs\footnote{For simplicity, in the following we will use the term ``EoS'' for actually referring to the resulting frequency/data point obtained from a calculation for this EoS.} which lie above the fit in one plot, typically lie above the fit in the other relations as well. The same holds for EoSs lying below the fit. The data apparently follows the same systematic behavior in all three relations. From the fact that $f_{\mathrm{peak}}(R^{90\%})$ shows the same trend (middle panel) we conclude that this general observation of similar frequency deviations is insensitive to the low-density regime of the star. Furthermore, we find a similar pattern in plots for other binary masses.

\begin{figure*}
    
    \subfloat[\label{fpeak135_R16_band}]{%
        \includegraphics[width=\columnwidth]{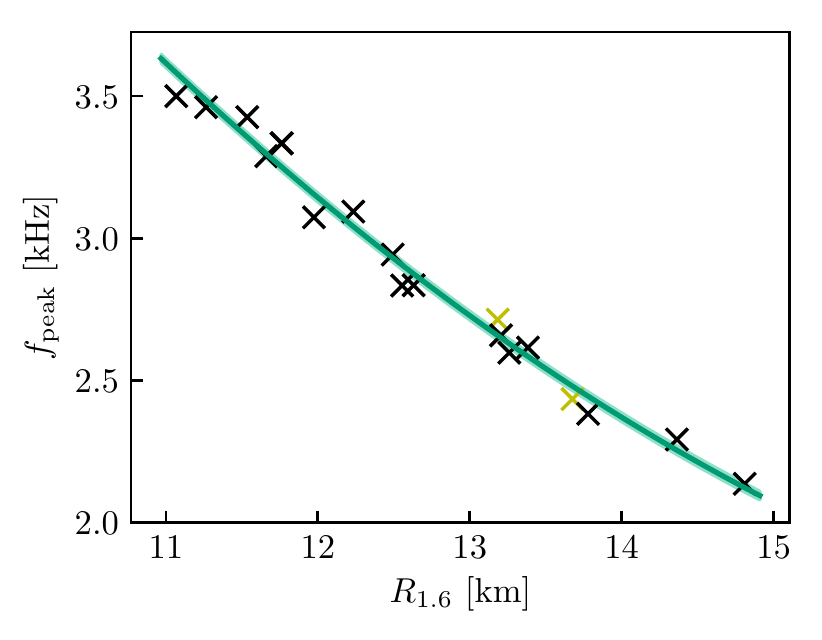}}
    \hfill
    \subfloat[\label{fpeak135_Lam135_band}]{%
        \includegraphics[width=\columnwidth]{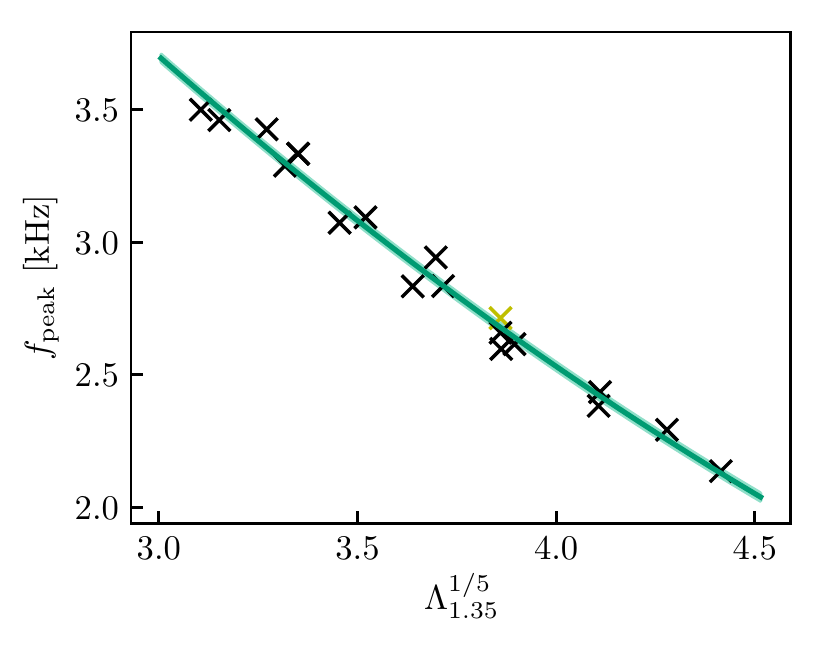}}

    \vspace*{8pt}%
    
    \subfloat[\label{fpert_R16_band}]{%
         \includegraphics[width=\columnwidth]{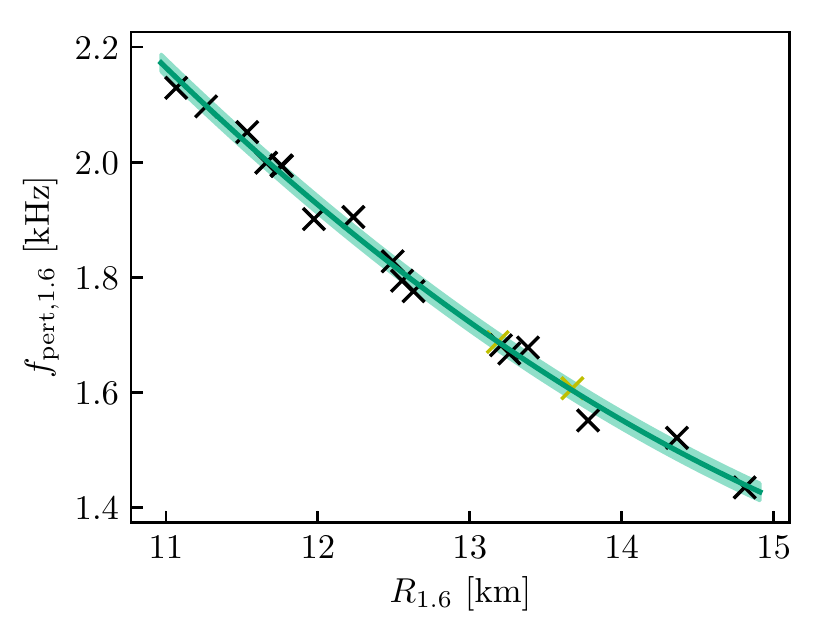}}
    \hfill
    \subfloat[\label{fpert_Lam135_band}]{%
        \includegraphics[width=\columnwidth]{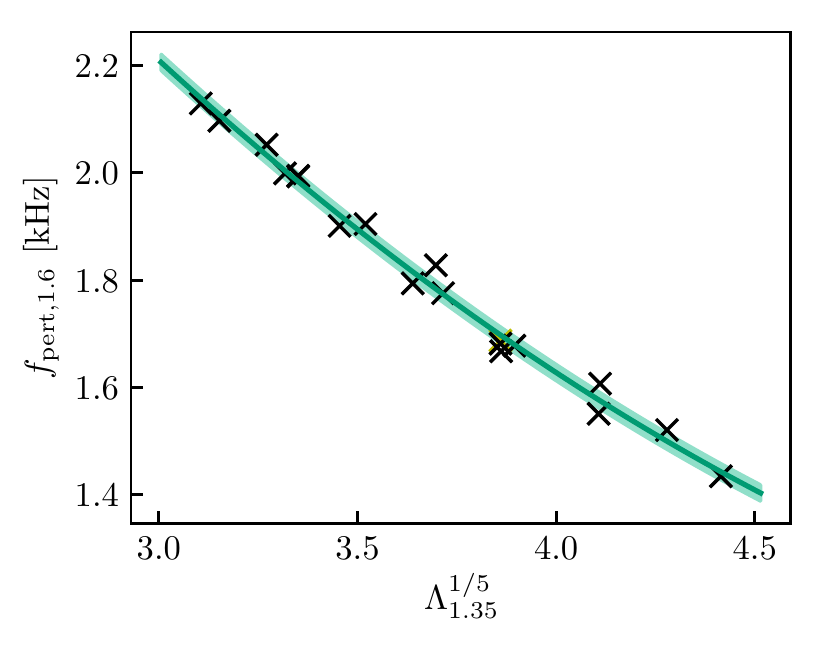}}
        
    \vspace*{8pt}%
    
    \subfloat[\label{Lam16_R16}]{%
         \includegraphics[width=\columnwidth]{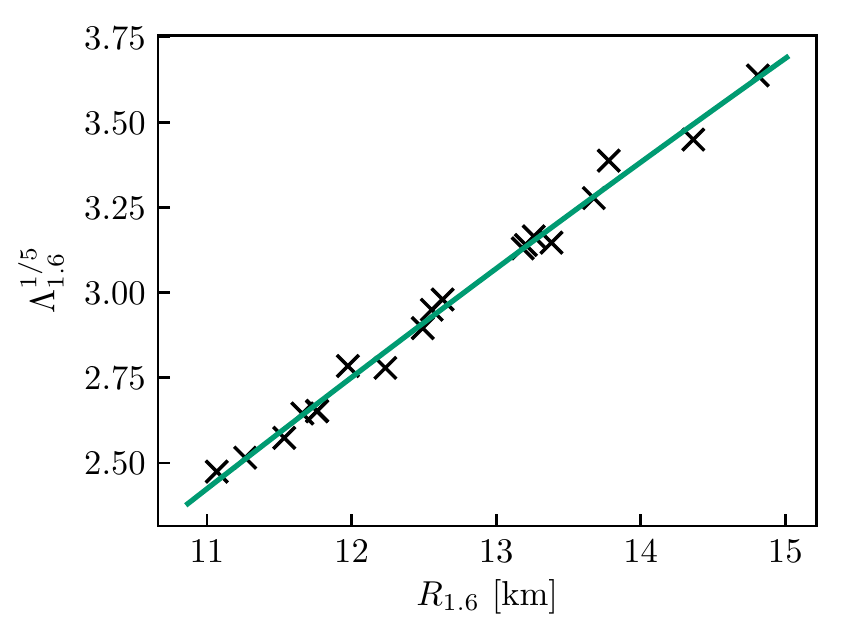}}
    \hfill
    \subfloat[\label{Lam16_Lam135}]{%
        \includegraphics[width=\columnwidth]{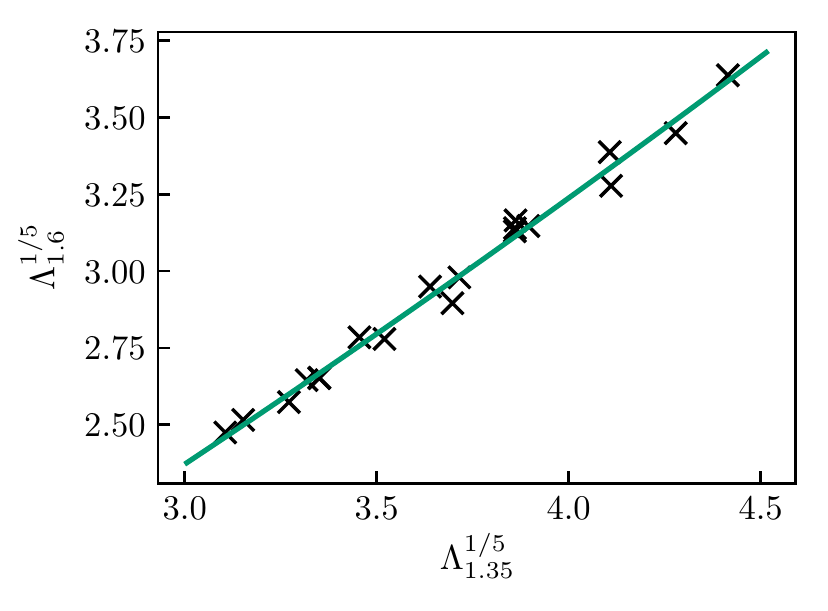}}

    \caption{Panels \protect\subref{fpeak135_R16_band} and \protect\subref{fpert_R16_band} show postmerger frequencies $f_{\mathrm{peak}}$ for $1.35+1.35~M_\odot$ binary systems and perturbative frequencies for $1.6~M_\odot$ stars, $f_{\mathrm{pert,1.6}}$, respectively versus the radius of static $1.6~M_\odot$ stars, $R_{1.6}$. Panels \protect\subref{fpeak135_Lam135_band} and \protect\subref{fpert_Lam135_band} display $f_{\mathrm{peak}}$ for $1.35+1.35~M_\odot$ binary systems and $f_{\mathrm{pert,1.6}}$ as function of $\Lambda^{1/5}_{1.35}$ for static models with $1.35~M_\odot$. Panels \protect\subref{Lam16_R16} and \protect\subref{Lam16_Lam135} provide $\Lambda^{1/5}_{1.6}$ versus $R_{1.6}$ and $\Lambda^{1/5}_{1.35}$, respectively. In all plots the solid curve shows a second-order fit to the data points. We plot a band with a total width of $30$~Hz around frequency fits. See the main text for an explanation of the symbols' colors.
    }\label{fpert_fpeak_R}
\end{figure*}

We now compare in more detail the point scatter in the data from merger simulations to that of perturbation calculations of static stars. In Fig.~\ref{fpert_fpeak_R} we show six plots. We first focus on the comparison between the panels on the left. The upper left panel displays postmerger frequencies for $1.35+1.35~M_\odot$ binary systems versus the radii $R_{1.6}$ of $1.6~M_\odot$ static stars. The middle left panel is a plot of $f_{\mathrm{pert}}$ versus $R_{1.6}$. Hence, we show both frequencies as function of the same independent variable.

We compare how individual points scatter around the fits by examining the location of each data point with respect to the corresponding fits. We depict EoSs which are on the same side of the fit in both plots as black crosses. Remarkably, most data points follow this behavior, altough the frequencies describe very different systems. We use yellow symbols for EoSs which lie on opposite sides of the respective fits in these two plots (upper and middle left panel) and which thus do not follow the systematic behavior. This is a rather strict classification, especially for points which lie relatively close to the fit. For instance, changing the set of EoS models to construct the fit or choosing another functional ansatz for the fit, would lead to another fit function and thus possibly change the character of the deviation. This is obvious for points which are very close to the fit. Hence, one should not classify such points as actual outliers, even if they do not formally fulfill the corresponding quantitative criterion. We thus refine the criterion to identify actual outliers.

We introduce green shaded bands around the fits. They extend $15$~Hz towards both sides of the fits, resulting in a total width of $30$~Hz. EoSs which lie within these bands in both plots are also displayed as black points and are not considered outliers. If however models lie outside these green shaded bands in at least one of the two plots, we mark them with either black or yellow symbols as described above.

Adding such a band is well justified. As shown in Table \ref{Fit_table_Mergers}, the mean deviation is $48$~Hz for the $f_{\mathrm{peak}}(R_{1.6})$ relation. Consequently, most points lie more than $15$~Hz away from the fit and they occur outside the green shaded region at least in the $f_{\mathrm{peak}}-R$ plot. Thus, only a small number of EoSs, lying very close to the fit, is captured by this criterion. For $1.35+1.35~M_\odot$ mergers, only $4$ points out of $19$ lie within the band in both plots with $R_{1.6}$.

Based on the described classification, $17$ out of $19$ EoSs spread in the same way with respect to the fits when comparing $f_{\mathrm{pert}}-R$ and $f_{\mathrm{peak}}-R$ relations shown in Fig.\ \ref{fpert_fpeak_R} (see also Fig.\ \ref{deltafpeak_deltafpert_M16}). We find a similar behavior for systems of other binary masses, which we summarize in Table \ref{Tab_Outliers}. We also refer to the later discussion of Fig.\ \ref{deltafpeak_deltafpert_M16} showing that in fact the deviations of {\it all} data points follow the same trend.

\begin{table*}
\caption{\label{Tab_Outliers}Second and third columns list relations for which we compare frequency deviations w.r.t.\ the fit (see main text). Fourth column provides the number of data points lying on opposite sides of the relations and outside the $30$~Hz band in both plots for the binary system given in the first column.}
\begin{ruledtabular}
\begin{tabular}{cccc}
\multicolumn{1}{c}{\textrm{Binary masses}}&
\textrm{Relation $1$}&
\textrm{Relation $2$}&
\textrm{Number of outliers}\\
\textrm{[$M_\odot$]} & & & \\
\colrule

 $1.2+1.2$ & $f_{pert,1.4}(R_{1.4})$                         & $f_{\mathrm{peak}}(R_{1.4})$               & $1/16$  \\
 $1.2+1.2$ & $f_{pert,1.4}(\Lambda^{1/5}_{1.2})$             & $f_{\mathrm{peak}}(\Lambda^{1/5}_{1.2})$   & $0/16$  \\
 $1.2+1.2$ & $f_{pert,1.4}(R_{1.2})$                         & $f_{\mathrm{peak}}(R_{1.2})$               & $1/16$  \\
 $1.2+1.2$ & $f_{\mathrm{peak}}(\Lambda^{1/5}_{1.2})$        & $f_{\mathrm{peak}}(R_{1.2})$               & $1/16$  \\  \vspace{0.5cm}
 $1.2+1.2$ & $f_{\mathrm{peak}}(R_{1.4})$                    & $f_{\mathrm{peak}}(R_{1.2})$               & $1/16$  \\

 $1.35+1.35$ & $f_{pert,1.6}(R_{1.6})$                       & $f_{\mathrm{peak}}(R_{1.6})$               & $2/19$  \\
 $1.35+1.35$ & $f_{pert,1.6}(\Lambda^{1/5}_{1.35})$          & $f_{\mathrm{peak}}(\Lambda^{1/5}_{1.35})$  & $1/19$  \\
 $1.35+1.35$ & $f_{pert,1.6}(R_{1.35})$                      & $f_{\mathrm{peak}}(R_{1.35})$              & $2/19$  \\
 $1.35+1.35$ & $f_{\mathrm{peak}}(\Lambda^{1/5}_{1.35})$     & $f_{\mathrm{peak}}(R_{1.35})$              & $1/19$  \\  \vspace{0.5cm}
 $1.35+1.35$ & $f_{\mathrm{peak}}(R_{1.6})$                  & $f_{\mathrm{peak}}(R_{1.35})$              & $1/19$  \\
 
 $1.4+1.4$ & $f_{pert,1.7}(R_{1.7})$                         & $f_{\mathrm{peak}}(R_{1.7})$               & $3/16$  \\
 $1.4+1.4$ & $f_{pert,1.7}(\Lambda^{1/5}_{1.4})$             & $f_{\mathrm{peak}}(\Lambda^{1/5}_{1.4})$   & $2/16$  \\
 $1.4+1.4$ & $f_{pert,1.7}(R_{1.4})$                         & $f_{\mathrm{peak}}(R_{1.4})$               & $2/16$  \\
 $1.4+1.4$ & $f_{\mathrm{peak}}(\Lambda^{1/5}_{1.4})$        & $f_{\mathrm{peak}}(R_{1.4})$               & $1/16$  \\  \vspace{0.5cm}
 $1.4+1.4$ & $f_{\mathrm{peak}}(R_{1.7})$                    & $f_{\mathrm{peak}}(R_{1.4})$               & $2/16$  \\
 
 $1.5+1.5$ & $f_{pert,1.75}(R_{1.75})$                       & $f_{\mathrm{peak}}(R_{1.75})$              & $1/6$   \\
 $1.5+1.5$ & $f_{pert,1.75}(\Lambda^{1/5}_{1.5})$            & $f_{\mathrm{peak}}(\Lambda^{1/5}_{1.5})$   & $0/6$   \\
 $1.5+1.5$ & $f_{pert,1.75}(R_{1.5})$                        & $f_{\mathrm{peak}}(R_{1.5})$               & $0/6$   \\
 $1.5+1.5$ & $f_{\mathrm{peak}}(\Lambda^{1/5}_{1.5})$        & $f_{\mathrm{peak}}(R_{1.5})$               & $0/6$   \\
 $1.5+1.5$ & $f_{\mathrm{peak}}(R_{1.75})$                   & $f_{\mathrm{peak}}(R_{1.5})$               & $2/6$   \\
\end{tabular}
\end{ruledtabular}
\end{table*}

Finally, we extend the comparison by considering data as function of $\Lambda^{1/5}_{1.35}$ in the right panels of  Fig.\ \ref{fpert_fpeak_R}. We show postmerger frequencies and perturbation frequencies of static stars in the upper and middle right panels of Fig.\ \ref{fpert_fpeak_R} with $\Lambda^{1/5}_{1.35}$ as independent variable. Note that we employ the perturbative frequency of a more massive star with $1.6~M_\odot$ (as in the middle left panel of Fig.\ \ref{fpert_fpeak_R}). Frequencies deviate in the same way in both relations (only one outlier). We also notice a very similar distribution of data points in the upper left and upper right panel as well as in the middle panels, i.e.\ in all four plots. A similar behavior is found by comparing the panels in Fig.\ \ref{AB_plots}.

We summarize the different comparisons in Table \ref{Tab_Outliers} employing the same criterion as described above to quantify the behavior of the scatter in these relations. We consider additional pairs of relations in Table \ref{Tab_Outliers} and determine the number of outliers for each of them. Throughout all pairs of relations and binary masses the number of outliers is very small. This corroborates our observation that data points referring to two different systems scatter in a similar way (see also Fig.\ \ref{deltafpeak_deltafpert_M16}).

The agreement is even more pronounced in cases where the independent variable ($R$ or $\Lambda^{1/5}$) refers to static stars with the same mass as the inspiraling stars. In these plots the data points on average deviate more from the respective fit. Hence, the location of data points with respect to the fit is less sensitive to small changes of the fit and the similarities in the frequency deviations become more evident for overall larger deviations. This further substantiates the observation that the location of individual data points with respect to the fits, which represent some kind of average behavior, follows a systematic pattern determined by the EoS.

In a broader sense, we find that in fact all points behave consistently in plots like Fig.\ \ref{fpert_fpeak_R}. Considering for instance clusters of points, we recognize very similar patterns of the distribution of points in the corresponding plots. This general consistency between the behavior in both sets of frequency data is indeed remarkable, considering statistical fluctuations and uncertainties, which stem from the complexity of merger simulations. 

The fact that we consider a large number of EoSs for different binary masses makes the observation even more remarkable. The fits are based on a significant number of EoS models, which essentially cover the full viable range in the $M-R$ diagram and arguably somewhat beyond. Including a few additional EoSs will not significantly alter the fit and thus will not strongly affect the distribution of the current data points with respect to it.

We emphasize once more that the agreement of frequency deviations with respect to the fits in Fig.\ \ref{fpert_fpeak_R} and Table \ref{Tab_Outliers} is very remarkable and by no means expected. $f_{\mathrm{pert}}$ refers to the frequencies from pertubative calculations of static, non-rotating stars with a mass of 1.6$~M_\odot$, whereas $f_{\mathrm{peak}}$ frequencies describe the dominant oscillation mode of rapidly rotating, hot merger remnants of significantly higher mass, which actually still undergo a dynamical evolution while $f_{\mathrm{peak}}$ is extracted. We would like to make two further remarks. 

\begin{enumerate}
 \item Notably, the merger frequencies are obtained from a three-dimensional relativistic hydrodynamical simulation code, which is computationally much more complex than solving the equations of linearized perturbations around a background equilibrium model (see Sections \ref{SubSecPertdata} and \ref{SubSecMergerdata}). Clearly, the latter code, in comparison, yields more robust and accurate results. Therefore, it is generally encouraging that the hydrodynamical simulations with the current resolution are apparently able to uncover the frequencies to a degree that the frequency deviations resolve some underlying physics. This does not necessarily mean that the accuracy of about $10$~Hz, i.e. the level of frequency deviations, reflects the full systematic uncertainties involved in the numerical model nor that the frequencies are fully converged with respect to the numerical resolution. This said, we comment that data points which do not follow the described behavior (yellow symbols), may well be attributed to numerical artifacts since the quoted frequency accuracy is certainly on the edge of what a code of this type can achieve. However, we argue below that also the outliers behave in some way consistently.
 
 \item The striking similarity of frequency deviations very likely points to an underlying mechanism responsible for the frequency shift in a certain direction. This implies that the frequency deviation on its own encodes additional information about the EoS, which is the only link between the two systems. In the following section we further investigate this point and identify which EoS properties, or equivalently NS parameters, are causing the frequencies to deviate in a certain way. We stress that, at least in principle, the deviations may be measurable. If the fits can be constructed based on simulations with sufficient precision, measurements of the frequency and the respective independent quantity inform about the frequency deviation from the fit. The radius or tidal deformability could be obtained  either from independent measurements or from the very same merger event providing $f_{\mathrm{peak}}$. Clearly, these ideas require a high measurement accuracy. It may also be possible that the frequency deviations correlate with other features of the GW signal of a NS merger. We note that secondary frequencies apparently deviate in the same way as the main peak (see Fig.\ 6 in \cite{Bauswein2015}).
\end{enumerate} 

We also remark that additional simulations for DD2F and SFHX with  binary mass ratios \footnote{We define the mass ratio as $q=M_1/M_2\le1$, where $M_1$ and $M_2$ are the masses of the individual stars in the binary system.} of $q=0.95$ and $q=0.9$ yield frequencies very similar to the ones from the equal-mass binary of the same total mass of 2.7~$M_\odot$ (for $q=0.95$ and $q=0.9$ the frequencies deviate from the respective equal-mass models by some 10~Hz, which seems to be dominated by statistical fluctuations). This suggests that at least within a relatively small range of $q$ the influence of the EoS on the frequency deviations is stronger than that of the mass ratio.

Finally, we assess the robustness of the frequency deviations performing additional simulations for 1.35-1.35~$M_\odot$ binaries with the DD2F and SFHX EoSs. Both EoS models result in a comparable $R_{1.6}$ of about 12~km. DD2F yields a frequency increase w.r.t.\ the fit, whereas SFHX results in a smaller $f_\mathrm{peak}$. We perform additional simulations with larger orbital separations (resulting in 4 and 5 orbits before merging respectively) and find that for both EoS models the resulting $f_\mathrm{peak}$ deviate as expected w.r.t.\ the fit. For DD2F we also run calculations with different resolution (about 100,000 and 600,000 SPH particles instead of our default choice of 300,000 particles) and observe that the frequency differences in simulations with different resolution are smaller than the frequency deviations from the fit. Although the different setups lead to small statistical fluctuations in $f_\mathrm{peak}$ of a few 10~Hz, the differences in $f_\mathrm{peak}$ between the two EoSs are larger than those fluctuations. Most importantly, calculations with another setup lie on the correct side of the fit generated by our standard setup. We do not find systematic trends with the SPH particle number or the initial orbit. We are thus confident that the frequency shifts are in fact mostly caused by the physics of the underlying EoS. We finally compare our data for 1.35-1.35~$M_\odot$ binaries to another set of simulations which we performed for 1.37-1.37~$M_\odot$ binaries with the same set of EoS models. In these calculations we used another SPH kernel function (the $C^6$ Wendland kernel function~\cite{Wendland1995, Dehnen2012}). Again we find that the individual data points scatter in the exact same way from a respective fit to the full data. This demonstrates the robustness against certain details of the numerical scheme. Similarly, the fact that a setup with a slightly different binary mass leads to the same pattern of frequency deviations reassures that frequency shifts are insensitive to numerics and represent a physical effect of the EoS. In conclusion, the numerical treatment does to some extent affect the exact frequencies, but we employ the same numerical setup for our calculations. We thus expect that the pattern of the frequency deviations is insensitive to the exact treatment and only suffers from statistical fluctuations which are apparently smaller than the typical frequency shifts.

\subsection{Physical explanation for frequency deviations and encoded EoS information}\label{SubSec_Physical_Explanation}
The fact that points scatter in a similar way in frequency versus radius/$\Lambda$ plots, for two very different systems, suggests that there is a physical reason behind it. It is clear that the EoS determines where individual points occur in the diagram. In order to investigate this aspect, we focus on perturbative frequencies. We already discussed that $f_{\mathrm{pert}}-R_x$ and $f_{\mathrm{peak}}-R_x$ diagrams show very similar patterns. The perturbative data refer to a simpler system, which is why we expect that relations for $f_{\mathrm{pert}}$ are more accurate and reliable. Hence, they are better suited to identify what causes points to occur at a certain location.

\begin{figure}
    \includegraphics{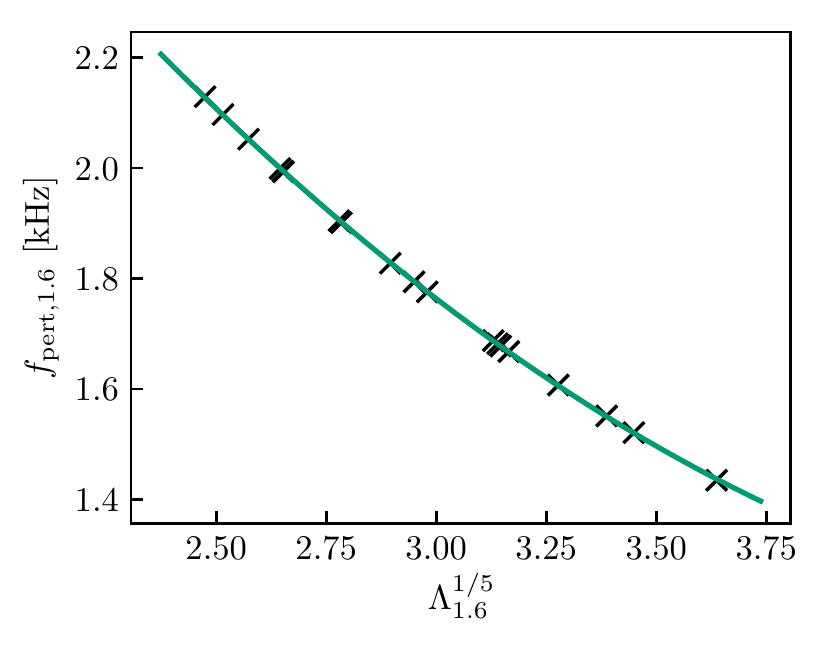}
    \caption{ $f_{\mathrm{pert}}$ versus $\Lambda^{1/5}_{1.6}$ for non-rotating $1.6~M_\odot$ stars. There is minimal scatter with a maximum deviation of $2.2$~Hz.}
    \label{MfLam_pointdist}
\end{figure}

For static stars with different masses there is a very tight relation between the mass-scaled $f_{\mathrm{pert}}$ and $\Lambda^{-1/5}$ (see Section~\ref{Sec_stat_stars}, Fig.~\ref{TidalDefFreq_Static} and Table~\ref{Fit_table}). The corresponding relation for a {\it fixed} mass shown in Fig.~\ref{MfLam_pointdist} is very tight with a maximum deviation of only $2.2$~Hz for $1.6~M_\odot$ stars (see Table \ref{Fit_table_fixed_mass}). We find a similarly high accuracy for relations with other fixed masses. Hence, one can consider $f_{\mathrm{pert}}$ and $\Lambda^{-1/5}$ being practically equivalent.

\begin{table*}
\caption{\label{Fit_table_fixed_mass}Relations between $f-$mode frequencies or tidal deformabilities $\Lambda^{1/5}$ of static stars and different stellar parameters for a fixed mass. Third and fourth column provide the average and maximum deviation between fit and underlying data. Frequencies are in kHz, radii in km and the tidal deformability $\Lambda$ is dimensionless. Deviations for relations involving $f_{pert,1.6}$ are in Hz, while deviations for the relation between the tidal deformabilities are dimensionless.}
\begin{ruledtabular}
\begin{tabular}{clcc}
\textrm{Position}&
\multicolumn{1}{c}{\textrm{Fit}}&
\textrm{Mean dev.}&
\textrm{Max dev.} \\
 & & \textrm{[Hz]} & \textrm{[Hz]}\\ 
\colrule \vspace{0.2cm}
Fig.\ (\ref{fpert_R16_band})     & $f_{pert,1.6} = 7.04  -0.631R_{1.6}                                   +1.708\times10^{-2}R^2_{1.6}$            & $15$ & $36$ \\  \vspace{0.1cm}
Fig.\ (\ref{fpert_Lam135_band})  & $f_{pert,1.6} = 5.11 -1.255\Lambda^{1/5}_{1.35}                       +9.618\times10^{-2}\Lambda^{2/5}_{1.35}$ & $17$ & $45$ \\  \vspace{0.1cm}
Fig.\ (\ref{MfLam_pointdist})    & $f_{pert,1.6} = 4.988 -1.539\Lambda^{1/5}_{1.6}                       +1.546\times10^{-1}\Lambda^{2/5}_{1.6}$  & $1$ & $2.2$ \\
Fig.\ (\ref{Lam16_Lam135})       & $\Lambda^{1/5}_{1.6} = 0.205 +0.614\Lambda^{1/5}_{1.35} +0.036\Lambda^{2/5}_{1.35}$                            & $0.073$   & $0.029$
\end{tabular}
\end{ruledtabular}
\end{table*}

Comparing Figs.~\ref{fpert_R16_band} and~\ref{MfLam_pointdist} we notice a drastically different distribution of points. In the plot involving the radius the points significantly scatter around the respective relation. Employing the tidal deformability instead the data points hardly exhibit any scatter~\footnote{We remark that the frequency deviations are not related to the low-density regime of the EoS, which affects radii stronger than the tidal deformability. In Section \ref{Sec_stat_stars}, we introduce a newly defined radius, $R^{90\%}$, such that it is insensitive to the EoS at lower densities, and we find significantly tighter relations between $f_{\mathrm{pert}}$ and this new measure. Still, the relations feature a sizable point-to-point scatter, from which we conclude that it does not entirely result from the low-density EoS. In this context the term ``low-density'' thus refers to the material in the outer shell of the star containing $10\%$ of its total mass. In addition, we extract the radius $R^{\mathrm{cc}}$ based on the crust-core transition density. Relations involving this radius also exhibit scatter, which further supports the argument that frequency deviations are influenced by the high-density EoS.}.

The comparison between these two figures indicates which EoS properties cause the frequency deviations. Figure \ref{MfLam_pointdist} shows that there is an essentially exact relation between $f_{\mathrm{pert}}$ and $\Lambda^{1/5}$ (for fixed mass) meaning that $f_{\mathrm{pert}}$ can be equivalently replaced by $\Lambda^{1/5}$ in the relations in Fig.\ \ref{fpert_fpeak_R}. This implies that deviations in the $f_{\mathrm{pert},1.6}-R_{1.6}$ plot (panel \subref{fpert_R16_band} in Fig.\ \ref{fpert_fpeak_R}) are tightly anti-correlated with deviations in a $\Lambda^{1/5}_{1.6}-R_{1.6}$ diagram (panel \subref{Lam16_R16} in Fig.\ \ref{fpert_fpeak_R}; compare also panel \subref{fpeak135_R16_band} and \subref{Lam16_R16}).

\begin{figure}
    \includegraphics[width=\columnwidth]{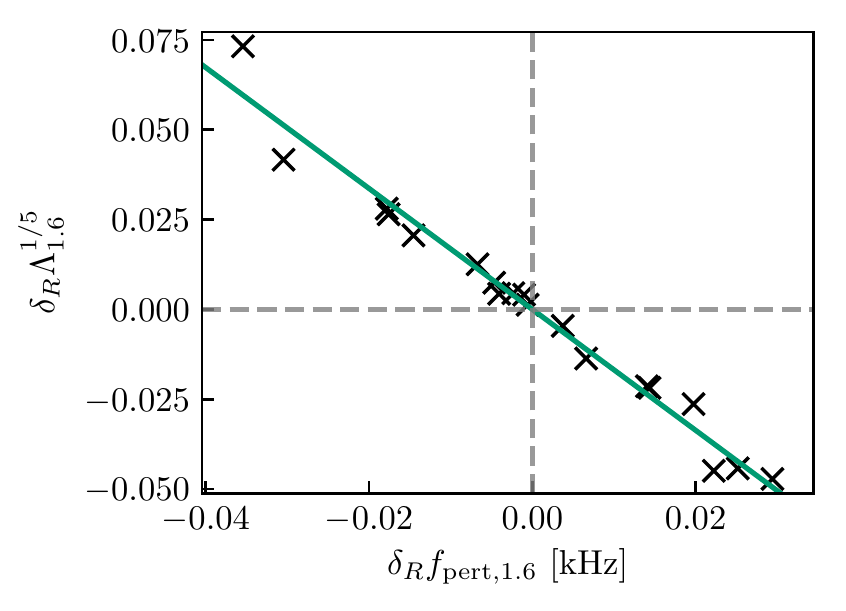}
    \caption{Deviations $\delta_R \Lambda^{1/5}$ between data points and a second-order fit in a $\Lambda^{1/5}_{1.6}-R_{1.6}$ diagram versus frequency deviations $\delta_R f_{\mathrm{pert,1.6}}$ in a $f_{\mathrm{pert}}(R_{1.6})$ relation (see panels \protect\subref{Lam16_R16} and \protect\subref{fpert_R16_band} of Fig.\ \ref{fpert_fpeak_R} respectively). Solid line displays a first-order fit to the data.}
    \label{deltaLam_deltafpert_R16}
\end{figure}

In Fig.\ \ref{deltaLam_deltafpert_R16} we verify that this is indeed the case for $1.6~M_\odot$ stars. We define deviations in terms of frequencies, denoted by $\delta_R f_{\mathrm{pert}}$, between data points and the second-order fit in panel \subref{fpert_R16_band} of Fig.\ \ref{fpert_fpeak_R}. Similarly, deviations in terms of $\Lambda^{1/5}_{1.6}$, denoted as $\delta_R \Lambda^{1/5}$ in Fig.\ \ref{deltaLam_deltafpert_R16}, are defined between data points and a second-order $\Lambda^{1/5}_{1.6}(R_{1.6})$ fit. The deviations are strongly anti-correlated and follow a linear trend. We find a similar behavior of the deviations for any other fixed mass within the mass range of static stars considered here.

The fact that deviations in $f_{\mathrm{pert}}(R_{1.6})$ and $\Lambda^{1/5}_{1.6}(R_{1.6})$ are tightly correlated implies that we can trace back and explain frequency deviations in $f_{\mathrm{pert}}$, and ultimately in $f_{\mathrm{peak}}$, by the difference between $R_{1.6}$ and $\Lambda^{1/5}_{1.6}$ (Fig.\ \ref{Lam16_R16} exhibits the same pattern of deviations as Fig.\ \ref{fpert_R16_band} and similarly for panels \subref{Lam16_Lam135} and \subref{fpert_Lam135_band}).

Therefore, the frequency deviations, i.e.\ the scatter in frequency plots in Fig.\ \ref{fpert_fpeak_R}, are directly linked to the tidal love number $k_2$, which describes the difference between $\Lambda$ and $R$ through $\Lambda=\frac{2}{3}k_2(\frac{c^2R}{GM})^5$. $k_2$ is known to \textit{roughly} correlate with the inverse compactness ($R/M$) (for instance the relation $\Lambda\simeq\alpha(\frac{c^2R}{GM})^6$ with $\alpha=0.0093\pm0.0007$ in \cite{2018PhRvL.121i1102D} implies an average $k_2^{\mathrm{av}}=\frac{3}{2}\alpha(\frac{c^2R}{GM})$). More specifically, the scatter in $\Lambda^{1/5}_{1.6}(R_{1.6})$, and thus the frequency deviations, are determined by how much $k_2$ deviates from an average $k_2^{\mathrm{av}}$ estimated based on the compactness. We thus directly link the frequency scatter to the detailed behavior of $k_2$. This in turn implies that observational constraints on the frequency deviation, possibly only its sign, informs about properties of $k_2$ e.g.\ by how much it deviates from an average $k_2$ given by the compactness and can be employed to break the degeneracy between $\Lambda$, $k_2$ and $R$.

\begin{figure}
    \includegraphics[width=\columnwidth]{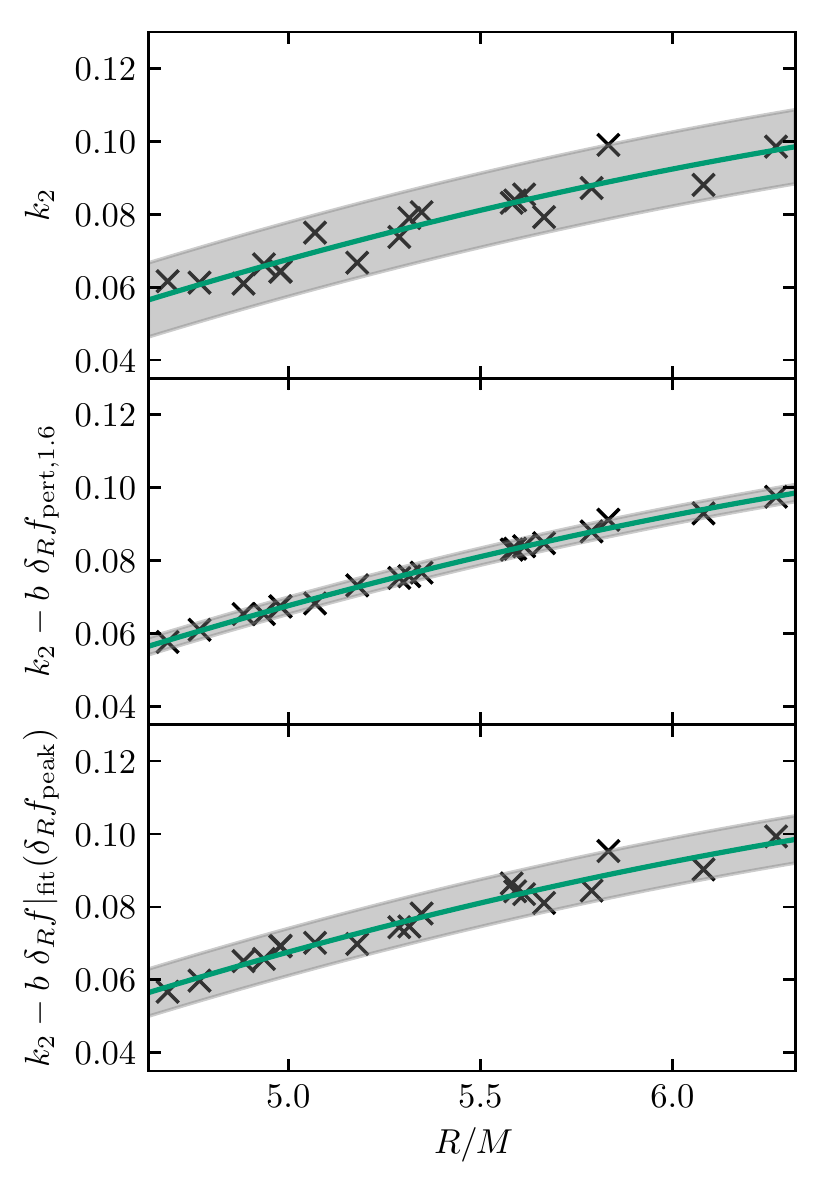}
    \caption{Upper panel shows $k_2$ as a function of $R/M$ for $1.6~M_\odot$ static stars. Middle panel displays ``corrected'' tidal Love number $k_2-b~\delta_R f_{\mathrm{pert,1.6}}$ as function of $R/M$. Bottom panel presents ``corrected'' $k_2$ using $\delta_R f_{\mathrm{peak}}$ values via a first-order fit between $\delta_R f_{\mathrm{pert,1.6}}$ and $\delta_R f_{\mathrm{peak}}$ (see Fig.\ \ref{deltafpeak_deltafpert_M16}). In the middle and bottom panels $b=-0.2206~\mathrm{kHz}^{-1}$. The gray shaded band represents the maximum deviation in each panel. Solid curve is a second-order fit to $k_2(R/M)$ and identical in all panels.}
    \label{k2_k2cor_InvComp_M16}
\end{figure}

In the upper panel of Fig.\ \ref{k2_k2cor_InvComp_M16} we plot $k_2$ versus $R/M$ for $1.6~M_\odot$ static models. We include a second-order fit to the data. Data points \textit{roughly} follow the fit, but they partially exhibit visible deviations from it. The gray shaded band shows the maximum deviation in each panel. As argued, the deviations are related to frequency deviations $\delta_R f_{\mathrm{pert}}$.

Following the above reasoning about the equivalence between frequency deviations $\delta_R f_{\mathrm{pert}}$ and differences between $\Lambda_{1.6}$ and $R_{1.6}$, we introduce a correction to $k_2$, which is proportional to $\delta_R f_{\mathrm{pert}}$. We obtain the proportionality constant $b$ by a single fit to the deviations in the upper panel in Fig.\ \ref{k2_k2cor_InvComp_M16}. The resulting relation is shown in the middle panel, which includes the same second-order fit from the top panel and exhibits a very tight correlation of the corrected $k_2-b\;\delta_R f_{\mathrm{pert}}$ with $R/M$. For this figure we find $b=-0.2206~\mathrm{kHz}^{-1}$ and observe a similar behavior for other masses in the range $1.1-1.9~M_\odot$. Obviously, we can also include the correction in $R/M$ by changing the independent variable, which becomes $R/M-b'\;\delta_R f_{\mathrm{pert}}$\footnote{Note that $b'\ne b$. In order to obtain it one needs to quantify horizontal deviations of data points from the fit in the upper panel of Fig. \ref{k2_k2cor_InvComp_M16}. Fitting $\delta_R f_{\mathrm{pert,1.6}}$ to these deviations determines $b'$.} and directly determines $k_2$.

\begin{figure}
    \includegraphics[width=\columnwidth]{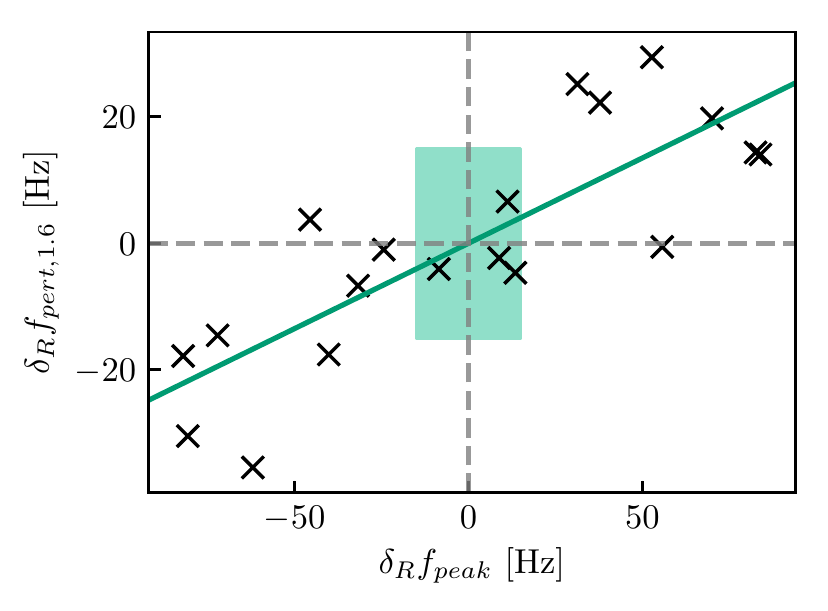}
    \caption{Frequency deviations as occurring in panels \protect\subref{fpeak135_R16_band} and \protect\subref{fpert_R16_band} of Fig.\ \ref{fpert_fpeak_R} respectively. The green shaded box has a side length of $30$~Hz and matches the bands introduced in Fig.\ \ref{fpert_fpeak_R}. The solid green line is a first-order fit to the data given by $\delta_R f_{\mathrm{pert,1.6}} = 0.2697~\delta_R f_{\mathrm{peak}}$.}
    \label{deltafpeak_deltafpert_M16}
\end{figure}

In Fig.\ \ref{deltafpeak_deltafpert_M16} we plot $\delta_R f_{\mathrm{peak}}$, the deviation of data points from the fit in terms of postmerger frequencies in panel \subref{fpeak135_R16_band} of Fig.\ \ref{fpert_fpeak_R}, versus $\delta_R f_{\mathrm{pert,1.6}}$. The green shaded area is a box with a width of $30$~Hz, which matches the band we introduced in Fig.\ \ref{fpert_fpeak_R}. The solid green line is a first-order fit to the data, which we refer to as $\delta_R f|_{\mathrm{fit}}$. Evidently, the points approximately follow the line, which can be used to obtain an estimate for $\delta_R f_{\mathrm{pert}}$ based on $\delta_R f_{\mathrm{peak}}$. This estimate can then be employed to obtain a better estimate for $k_2$, as shown in the bottom panel of Fig.\ \ref{k2_k2cor_InvComp_M16}. Data points shifted by $b~\delta_R f|_{\mathrm{fit}} (\delta_R f_{\mathrm{peak}})$ deviate less from the $k_2(R/M)$ fit with the average and maximum deviation reduced by $33\%$ and $36\%$ respectively. In particular, the improvement is significant for most points, especially those with $R/M<5.5$. A single point with $R/M\simeq 5.83$ is the only one which still arguably deviates from the fit. Furthermore, we find that using deviations defined on $f_{\mathrm{peak}}$ versus radius plots for various different choices of the mass to which the radius refers, also leads to improved relations for $k_2$. In particular, $\delta_{R_{1.35}} f_{\mathrm{peak}}$ produces even better results than the bottom panel of Fig.\ \ref{k2_k2cor_InvComp_M16}, which is rather interesting as $R_{1.35}$ can potentially be extracted from the analysis of the inspiral.

Although estimating $\delta_R f_{\mathrm{pert}}$ through the linear fit is not accurate and measuring $\delta_R f_{\mathrm{peak}}$ may be challenging, knowing whether $\delta_R f_{\mathrm{pert}}$ is positive or negative is already useful: The sign of $\delta_R f_{\mathrm{pert}}$ informs whether the corresponding point lies above or below the respective $k_2(R/M)$ fit. This suffices to reduce the error in determining $k_2$ through the respective fit by half. In this context we recall that frequencies of postmerger oscillations can be recovered with $\sim10$~Hz accuracy with sufficient signal-to-noise ratio (SNR) with future ground-based detector configurations, which has been shown by simulated injections \cite{2014PhRvD..90f2004C, 2017PhRvD..96l4035C}. Hence, the prospects to infer frequency deviations rely mostly on the challenge to construct by calculations reliable theoretical relations between frequency and TOV properties, to which measured frequencies can be compared.

In summary, these relations show that $\delta_R f_{\mathrm{pert}}$ or $\delta_R f_{\mathrm{peak}}$ can be used for a more accurate estimate of $k_2$ (beyond a relation with the compactness C) and thus to establish the exact relationship between tidal deformability and radius, which is for instance important for EoS constraints from the GW inspiral. As we already mentioned quantifying the exact frequency deviation for the merger data is challenging and may also explain the few outliers in Fig.\ \ref{fpert_fpeak_R} and Table \ref{Tab_Outliers}. In this respect we also refer to Fig.\ \ref{deltafpeak_deltafpert_M16}, where one can clearly see that in fact {\it all} data points do follow the same trend including the two outliers. This exemplifies that our criterion for defining outliers above is arguably too conservative and could in principle be replaced by a better classification scheme. At any rate, the consistent behavior of all data points in Fig.\ \ref{deltafpeak_deltafpert_M16} corroborates our observation that frequency deviations are correlated.

\subsection{Frequency deviations and the tidal deformability of high-mass neutron stars} 
Finally, we connect frequency deviations with the behavior of the tidal deformability $\Lambda(M)$ as function of mass. In Fig.~\ref{fpert_Lam135_band} we plot the perturbative frequency $f_{\mathrm{pert,1.6}}$ versus the tidal deformability $\Lambda_{1.35}^{1/5}$. However, as already discussed, $f_{\mathrm{pert,1.6}}$ scales very tightly with the tidal deformability of the stellar system with the same mass $\Lambda_{1.6}^{1/5}$ (see Fig.\ \ref{MfLam_pointdist}). Hence, Fig.~\ref{fpert_Lam135_band} practically displays the relation between $\Lambda^{1/5}$ referring to two  different masses, namely $\Lambda_{1.35}^{1/5}$ and $\Lambda_{1.6}^{1/5}$. The difference between these two values of $\Lambda^{1/5}$ approximates the derivative of $\Lambda^{1/5}$ w.r.t.\ the mass. 

\begin{figure}
    \includegraphics{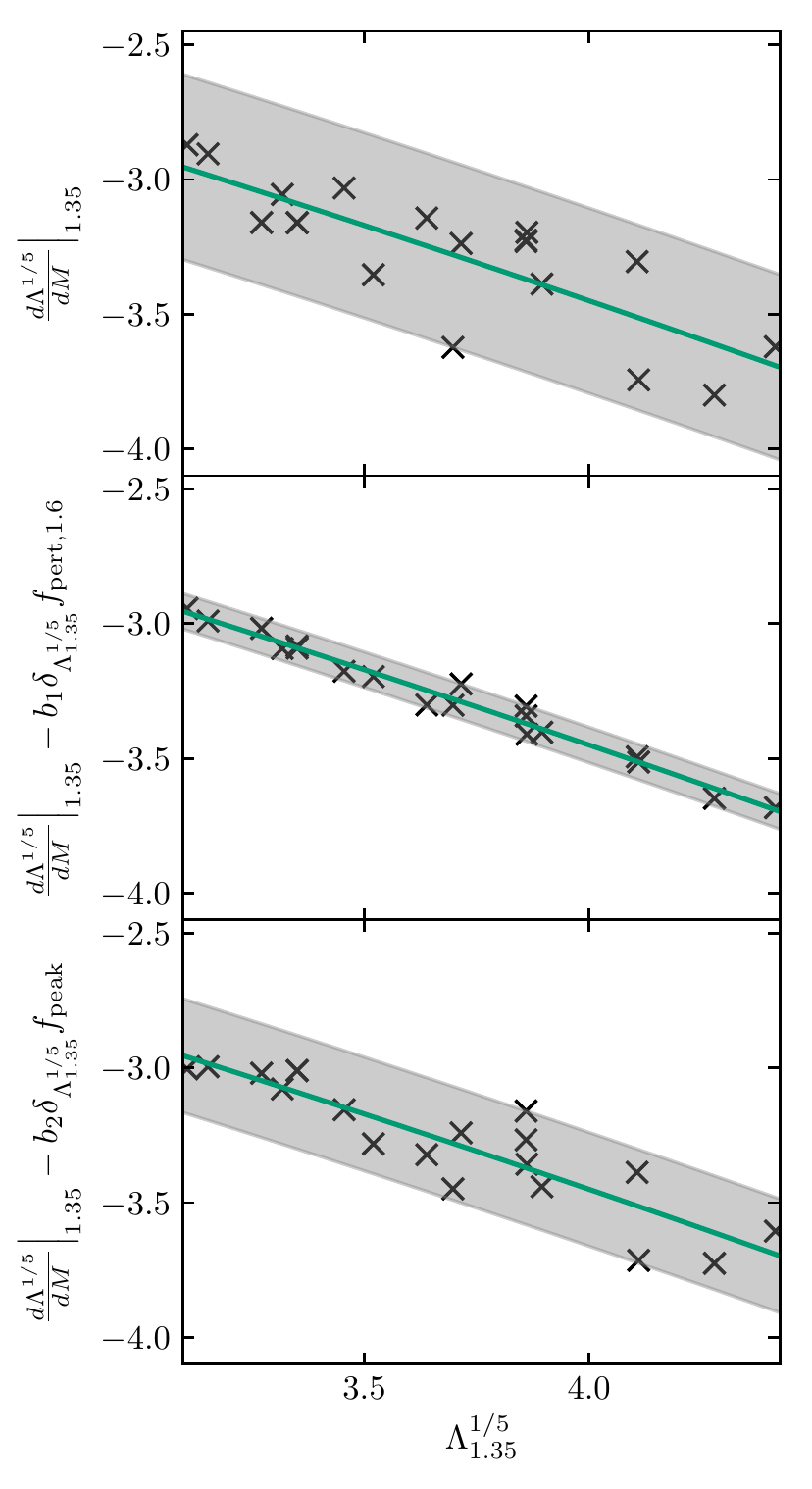}
    \caption{Upper panel shows the derivative $d\Lambda^{1/5}/dM$ at a fixed mass equal to $1.35~M_\odot$ as a function of $\Lambda^{1/5}_{1.35}$. Middle panel presents a "corrected" derivative using the deviations $\delta_{\Lambda^{1/5}_{1.35}} f_{\mathrm{pert,1.6}}$. Bottom panel displays a corrected derivative through deviations $\delta_{\Lambda^{1/5}_{1.35}} f_{\mathrm{peak}}$ (see Fig. \ref{fpeak135_Lam135_band}). The values of the fit parameters are $b_1=-7.293~\mathrm{kHz}^{-1}$ and $b_2=-2.029~\mathrm{kHz}^{-1}$ respectively. The gray shaded area represents the maximum deviation in each panel. Solid curve is a second-order fit to the data points in the upper panel and identical in all panels.}
    \label{dLamdM_corrections_panels}
\end{figure}

The upper panel of Fig.\ \ref{dLamdM_corrections_panels} shows the derivative $d\Lambda^{1/5}/dM$ at $M=1.35~M_\odot$ versus $ \Lambda_{1.35}^{1/5}$. The data points in the upper panel follow a coarse trend described by a fit (green curve), but they exhibit some sizable scatter because the derivative may still be different for the same $\Lambda_{1.35}^{1/5}$. However, it is clear that the tidal deformability of some higher mass NS does carry information about the behavior of the slope of $\Lambda(M)$. We thus anticipate that the frequency deviations in Fig.~\ref{fpert_Lam135_band} (or any other frequency deviation correlated to it like $\delta_{\Lambda^{1/5}_{1.35}} f_{\mathrm{peak}}$) can be employed to remove the significant scatter in the upper panel.

Following a very similar procedure as in Fig.~\ref{k2_k2cor_InvComp_M16} the additional information encoded in the frequency deviations can be included. Employing either $\delta_{\Lambda^{1/5}_{1.35}} f_{\mathrm{pert,1.6}}$ or $\delta_{\Lambda^{1/5}_{1.35}} f_{\mathrm{peak}}$ leads to tighter relations for the derivative $d\Lambda^{1/5}/dM$ (see middle and bottom panel of Fig.\ \ref{dLamdM_corrections_panels} respectively). In particular, in the case of $\delta_{\Lambda^{1/5}_{1.35}} f_{\mathrm{pert,1.6}}$ the accuracy of the relation improves significantly. The maximum deviation is reduced by $80\%$. 

The importance of Fig.\ \ref{dLamdM_corrections_panels} is that the observation of a single BNS is in principle sufficient to determine both the tidal deformability and its derivative w.r.t.\ mass. This means the properties of $\Lambda(M)$ at higher masses are accessible without explicitly measuring the tidal deformability at higher masses if information on the frequency deviations is available (possibly from the same event). 

We remark that it is not strictly necessary to pick the mass $1.6~M_\odot$ for the deviations $\delta_{\Lambda^{1/5}_{1.35}} f_{\mathrm{pert,1.6}}$. In principle measuring $f_{\mathrm{pert}}$ of any mass can be used to obtain a corrected value for the derivative. In practice, values closer to $1.35~M_\odot$ (or generally the mass of the inspiraling stars) may even lead to a more significant improvement. We also note that similar figures can be obtained for other binary masses. Furthermore, we comment that the reasoning in this subsection may also be reversed. It may be conceivable to use information on the derivative $d\Lambda^{1/5}/dM$, e.g. from measuring $\Lambda$ in two BNS events with different mass, to provide a more accurate prediction of the postmerger frequency.

\section{Direct relations between frequencies of static stars and merger remnants}\label{SecTheorRel}
Both perturbative frequencies $f_{\mathrm{pert}}$ and postmerger frequencies $f_{\mathrm{peak}}$ scale tightly with stellar parameters of static stars such as the radius $R$ and the tidal deformability $\Lambda$ (see Figs.\ \ref{AB_plots}, \ref{fpert_fpeak_R} and \ref{MfLam_pointdist}). Furthermore, as discussed in Sections \ref{PointScatterObs} and \ref{SubSec_Physical_Explanation}, data points deviate from such relations in a very similar way for $f_{\mathrm{pert}}$ and $f_{\mathrm{peak}}$. This implies that there should also exist a direct correlation between the $f-$mode frequency $f_{\mathrm{pert}}$ and the dominant postmerger oscillation frequency $f_{\mathrm{peak}}$. One may expect such relations to become particularly tight, because the frequency deviations in $f_{\mathrm{pert}}$ and $f_{\mathrm{peak}}$, which we found to be correlated, may to some extent cancel/compensate each other.

\begin{figure}
    \includegraphics{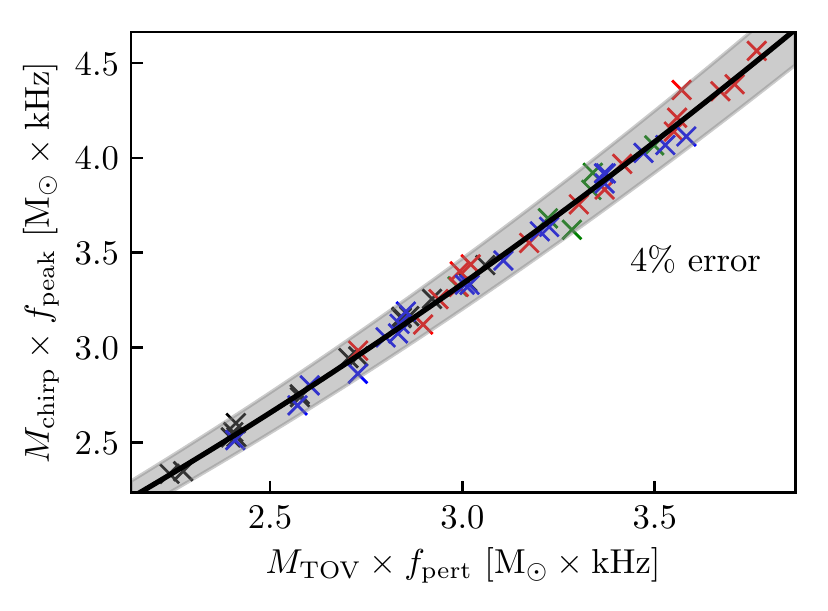}
    \caption{Dominant postmerger oscillation frequency $f_{\mathrm{peak}}(M_{\mathrm{tot}})$ scaled by the chirp mass $M_{\mathrm{chirp}}(M_{\mathrm{tot}})$ as a function of the perturbative frequency $f_{\mathrm{pert}}(M_{\mathrm{TOV}})$ scaled by the mass of the corresponding stellar configuration $M_{\mathrm{TOV}}$. The quantities $M_{\mathrm{TOV}}$ and $f_{\mathrm{pert}}$ refer to a stellar model with mass $1.23\times M_{\mathrm{tot}}/2$ for binary systems with mass $M_{\mathrm{tot}}$. Black symbols refer to $1.2+1.2~M_\odot$, blue to $1.35+1.35~M_\odot$, red to $1.4+1.4~M_\odot$ and green to $1.5+1.5~M_\odot$ systems. The solid black curve is a second-order fit to the data. The gray shaded area illustrates the $4\%$ error band. All points lie within the band.}
    \label{fpeak_fpert_mass}
\end{figure}

\begin{figure}
    \includegraphics{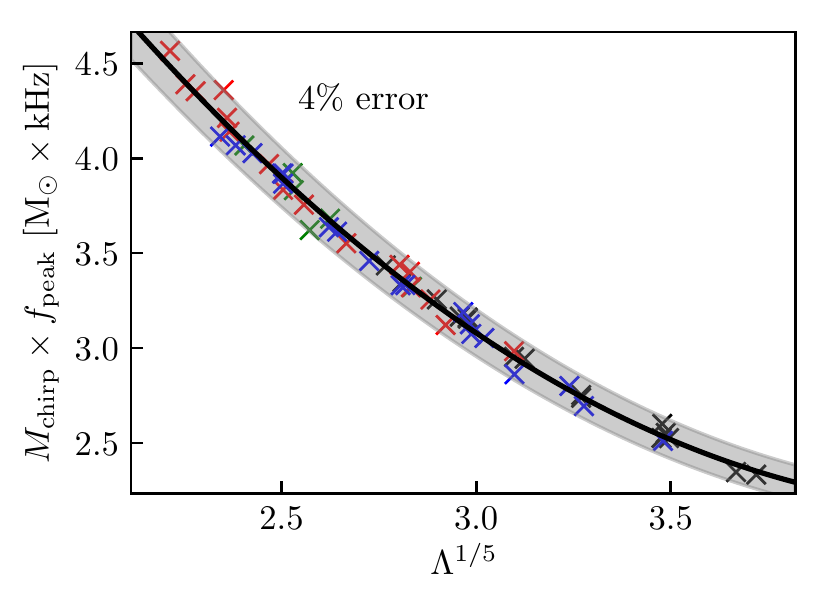}
    \caption{Dominant postmerger oscillation frequency $f_{\mathrm{peak}}(M_{\mathrm{tot}})$ scaled by the chirp mass $M_{\mathrm{chirp}}(M_{\mathrm{tot}})$ versus the tidal deformability $\Lambda^{1/5}(M_{\mathrm{TOV}})$ of static stars. The masses of static stars and symbol colors are as in Fig.\ \ref{fpeak_fpert_mass}. The solid black curve is a second-order fit to the data. All points lie within the gray $4\%$ error band.} 
    \label{Mchirpfpeak_Lambda_mass}
\end{figure}

Figure \ref{fpeak_fpert_mass} presents a mass-independent relation between $f_{\mathrm{peak}}(M_{\mathrm{tot}})$ scaled by the chirp mass $M_{\mathrm{chirp}}$\footnote{For binary systems with individual star masses $M_1$ and $M_2$, the chirp mass is defined as $M_{\mathrm{chirp}}=\frac{(M_1 M_2)^{3/5}}{(M_1+M_2)^{1/5}}$. Note that the chirp mass is fully equivalent to the total mass for equal-mass binaries.} and $f_{\mathrm{pert}}(M_{\mathrm{TOV}})$ scaled by the mass of the corresponding static star $M_{\mathrm{TOV}}$ for all $57$ equal-mass systems considered in this work. We relate each binary configuration to a static NS by choosing the mass of the static star such that the densities in both systems are comparable. For instance, we find that the choice $M_{\mathrm{TOV}}=1.23\times M_{\mathrm{tot}}/2$ serves this purpose for the binary systems considered here based on an analysis similar to the one we present in Appendix~\ref{Appendix_A}. We obtain $f_{\mathrm{pert}}(M_{\mathrm{TOV}})$ (and $\Lambda^{1/5}(M_{\mathrm{TOV}})$ in Fig. \ref{Mchirpfpeak_Lambda_mass}) for any mass $M_{\mathrm{TOV}}$ by a cubic spline fit to our perturbative data.

We find a highly accurate relation between $M_{\mathrm{chirp}} f_{\mathrm{peak}}(M_{\mathrm{tot}})$ and $M_{\mathrm{TOV}} f_{\mathrm{pert}}(M_{\mathrm{TOV}})$. The average and maximum deviation of the data from the fit are $30$~Hz and $134$~Hz respectively. This is very small considering that this is a mass-independent relation, while the relations in Table \ref{Fit_table_Mergers} are in comparison only slightly more accurate. The high accuracy of this relation further highlights the strong connection between $f_{\mathrm{peak}}$ and $f_{\mathrm{pert}}$ over the whole range of densities realized in postmerger remnants.

Figure \ref{fpeak_fpert_mass} further indicates that there is a tight mass-independent relation between $M_{\mathrm{chirp}} f_{\mathrm{peak}}$ and the tidal deformability of static stars. As discussed in Section \ref{Sec_stat_stars} (in particular Fig.\ \ref{TidalDefFreq_Static}), the mass-scaled $f_{\mathrm{pert}}$ correlates extremely tightly with $\Lambda^{-1/5}$. Hence, we expect that a similarly tight relation $M_{\mathrm{chirp}} f_{\mathrm{peak}}(\Lambda^{1/5})$ exists.

In Fig.\ \ref{Mchirpfpeak_Lambda_mass} we replace $M_{\mathrm{TOV}} f_{\mathrm{pert}}(M_{\mathrm{TOV}})$ by $\Lambda^{1/5}(M_{\mathrm{TOV}})$ of the corresponding static model. As expected, we find a tight correlation between the data. The average and maximum deviation of the second-order fit to the data is $31$~Hz and $138$~Hz respectively, which is perfectly in line with the deviations of the $M_{\mathrm{chirp}} f_{\mathrm{peak}}(M_{\mathrm{TOV}} f_{\mathrm{pert}})$ relation. These deviations in the mass-independent relation correspond to mass-scaled deviations of $37~\mathrm{M_\odot\times Hz}$ and $168~\mathrm{M_\odot\times Hz}$ respectively, which is significantly more accurate than relation (4) in \cite{2020PhRvD.102l3023B}. The fit parameters for both relations are summarized in Table \ref{Fit_table_fpeak_fpert}.

\begin{table*}
\caption{\label{Fit_table_fpeak_fpert}Mass-independent relations between $f_{\mathrm{peak}}$ or mass-scaled $M_{\mathrm{chirp}}f_{\mathrm{peak}}$ and static star properties. First column lists the figures presenting the corresponding relation, while third and fourth columns provide the average and maximum deviation of each relation in Hz respectively. The frequencies are in kHz, masses in $M_\odot$ and the tidal deformability is dimensionless.}
\begin{ruledtabular}
\begin{tabular}{cclcc}
\textrm{Fig.}&
\textrm{Systems}&
\multicolumn{1}{c}{\textrm{Fit}}&
\textrm{Mean dev.}&
\textrm{Max dev.}\\
 & & & \textrm{[Hz]} & \textrm{[Hz]}\\
\colrule

\ref{fpeak_fpert_mass}        & Equal-mass & $M_{\mathrm{chirp}}f_{\mathrm{peak}}       =  0.299 + 0.595M_{\mathrm{TOV}}f_{\mathrm{pert}} + 1.392\times10^{-1}\left(M_{\mathrm{TOV}}f_{\mathrm{pert}}\right)^2$ & $30$ & $134$\\
\ref{Mchirpfpeak_Lambda_mass} & Equal-mass &  $M_{\mathrm{chirp}}f_{\mathrm{peak}}      =  11.846      -4.464\Lambda^{1/5}   +5.139\times10^{-1}\Lambda^{2/5}$ & $31$ & $138$\\
\ref{fpeak_fpert_mass_errorbands_All} & All & $M_{\mathrm{chirp}}f_{\mathrm{peak}}      =  0.013 + 0.794M_{\mathrm{TOV}}f_{\mathrm{pert}} + 1.042\times10^{-1}\left(M_{\mathrm{TOV}}f_{\mathrm{pert}}\right)^2$ & $35$ & $150$\\
-                                     & All &  $M_{\mathrm{chirp}}f_{\mathrm{peak}}     =  11.536      -4.261\Lambda^{1/5}   +4.806\times10^{-1}\Lambda^{2/5}$ & $36$ & $151$
\end{tabular}
\end{ruledtabular}
\end{table*}

In order to consider a broader parameter range, we extend our data set by also including unequal-mass binaries and constructing relations of the same type. We directly import the unequal mass data from Table II in~\cite{2020PhRvD.101h4039V} (except for one EoS which is not considered in this study). We include a total of 40 unequal-mass binary systems, with mass ratios as low as 0.67. Each binary configuration is related to a static star through

\begin{equation}\label{Mtov_general}
M_{\mathrm{TOV}}= \left[ a + b\times(1-q)^2 \right] \times \frac{M_{\mathrm{tot}}}{2},
\end{equation}
with $a=1.23$ and $b=-0.67$. The term in brackets introduces a mild dependence on $q$ since we realized that unequal-mass results are better captured by a slightly smaller $M_{\mathrm{TOV}}$. For $q=1$ the value of $a$ reproduces the relation between $M_{\mathrm{TOV}}$ and $M_{\mathrm{tot}}$ which we introduced above. We fix $b$ by an analysis similar to the one in Appendix~\ref{Appendix_A} to approximately minimize the frequency deviations for unequal-mass binaries.

\begin{figure}
    \includegraphics{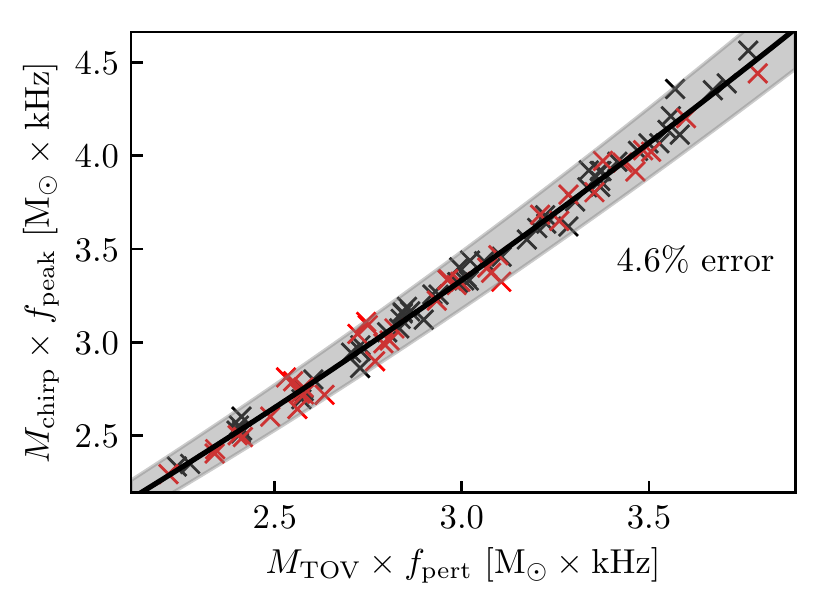}
    \caption{Same as in Fig.\ \ref{fpeak_fpert_mass}, but black symbols represent equal-mass systems and red symbols denote unequal-mass symbols. $M_{\mathrm{TOV}}$ follows from Eq. \eqref{Mtov_general}.  The solid black curve is a second-order fit to the data. All data points lie within the gray $4.6\%$ error band.} 
    \label{fpeak_fpert_mass_errorbands_All}
\end{figure}

Figure \ref{fpeak_fpert_mass_errorbands_All} displays all equal-mass data present from Fig.\ \ref{fpeak_fpert_mass} as black symbols, alongside unequal-mass data denoted by red symbols. We find a highly accurate relation, practically as tight as the one for equal-mass binaries. As expected, an identically accurate relation exists between $M_{\mathrm{chirp}} f_{\mathrm{peak}}(M_{\mathrm{tot}})$ and $\Lambda^{1/5}(M_{\mathrm{TOV}})$. We include the expressions for both relations and their respective average and maximum deviations in Table \ref{Fit_table_fpeak_fpert}.

\section{Summary and discussion}
In this study we consider the frequency of the fundamental quadrupolar fluid mode in isolated NSs and the dominant oscillation of postmerger remnants. We compute the oscillation frequencies $f_{\mathrm{pert}}$ of isolated NSs with a perturbative method. In contrast, we obtain the frequency $f_{\mathrm{peak}}$ of the dominant postmerger oscillation from a full dynamical simulation. We consider a large sample of different high-density EoSs for both stellar systems and vary the masses in a considerable range.

Considering these frequency data separately for both types of objects we construct fits, which relate the frequency to stellar parameters of non-rotating NSs that we choose to characterize the EoS. We employ different stellar parameters like radii and the tidal deformability as independent variables and assess the accuracy of these relations by quantifying the maximum and average deviations of the individual data points from the least-square fit to all data points. Some of those relations have been proposed previously in the literature and by employing the same set of data we can consistently compare between these fits and evaluate their accuracy. By constructing second-order fits we find that the relation involving the moment of inertia $I$ is the most accurate, while relations with the tidal deformability as independent variable are only slightly less tight. Extending these relations to higher order, in particular the relations between the mass-scaled perturbative frequency and the tidal deformability become even tighter and essentially exact for all practical purposes. For fixed masses, second-order relations of the form $f_{\mathrm{pert}}(\Lambda^{1/5})$ are practically exact throughout the whole mass range and thus one can use $f_{\mathrm{pert}}$ and $\Lambda^{1/5}$ interchangeably.

Furthermore, we introduce a newly defined stellar radius $R^{90\%}$, where we disregard the outer mass shells containing $10\%$ of the total mass. By doing this we obtain a measure for the stellar compactness, which is largely insensitive to the low-density regime of the EoS (below approximately $(1.27-4.88)\times 10^{14}~\mathrm{g/cm^3}$). Employing $R^{90\%}$, we observe that relations for isolated stars as well as for postmerger remnants become generally tighter with regard to the mean and maximum deviations. For perturbative results of isolated stars the deviations are more comparable to those with $\Lambda$. These results indicate that oscillation frequencies in both systems are predominantly determined by the high-density regime of the EoS. Like the commonly defined radius $R$ at the stellar surface, $R^{90\%}$ is uniquely linked to the EoS, but unbiased by the low-density part, which presumably has a smaller influence on the oscillation frequencies. Thus, a determination of $R^{90\%}$ is likely more informative about the high-density EoS than $R$, as the latter may be ``biased'' by the low-density EoS. Employing $R^{90\%}$ relations may thus be preferable in GW asteroseismology since it results in a more accurate determination if the scatter in the fit formulae is taken into account as source of error and since it represents a more direct measure of the EoS properties in the relevant density regime.

Along these lines we also consider stellar configurations with the radius $R^{\mathrm{cc}}$ truncated at the crust-core transition. For isolated stars, the relation involving the mean density defined via $R^{\mathrm{cc}}$ becomes tighter compared to relations with the actual stellar radius and mass, while there are no significant differences in a relation involving the compactness. Relations for postmerger remnants also become tighter, but only if $R^{\mathrm{cc}}$ refers to a stellar model with high fiducial mass (see Appendix~\ref{Appendix_A}). Overall, we observe that relations between perturbative frequencies and stellar parameters characterizing the core of the stellar model still exhibit some scatter. This further supports the argument that the frequencies, and in particular the scatter of points in such frequency relations, is at least partially influenced by the high-density regime of the EoS.

However, we also notice that there are finite frequency deviations in the $f_{\mathrm{peak}}$ relations for any of the independent variable we tested, i.e $R$, $R^{90\%}$, $R^{\mathrm{cc}}$ or the tidal deformability. With regard to this scatter, the main
finding of this study is that frequency deviations follow the very same behavior in isolated NSs and in postmerger remnants if frequencies are considered with respect to the same independent variable: If $f_{\mathrm{pert}}$ for a given EoS model is slightly increased with respect to the fit to all data points of the perturbative calculations of isolated stars, the
postmerger oscillation frequency for this EoS also occurs at slightly higher frequency compared to fit to all merger simulations. Similarly, data points for other EoS models exhibit slightly reduced frequencies in both stellar systems.

The consistent behavior of frequency deviations in relations describing isolated NSs on one hand and relations for merger remnants on the other hand is very remarkable: We compare the frequency of a cold, isolated, non-rotating NS to oscillations of a hot, rapidly rotating, non-stationary, massive merger remnant. We observe the correspondence of frequency deviations in various relations for different independent variables characterizing the EoS, and for different (binary) masses. We
identify, if at all, only a very small number of outliers with respect to this behavior, which is why it is unlikely that we describe a mere coincidence. Instead, the agreement of the frequency scatter points to some underlying physical mechanism which is mediated by the EoS as the only common ingredient in both types of calculations. Also, the relatively large number of tested EoS models supports the argument of additional EoS information being encoded in the frequency beyond the gross scaling of universal relations.

In this regard, we stress that we compare frequencies from perturbative calculations for isolated NSs, which one should consider as rather robust and converged results, and frequencies which are extracted from complex, three-dimensional hydrodynamical simulations of the full merger process using a different numerical code. Also, the merger remnant has not yet reached a stationary configuration when the dominant frequency peak of the GW emission is shaped. We note that the magnitude of
frequency deviations is typically of the order of some $10$~Hz. It is thus remarkable that the hydrodynamical simulations apparently resolve some systematic behavior of the frequency deviations, which are of this magnitude. Since this level of precision is certainly challenging for a hydrodynamical code of this type, we may even speculate that the few
outliers we observed can be attributed to inaccuracies of the merger simulations and that frequency deviations follow the indicated trends even more closely.

We further investigate the source of frequency deviations in GW asteroseismology relations like for instance $f(R)$. To this end we exploit the correspondence between the frequency increase or decrease in isolated NSs and merger remnants, and thus focus on explaining the slight frequency shifts for static stars. Moreover, we employ the fact that for static NSs there is a practically exact relation between the f-mode frequency and the tidal deformability. This implies that frequency deviations in $f(R)$ are fully equivalent to deviations in $\Lambda(R)$. Hence, we can attribute frequency shifts to the scatter in the relationship between the tidal deformability and the stellar radius, which by definition is given by the tidal Love number $k_2$. The frequency deviations thus encode by how much the tidal Love number deviates from an approximate scaling of $k_2$ with the stellar radius.

This indicates new directions to exploit this result in future measurements and theoretical studies particularly in the context of merger remnants, where oscillation frequencies might be more likely to be measured, although $f-$mode frequencies of isolated stars may play a role during the inspiral phase \cite{2019PhRvD.100b1501S, 2019JPhG...46l3002G, 2020NatCo..11.2553P, 2020GReGr..52..109C} and in other astrophysical systems. At least in principle frequency deviations from an expected universal relation, reflecting the average behavior of a large class of EoS models, can be measured. As an example, measuring the magnitude or at least the sign of a frequency deviation from a universal relation can be employed to break the degeneracy between radius and tidal deformability and can thus lead to a more precise determination of the tidal Love number and ultimately properties of the EoS. We show an explicit case where $k_2$ is determined more precisely if additional information for the frequency deviation is available. Also, understanding the link between frequency shifts and stellar properties can be used to construct tighter universal relations between GW frequencies like $f_{\mathrm{peak}}$ and stellar parameters by removing the frequency shifts. Hence, more information can be extracted from a measurement if more accurate asteroseismology relations are available.

Along the same lines we explicitly show that a measurement of the postmerger frequency and a measurement of the tidal deformability in the same event can be combined to yield information on the slope of $\Lambda(M)$. Here, we again consider the deviation between the measured postmerger frequency and the one expected from a universal relation for the given tidal deformability. This reflects the additional information about properties at higher densities being encoded in the postmerger remnant. This is in line with the observation that the dominant postmerger frequency shows a particularly tight correlation with the tidal deformability of a NS with a higher mass compared to that of the inspiralling stars.

In this respect we also refer to the extensive analysis of $f_\mathrm{peak}(\Lambda)$ and $f_\mathrm{peak}(R)$ relations in Appendix~\ref{Appendix_A}. In particular, we point out  that $f_\mathrm{peak}$ relations for a fixed binary mass $M_\mathrm{tot}$ are tighter if one relates $f_\mathrm{peak}$ to the tidal deformability of a more massive fiducial star with $M>M_\mathrm{tot}/2$, i.e. a mass larger than that of the inspiraling star, similar to what has been observed for frequency-radius relations~\cite{2012PhRvD..86f3001B}. This is  summarized by the comparison in Tab.~\ref{Fit_table_Mergers}. 

Since frequency deviations in static stars and merger remnants are correlated, one can employ this correspondence to partially remove the scatter in plots which directly relate the perturbative frequency of static stars and postmerger GW frequencies. In fact, we find very accurate mass-independent relations. We emphasize that for such type of relationships there is the freedom to choose a fiducial mass of the static model corresponding to a given binary mass. We identify a simple, analytic mapping $M_{\mathrm{TOV}}=1.23\times M_{\mathrm{tot}}/2$ between both masses, which yields particularly tight relations with an accuracy nearly comparable to that of correlations for fixed masses. Exploiting the practically exact relationship between the f-mode frequency of static stars and their tidal deformability, the mapping equivalently implies a highly accurate mass-independent relation between the postmerger frequency and $\Lambda$. We extend the analysis by including data for unequal mass binaries and verify that similar  accurate relations hold even when considering a large range of mass ratios.

Finally, we remark that the striking similarity of the frequency scatter in relations for $f-$modes of isolated NSs and in relations for the dominant oscillation frequency of merger remnants provides additional evidence that the dominant oscillation in postmerger objects is linked to the fundamental quadrupolar fluid mode in line with previous arguments \cite{Stergioulas2011, Bauswein2015, Bauswein2016, 2020MNRAS.497.5480C}.

Future work should confirm that other hydrodynamical codes find a similar behavior of the frequency deviations in $f_{\mathrm{peak}}$. As mentioned one should keep in mind that resolving $f_{\mathrm{peak}}$ with this accuracy is certainly
challenging and that the frequency deviations are small in comparison to the typical FWHM of a few $100$~Hz of postmerger GW peaks. Other simulations not finding similar frequency patterns would not automatically imply that systematic frequency deviations are not real but instead that these numerical models are possibly more affected by numerical uncertainties. In future studies one may check for consistency between the frequency deviations of merger simulations and the frequency deviations of static stars from either perturbative calculations or simply from the expected frequencies employing the very tight relations between f-mode frequency and tidal deformability. By this one may benchmark the quality of simulation data in larger surveys. Moreover, we speculate that in future more accurate merger models may yield frequency deviations that more closely follow the quantitative dependencies, which we observed in this study, similar to those for perturbative frequencies. This aspect may also be addressed
by perturbative calculations of differentially rotating NSs in equilibrium resembling merger remnants \cite{2013PhRvD..88d4052D,2020PhRvL.125k1106K}.

By purpose we did not include EoS models with a strong phase transition in this study, which should be considered in future work. The significant and sudden softening of the EoS by a strong phase transition will lead to a strongly increased postmerger frequency, i.e. an extreme frequency deviation of some 100~Hz \cite{2019PhRvL.122f1102B,2020PhRvL.124q1103W,2020EPJST.229.3595B}. The effect on the different relations presented here will however very sensitively depend on the onset density of the phase transition and at which mass the stellar structure is affected. Thus choice of the dependent and independent variables is critical (in an extreme case one quantity would be affected by a phase transition, while another variable only being sensitive to lower densities does not carry any information about the EoS softening). Considering phase transitions would introduce several new effective degrees of freedom like the onset density, the density jump across the transition and the stiffening of the EoS beyond the phase transition. Such a variety can hardly be covered by a few models to allow a comprehensive study. We thus omit such models since they would severely affect the different fits representing an average behavior and thus the quantification of frequency deviations of the purely hadronic models. Physically, this approach is very well justified because the extreme frequency deviations by a strong phase transition would unambiguously indicate the presence of exotic forms of matter as argued in \cite{2019PhRvL.122f1102B} and thus caution that the considerations of the present study may not be applicable. Similarly, evidence for a phase transition may be provided by other independent measurements or observations.

More work should also be spend on concrete methods to implement the findings of our study. This includes extracting frequency deviations from GW signals and developing improved relations for GW asteroseismology where the scatter is reduced by taking into account the particular dependencies of the frequency deviations on EoS properties. Other aspects involve the frequency scatter of individual models in mergers of unequal mass, which we did not cover in great detail, and the behavior of subdominant GW peaks, which we only briefly mentioned to follow a similar trend.

\begin{acknowledgments}
 We are grateful to Katerina Chatziioannou, Reed Essick, Brynmor Haskell, Jocelyn Read and Stefan Typel for useful comments and discussions. G.L.\ and A.B.\ acknowledge support by the European Research Council (ERC) under the European Union’s Horizon 2020 research  and  innovation  programme  under  grant  agreement  No.\ 759253. A.B.\ acknowledges support by Deutsche Forschungsgemeinschaft (DFG, German Research Foundation) - Project-ID 279384907 - SFB 1245 and DFG - Project-ID  138713538 - SFB  881  (“The  Milky  Way  System”, subproject  A10). N.S.\ acknowledges support by the ARIS facility of GRNET in Athens (SIMGRAV, SIMDIFF and BNSMERGE allocations) and the “Aristoteles Cluster” at  AUTh, as well as by the COST actions CA16214 “PHAROS”, CA16104 “GWVerse”, CA17137 “G2Net”and CA18108 “QG-MM”. NS gratefully acknowledges the Italian Istituto Nazionale di Fisica Nucleare (INFN), the French Centre National de la Recherche Scientifique (CNRS) and the Netherlands Organization for Scientific Research, for the construction and operation of the Virgo detector and the creation and support of the EGO consortium.
\end{acknowledgments}

\appendix

\section{Accuracy of relations between $f_{\mathrm{peak}}$ and static stellar properties}\label{Appendix_A}
Throughout this work we discuss relations between $f_{\mathrm{peak}}$ for different binary systems and stellar properties of static stars with a fixed fiducial mass (e.g.\ Table \ref{Fit_table_Mergers} and Figures \ref{AB_plots}, \ref{fpert_fpeak_R}, \ref{fpeak_fpert_mass} and \ref{Mchirpfpeak_Lambda_mass}). In principle, the fiducial mass of the static models can be chosen freely. However, different choices for the fiducial mass lead to relations of different accuracy (see also~\cite{2012PhRvD..86f3001B}), and the choice of the mass of the static model should be justified.

\begin{figure}[h]
    \includegraphics{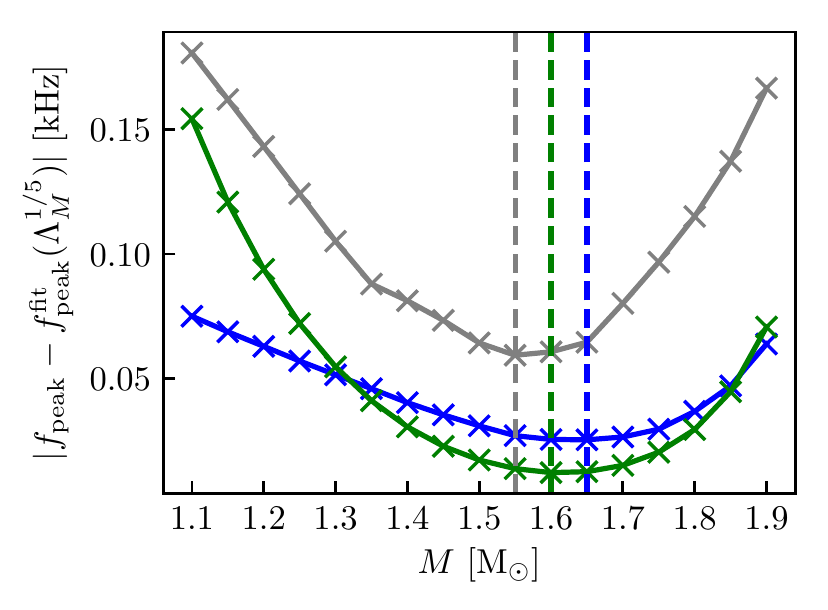}
    \caption{Different figures of merit to quantify the accuracy of $f_{\mathrm{peak}}(\Lambda^{1/5}_M)$ relations for $1.35+1.35~M_\odot$ systems as function of chosen fiducial masses $M$. The gray curve illustrates the maximum deviation, blue curve depicts the average deviation and green curve displays the normalized sum of squared residuals of the least-squares fit. Dashed lines indicate minima of the curves of the respective color.} 
    \label{fpeak135_Lambda_Mass_deviations}
\end{figure}

We consider relations between $f_{\mathrm{peak}}$ and four independent variables: the radius $R$, the radius $R^{90\%}$ referring to $90\%$ of the mass, the radius $R^{\mathrm{cc}}$ defined based on the crust-core transition density and the fifth-root of the tidal deformability $\Lambda^{1/5}$. In order to quantify the accuracy of the respective relations, we examine three different figures of merit. Specifically, the average deviations, maximum deviations and the sum of squared residuals of the least-squares fit. 

In Fig.~\ref{fpeak135_Lambda_Mass_deviations} we present the three accuracy metrics for the relation $f_{\mathrm{peak}}(\Lambda^{1/5}_M)$ for 1.35+1.35~$M_\odot$ binaries as function of the fiducial mass $M$. All three figures of merit are minimized within the mass range $1.55-1.65~M_\odot$. Thus, we identify this mass range as the optimal for this particular binary system and type of relation.

\begin{table}
\caption{\label{Optimal_fiducial_masses} Mass ranges of fiducial masses which minimize frequency deviations in relations between postmerger frequencies and stellar parameters of static stars for different binary systems. First column lists the masses of the binary systems. Second column provides the independent variable, i.e. the stellar parameter of a fiducial NS, which is employed in the respective fit. Third column gives the mass range over which the three considered figures of merit are minimized (see text). Fourth column lists the maximum values that the average and maximum deviations assume in the corresponding mass range.}
\begin{ruledtabular}
\begin{tabular}{cccl}
\textrm{Binary masses}&
\textrm{Independent}&
\textrm{Optimal mass}&
\multicolumn{1}{c}{\textrm{Mean/Max}}\\
 $[M_\odot]$ & \textrm{variable} & \textrm{range $[M_\odot]$} & \multicolumn{1}{c}{\textrm{dev. [Hz]}}\\
\colrule

 $1.2+1.2$& $R$ & $1.6-1.75$ & $<(33,82)$\\
 $1.2+1.2$& $R^{90\%}$ & $1.5-1.6$ & $<(20,49)$\\
 $1.2+1.2$& $R^{\mathrm{cc}}$ & $1.7-1.75$ & $<(18,57)$\\ \vspace{0.5cm}
 $1.2+1.2$& $\Lambda^{1/5}$ & $1.4-1.45$ & $<(18,50)$\\
 
 $1.35+1.35$& $R$ & $1.7-1.8$ & $<(38,91)$\\
 $1.35+1.35$& $R^{90\%}$ & $1.7-1.75$ & $<(22,62)$\\
 $1.35+1.35$& $R^{\mathrm{cc}}$ & $1.75-1.85$ & $<(28,69)$\\ \vspace{0.5cm}
 $1.35+1.35$& $\Lambda^{1/5}$ & $1.55-1.65$ & $<(27,65)$\\
 
 $1.4+1.4$& $R$ & $1.85-1.9$ & $<(42,105)$\\
 $1.4+1.4$& $R^{90\%}$ & $1.8-1.85$ & $<(30,99)$\\
 $1.4+1.4$& $R^{\mathrm{cc}}$ & $1.85-1.9$ & $<(28,98)$\\ \vspace{0.5cm}
 $1.4+1.4$& $\Lambda^{1/5}$ & $1.75-1.8$ & $<(35,109)$ \\
 
 $1.5+1.5$& $R$ & $1.75-1.8$ & $<(33,76)$\\
 $1.5+1.5$& $R^{90\%}$ & $1.75-1.8$ & $<(26,64)$\\
 $1.5+1.5$& $R^{\mathrm{cc}}$ & $1.9$ & $<(17,43)$\\ 
 $1.5+1.5$& $\Lambda^{1/5}$ & $1.65-1.75$ & $<(30,73)$\\
\end{tabular}
\end{ruledtabular}
\end{table}

We summarize the analysis for other binary masses and other relations in Table~\ref{Optimal_fiducial_masses}. We list the mass ranges of the fiducial stellar model for which relations between $f_{\mathrm{peak}}$ and the different independent variables become tightest. Evidently, for a fixed binary mass, relations w.r.t.\ different independent variables become tighter for slightly different fiducial masses. In particular, relations involving the radius tend to become more accurate for higher fiducial masses than relations w.r.t.\ $\Lambda^{1/5}$. Obviously, the ``optimal'' fiducial mass, in the sense of minimizing the deviations in frequency relations, is higher for more massive binaries\footnote{This is not the case for $1.5+1.5~M_\odot$ systems, because the data set is significantly smaller since many EoS models promptly collapse to a black hole.}. In all cases the optimal fiducial mass is higher than the mass of the inspiralling stars. This reflects the fact that merger remnants are in comparison more massive and that densities in the merger remnant are higher because of compression.

\begin{figure}
    \includegraphics{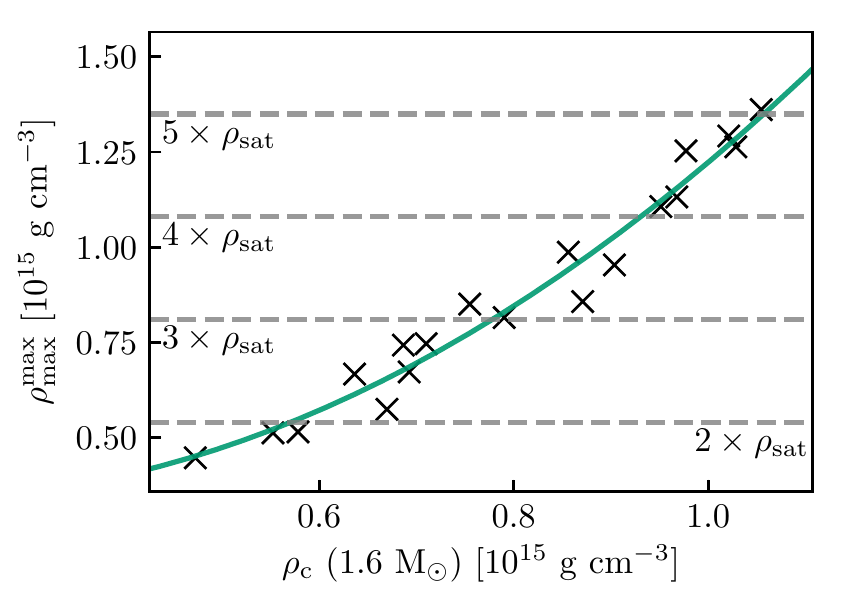}
    \caption{Maximum rest-mass density $\rho_{\mathrm{max}}^{\mathrm{max}}$ in the remnant during the first few milliseconds after merging for $1.35+1.35~M_\odot$ systems versus the central rest-mass density $\rho_{\mathrm{c}}$ of a $1.6~M_{\odot}$ star described by the same EoS. The solid green curve is a second-order fit to the data given by $\rho_{\mathrm{max}}^{\mathrm{max}}=0.372 -0.446\; \rho_{\mathrm{c}} + 1.296\;\rho_{\mathrm{c}}^2$.}
    \label{rhomaxmerger_rhoc_res135135}
\end{figure}

Based on Table~\ref{Optimal_fiducial_masses}, for each binary system there exists a fiducial mass range of about $0.25~M_\odot$ for which $f_{\mathrm{peak}}$ relations become particularly tight. In order to understand this observation, we consider the central rest-mass densities $\rho_\mathrm{c}$ of the fiducial static models and the maximum rest-mass densities in the merger remnants during the first few milliseconds after merging. There is no unique way to define a characteristic density of the remnant because it is strongly oscillating and dynamically evolving. We pick the maximum value of the maximum density $\rho_\mathrm{max}^\mathrm{max}$ which occurs over the first few oscillation cycles after merging (see \cite{2019PhRvL.122f1102B, 2020PhRvD.102l3023B})\footnote{An alternative definition is to extract an average density over the initial few milliseconds of the postmerger evolution. The situation is rather similar to Fig.\ \ref{rhomaxmerger_rhoc_res135135} in that case as well.}. In Figure~\ref{rhomaxmerger_rhoc_res135135} we plot these densities for binary systems of a total mass of $2.7~M_\odot$ and a static star with a mass of $1.6~M_\odot$. The choice of static star mass is motivated by the mass ranges in Table~\ref{Optimal_fiducial_masses}. Overall, we notice an agreement between the two densities, which explains why such a choice for the fiducial mass appears to be optimal. For softer EoSs, i.e. at higher densities, the densities in the remnant are in relation to those in the static stars slightly higher. This is in agreement with the results in Table~\ref{Optimal_fiducial_masses} showing that a higher fiducial mass represents the optimal description of matter in $1.35+1.35~M_\odot$ systems. Moreover, it indicates that the compression during the merger process is more pronounced for softer EoSs.

In summary, we conclude that relations between $f_{\mathrm{peak}}$ and different independent variables referring to stellar properties of static stars become most accurate for different values of the mass of the static star. Using different figures of merit we can identify the mass range which leads to the tightest relations. Typically this mass range refers to static stars with central densities comparable to typical densities realized in the merger remnants during the first few milliseconds after merging. Finally, we remark that the exact distribution of frequency deviations will depend on the chosen set of candidate EoSs and may also be affected by the numerical model. Thus, the optimal values of the fiducial mass might be slightly different in other surveys. In any case the extrema in Fig.~\ref{fpeak135_Lambda_Mass_deviations} are relatively broad. Hence, the exact choice of the fiducial mass is not essential, and we expect that the ranges given in Table~\ref{Optimal_fiducial_masses} are robust and representative.

\bibliography{bibliography}

\providecommand{\noopsort}[1]{}\providecommand{\singleletter}[1]{#1}%
\begin{thebibliography}{110}%
\makeatletter
\providecommand \@ifxundefined [1]{%
 \@ifx{#1\undefined}
}%
\providecommand \@ifnum [1]{%
 \ifnum #1\expandafter \@firstoftwo
 \else \expandafter \@secondoftwo
 \fi
}%
\providecommand \@ifx [1]{%
 \ifx #1\expandafter \@firstoftwo
 \else \expandafter \@secondoftwo
 \fi
}%
\providecommand \natexlab [1]{#1}%
\providecommand \enquote  [1]{``#1''}%
\providecommand \bibnamefont  [1]{#1}%
\providecommand \bibfnamefont [1]{#1}%
\providecommand \citenamefont [1]{#1}%
\providecommand \href@noop [0]{\@secondoftwo}%
\providecommand \href [0]{\begingroup \@sanitize@url \@href}%
\providecommand \@href[1]{\@@startlink{#1}\@@href}%
\providecommand \@@href[1]{\endgroup#1\@@endlink}%
\providecommand \@sanitize@url [0]{\catcode `\\12\catcode `\$12\catcode
  `\&12\catcode `\#12\catcode `\^12\catcode `\_12\catcode `\%12\relax}%
\providecommand \@@startlink[1]{}%
\providecommand \@@endlink[0]{}%
\providecommand \url  [0]{\begingroup\@sanitize@url \@url }%
\providecommand \@url [1]{\endgroup\@href {#1}{\urlprefix }}%
\providecommand \urlprefix  [0]{URL }%
\providecommand \Eprint [0]{\href }%
\providecommand \doibase [0]{https://doi.org/}%
\providecommand \selectlanguage [0]{\@gobble}%
\providecommand \bibinfo  [0]{\@secondoftwo}%
\providecommand \bibfield  [0]{\@secondoftwo}%
\providecommand \translation [1]{[#1]}%
\providecommand \BibitemOpen [0]{}%
\providecommand \bibitemStop [0]{}%
\providecommand \bibitemNoStop [0]{.\EOS\space}%
\providecommand \EOS [0]{\spacefactor3000\relax}%
\providecommand \BibitemShut  [1]{\csname bibitem#1\endcsname}%
\let\auto@bib@innerbib\@empty
\bibitem [{\citenamefont {{Kokkotas}}\ and\ \citenamefont
  {{Schmidt}}(1999)}]{1999LRR.....2....2K}%
  \BibitemOpen
  \bibfield  {author} {\bibinfo {author} {\bibfnamefont {K.~D.}\ \bibnamefont
  {{Kokkotas}}}\ and\ \bibinfo {author} {\bibfnamefont {B.~G.}\ \bibnamefont
  {{Schmidt}}},\ }\bibfield  {title} {\bibinfo {title} {{Quasi-Normal Modes of
  Stars and Black Holes}},\ }\href {https://doi.org/10.12942/lrr-1999-2}
  {\bibfield  {journal} {\bibinfo  {journal} {Living Reviews in Relativity}\
  }\textbf {\bibinfo {volume} {2}},\ \bibinfo {eid} {2} (\bibinfo {year}
  {1999})},\ \Eprint {https://arxiv.org/abs/gr-qc/9909058} {arXiv:gr-qc/9909058
  [gr-qc]} \BibitemShut {NoStop}%
\bibitem [{\citenamefont {{Xing}}\ \emph {et~al.}(1994)\citenamefont {{Xing}},
  \citenamefont {{Centrella}},\ and\ \citenamefont
  {{McMillan}}}]{1994PhRvD..50.6247X}%
  \BibitemOpen
  \bibfield  {author} {\bibinfo {author} {\bibfnamefont {Z.}~\bibnamefont
  {{Xing}}}, \bibinfo {author} {\bibfnamefont {J.~M.}\ \bibnamefont
  {{Centrella}}},\ and\ \bibinfo {author} {\bibfnamefont {S.~L.~W.}\
  \bibnamefont {{McMillan}}},\ }\bibfield  {title} {\bibinfo {title}
  {{Gravitational radiation from coalescing binary neutron stars}},\ }\href
  {https://doi.org/10.1103/PhysRevD.50.6247} {\bibfield  {journal} {\bibinfo
  {journal} {\prd}\ }\textbf {\bibinfo {volume} {50}},\ \bibinfo {pages} {6247}
  (\bibinfo {year} {1994})},\ \Eprint {https://arxiv.org/abs/gr-qc/9411029}
  {arXiv:gr-qc/9411029 [gr-qc]} \BibitemShut {NoStop}%
\bibitem [{\citenamefont {Shibata}(2005)}]{PhysRevLett.94.201101}%
  \BibitemOpen
  \bibfield  {author} {\bibinfo {author} {\bibfnamefont {M.}~\bibnamefont
  {Shibata}},\ }\bibfield  {title} {\bibinfo {title} {Constraining nuclear
  equations of state using gravitational waves from hypermassive neutron
  stars},\ }\href {https://doi.org/10.1103/PhysRevLett.94.201101} {\bibfield
  {journal} {\bibinfo  {journal} {Phys. Rev. Lett.}\ }\textbf {\bibinfo
  {volume} {94}},\ \bibinfo {pages} {201101} (\bibinfo {year}
  {2005})}\BibitemShut {NoStop}%
\bibitem [{\citenamefont {{Shibata}}\ \emph {et~al.}(2005)\citenamefont
  {{Shibata}}, \citenamefont {{Taniguchi}},\ and\ \citenamefont
  {{Ury{\={u}}}}}]{2005PhRvD..71h4021S}%
  \BibitemOpen
  \bibfield  {author} {\bibinfo {author} {\bibfnamefont {M.}~\bibnamefont
  {{Shibata}}}, \bibinfo {author} {\bibfnamefont {K.}~\bibnamefont
  {{Taniguchi}}},\ and\ \bibinfo {author} {\bibfnamefont {K.}~\bibnamefont
  {{Ury{\={u}}}}},\ }\bibfield  {title} {\bibinfo {title} {{Merger of binary
  neutron stars with realistic equations of state in full general
  relativity}},\ }\href {https://doi.org/10.1103/PhysRevD.71.084021} {\bibfield
   {journal} {\bibinfo  {journal} {\prd}\ }\textbf {\bibinfo {volume} {71}},\
  \bibinfo {eid} {084021} (\bibinfo {year} {2005})},\ \Eprint
  {https://arxiv.org/abs/gr-qc/0503119} {arXiv:gr-qc/0503119 [gr-qc]}
  \BibitemShut {NoStop}%
\bibitem [{\citenamefont {{Oechslin}}\ and\ \citenamefont
  {{Janka}}(2007)}]{2007PhRvL..99l1102O}%
  \BibitemOpen
  \bibfield  {author} {\bibinfo {author} {\bibfnamefont {R.}~\bibnamefont
  {{Oechslin}}}\ and\ \bibinfo {author} {\bibfnamefont {H.~T.}\ \bibnamefont
  {{Janka}}},\ }\bibfield  {title} {\bibinfo {title} {{Gravitational Waves from
  Relativistic Neutron-Star Mergers with Microphysical Equations of State}},\
  }\href {https://doi.org/10.1103/PhysRevLett.99.121102} {\bibfield  {journal}
  {\bibinfo  {journal} {\prl}\ }\textbf {\bibinfo {volume} {99}},\ \bibinfo
  {eid} {121102} (\bibinfo {year} {2007})},\ \Eprint
  {https://arxiv.org/abs/astro-ph/0702228} {arXiv:astro-ph/0702228 [astro-ph]}
  \BibitemShut {NoStop}%
\bibitem [{\citenamefont {{Stergioulas}}\ \emph {et~al.}(2011)\citenamefont
  {{Stergioulas}}, \citenamefont {{Bauswein}}, \citenamefont {{Zagkouris}},\
  and\ \citenamefont {{Janka}}}]{Stergioulas2011}%
  \BibitemOpen
  \bibfield  {author} {\bibinfo {author} {\bibfnamefont {N.}~\bibnamefont
  {{Stergioulas}}}, \bibinfo {author} {\bibfnamefont {A.}~\bibnamefont
  {{Bauswein}}}, \bibinfo {author} {\bibfnamefont {K.}~\bibnamefont
  {{Zagkouris}}},\ and\ \bibinfo {author} {\bibfnamefont {H.-T.}\ \bibnamefont
  {{Janka}}},\ }\bibfield  {title} {\bibinfo {title} {{Gravitational waves and
  non-axisymmetric oscillation modes in mergers of compact object binaries}},\
  }\href {https://doi.org/10.1111/j.1365-2966.2011.19493.x} {\bibfield
  {journal} {\bibinfo  {journal} {Mon. Not. R. Astron. Soc}\ }\textbf {\bibinfo
  {volume} {418}},\ \bibinfo {pages} {427} (\bibinfo {year}
  {2011})}\BibitemShut {NoStop}%
\bibitem [{\citenamefont {{Bauswein}}\ and\ \citenamefont
  {{Janka}}(2012)}]{2012PhRvL.108a1101B}%
  \BibitemOpen
  \bibfield  {author} {\bibinfo {author} {\bibfnamefont {A.}~\bibnamefont
  {{Bauswein}}}\ and\ \bibinfo {author} {\bibfnamefont {H.~T.}\ \bibnamefont
  {{Janka}}},\ }\bibfield  {title} {\bibinfo {title} {{Measuring Neutron-Star
  Properties via Gravitational Waves from Neutron-Star Mergers}},\ }\href
  {https://doi.org/10.1103/PhysRevLett.108.011101} {\bibfield  {journal}
  {\bibinfo  {journal} {\prl}\ }\textbf {\bibinfo {volume} {108}},\ \bibinfo
  {eid} {011101} (\bibinfo {year} {2012})},\ \Eprint
  {https://arxiv.org/abs/1106.1616} {arXiv:1106.1616 [astro-ph.SR]}
  \BibitemShut {NoStop}%
\bibitem [{\citenamefont {{Bauswein}}\ \emph {et~al.}(2012)\citenamefont
  {{Bauswein}}, \citenamefont {{Janka}}, \citenamefont {{Hebeler}},\ and\
  \citenamefont {{Schwenk}}}]{2012PhRvD..86f3001B}%
  \BibitemOpen
  \bibfield  {author} {\bibinfo {author} {\bibfnamefont {A.}~\bibnamefont
  {{Bauswein}}}, \bibinfo {author} {\bibfnamefont {H.~T.}\ \bibnamefont
  {{Janka}}}, \bibinfo {author} {\bibfnamefont {K.}~\bibnamefont {{Hebeler}}},\
  and\ \bibinfo {author} {\bibfnamefont {A.}~\bibnamefont {{Schwenk}}},\
  }\bibfield  {title} {\bibinfo {title} {{Equation-of-state dependence of the
  gravitational-wave signal from the ring-down phase of neutron-star
  mergers}},\ }\href {https://doi.org/10.1103/PhysRevD.86.063001} {\bibfield
  {journal} {\bibinfo  {journal} {\prd}\ }\textbf {\bibinfo {volume} {86}},\
  \bibinfo {eid} {063001} (\bibinfo {year} {2012})},\ \Eprint
  {https://arxiv.org/abs/1204.1888} {arXiv:1204.1888 [astro-ph.SR]}
  \BibitemShut {NoStop}%
\bibitem [{\citenamefont {{Hotokezaka}}\ \emph {et~al.}(2013)\citenamefont
  {{Hotokezaka}}, \citenamefont {{Kiuchi}}, \citenamefont {{Kyutoku}},
  \citenamefont {{Muranushi}}, \citenamefont {{Sekiguchi}}, \citenamefont
  {{Shibata}},\ and\ \citenamefont {{Taniguchi}}}]{2013PhRvD..88d4026H}%
  \BibitemOpen
  \bibfield  {author} {\bibinfo {author} {\bibfnamefont {K.}~\bibnamefont
  {{Hotokezaka}}}, \bibinfo {author} {\bibfnamefont {K.}~\bibnamefont
  {{Kiuchi}}}, \bibinfo {author} {\bibfnamefont {K.}~\bibnamefont {{Kyutoku}}},
  \bibinfo {author} {\bibfnamefont {T.}~\bibnamefont {{Muranushi}}}, \bibinfo
  {author} {\bibfnamefont {Y.}~\bibnamefont {{Sekiguchi}}}, \bibinfo {author}
  {\bibfnamefont {M.}~\bibnamefont {{Shibata}}},\ and\ \bibinfo {author}
  {\bibfnamefont {K.}~\bibnamefont {{Taniguchi}}},\ }\bibfield  {title}
  {\bibinfo {title} {{Remnant massive neutron stars of binary neutron star
  mergers: Evolution process and gravitational waveform}},\ }\href
  {https://doi.org/10.1103/PhysRevD.88.044026} {\bibfield  {journal} {\bibinfo
  {journal} {\prd}\ }\textbf {\bibinfo {volume} {88}},\ \bibinfo {eid} {044026}
  (\bibinfo {year} {2013})},\ \Eprint {https://arxiv.org/abs/1307.5888}
  {arXiv:1307.5888 [astro-ph.HE]} \BibitemShut {NoStop}%
\bibitem [{\citenamefont {{Takami}}\ \emph {et~al.}(2015)\citenamefont
  {{Takami}}, \citenamefont {{Rezzolla}},\ and\ \citenamefont
  {{Baiotti}}}]{2015PhRvD..91f4001T}%
  \BibitemOpen
  \bibfield  {author} {\bibinfo {author} {\bibfnamefont {K.}~\bibnamefont
  {{Takami}}}, \bibinfo {author} {\bibfnamefont {L.}~\bibnamefont
  {{Rezzolla}}},\ and\ \bibinfo {author} {\bibfnamefont {L.}~\bibnamefont
  {{Baiotti}}},\ }\bibfield  {title} {\bibinfo {title} {{Spectral properties of
  the post-merger gravitational-wave signal from binary neutron stars}},\
  }\href {https://doi.org/10.1103/PhysRevD.91.064001} {\bibfield  {journal}
  {\bibinfo  {journal} {\prd}\ }\textbf {\bibinfo {volume} {91}},\ \bibinfo
  {eid} {064001} (\bibinfo {year} {2015})},\ \Eprint
  {https://arxiv.org/abs/1412.3240} {arXiv:1412.3240 [gr-qc]} \BibitemShut
  {NoStop}%
\bibitem [{\citenamefont {{Bernuzzi}}\ \emph {et~al.}(2015)\citenamefont
  {{Bernuzzi}}, \citenamefont {{Dietrich}},\ and\ \citenamefont
  {{Nagar}}}]{2015PhRvL.115i1101B}%
  \BibitemOpen
  \bibfield  {author} {\bibinfo {author} {\bibfnamefont {S.}~\bibnamefont
  {{Bernuzzi}}}, \bibinfo {author} {\bibfnamefont {T.}~\bibnamefont
  {{Dietrich}}},\ and\ \bibinfo {author} {\bibfnamefont {A.}~\bibnamefont
  {{Nagar}}},\ }\bibfield  {title} {\bibinfo {title} {{Modeling the Complete
  Gravitational Wave Spectrum of Neutron Star Mergers}},\ }\href
  {https://doi.org/10.1103/PhysRevLett.115.091101} {\bibfield  {journal}
  {\bibinfo  {journal} {\prl}\ }\textbf {\bibinfo {volume} {115}},\ \bibinfo
  {eid} {091101} (\bibinfo {year} {2015})},\ \Eprint
  {https://arxiv.org/abs/1504.01764} {arXiv:1504.01764 [gr-qc]} \BibitemShut
  {NoStop}%
\bibitem [{\citenamefont {{Clark}}\ \emph {et~al.}(2016)\citenamefont
  {{Clark}}, \citenamefont {{Bauswein}}, \citenamefont {{Stergioulas}},\ and\
  \citenamefont {{Shoemaker}}}]{2016CQGra..33h5003C}%
  \BibitemOpen
  \bibfield  {author} {\bibinfo {author} {\bibfnamefont {J.~A.}\ \bibnamefont
  {{Clark}}}, \bibinfo {author} {\bibfnamefont {A.}~\bibnamefont {{Bauswein}}},
  \bibinfo {author} {\bibfnamefont {N.}~\bibnamefont {{Stergioulas}}},\ and\
  \bibinfo {author} {\bibfnamefont {D.}~\bibnamefont {{Shoemaker}}},\
  }\bibfield  {title} {\bibinfo {title} {{Observing gravitational waves from
  the post-merger phase of binary neutron star coalescence}},\ }\href
  {https://doi.org/10.1088/0264-9381/33/8/085003} {\bibfield  {journal}
  {\bibinfo  {journal} {Classical and Quantum Gravity}\ }\textbf {\bibinfo
  {volume} {33}},\ \bibinfo {eid} {085003} (\bibinfo {year} {2016})},\ \Eprint
  {https://arxiv.org/abs/1509.08522} {arXiv:1509.08522 [astro-ph.HE]}
  \BibitemShut {NoStop}%
\bibitem [{\citenamefont {{Bauswein}}\ \emph {et~al.}(2016)\citenamefont
  {{Bauswein}}, \citenamefont {{Stergioulas}},\ and\ \citenamefont
  {{Janka}}}]{Bauswein2016}%
  \BibitemOpen
  \bibfield  {author} {\bibinfo {author} {\bibfnamefont {A.}~\bibnamefont
  {{Bauswein}}}, \bibinfo {author} {\bibfnamefont {N.}~\bibnamefont
  {{Stergioulas}}},\ and\ \bibinfo {author} {\bibfnamefont {H.-T.}\
  \bibnamefont {{Janka}}},\ }\bibfield  {title} {\bibinfo {title} {{Exploring
  properties of high-density matter through remnants of neutron-star
  mergers}},\ }\href {https://doi.org/10.1140/epja/i2016-16056-7} {\bibfield
  {journal} {\bibinfo  {journal} {European Physical Journal A}\ }\textbf
  {\bibinfo {volume} {52}},\ \bibinfo {eid} {56} (\bibinfo {year}
  {2016})}\BibitemShut {NoStop}%
\bibitem [{\citenamefont {{Bauswein}}\ and\ \citenamefont
  {{Stergioulas}}(2019)}]{2019JPhG...46k3002B}%
  \BibitemOpen
  \bibfield  {author} {\bibinfo {author} {\bibfnamefont {A.}~\bibnamefont
  {{Bauswein}}}\ and\ \bibinfo {author} {\bibfnamefont {N.}~\bibnamefont
  {{Stergioulas}}},\ }\bibfield  {title} {\bibinfo {title} {{Spectral
  classification of gravitational-wave emission and equation of state
  constraints in binary neutron star mergers}},\ }\href
  {https://doi.org/10.1088/1361-6471/ab2b90} {\bibfield  {journal} {\bibinfo
  {journal} {Journal of Physics G Nuclear Physics}\ }\textbf {\bibinfo {volume}
  {46}},\ \bibinfo {eid} {113002} (\bibinfo {year} {2019})},\ \Eprint
  {https://arxiv.org/abs/1901.06969} {arXiv:1901.06969 [gr-qc]} \BibitemShut
  {NoStop}%
\bibitem [{\citenamefont {{Baiotti}}(2019)}]{2019PrPNP.10903714B}%
  \BibitemOpen
  \bibfield  {author} {\bibinfo {author} {\bibfnamefont {L.}~\bibnamefont
  {{Baiotti}}},\ }\bibfield  {title} {\bibinfo {title} {{Gravitational waves
  from neutron star mergers and their relation to the nuclear equation of
  state}},\ }\href {https://doi.org/10.1016/j.ppnp.2019.103714} {\bibfield
  {journal} {\bibinfo  {journal} {Progress in Particle and Nuclear Physics}\
  }\textbf {\bibinfo {volume} {109}},\ \bibinfo {eid} {103714} (\bibinfo {year}
  {2019})},\ \Eprint {https://arxiv.org/abs/1907.08534} {arXiv:1907.08534
  [astro-ph.HE]} \BibitemShut {NoStop}%
\bibitem [{\citenamefont {{Friedman}}\ and\ \citenamefont
  {{Stergioulas}}(2020)}]{2020IJMPD..2941015F}%
  \BibitemOpen
  \bibfield  {author} {\bibinfo {author} {\bibfnamefont {J.~L.}\ \bibnamefont
  {{Friedman}}}\ and\ \bibinfo {author} {\bibfnamefont {N.}~\bibnamefont
  {{Stergioulas}}},\ }\bibfield  {title} {\bibinfo {title} {{Astrophysical
  implications of neutron star inspiral and coalescence}},\ }\href
  {https://doi.org/10.1142/S0218271820410151} {\bibfield  {journal} {\bibinfo
  {journal} {International Journal of Modern Physics D}\ }\textbf {\bibinfo
  {volume} {29}},\ \bibinfo {eid} {2041015-632} (\bibinfo {year} {2020})},\
  \Eprint {https://arxiv.org/abs/2005.14135} {arXiv:2005.14135 [astro-ph.HE]}
  \BibitemShut {NoStop}%
\bibitem [{\citenamefont {{Bernuzzi}}(2020)}]{2020GReGr..52..108B}%
  \BibitemOpen
  \bibfield  {author} {\bibinfo {author} {\bibfnamefont {S.}~\bibnamefont
  {{Bernuzzi}}},\ }\bibfield  {title} {\bibinfo {title} {{Neutron star merger
  remnants}},\ }\href {https://doi.org/10.1007/s10714-020-02752-5} {\bibfield
  {journal} {\bibinfo  {journal} {General Relativity and Gravitation}\ }\textbf
  {\bibinfo {volume} {52}},\ \bibinfo {eid} {108} (\bibinfo {year} {2020})},\
  \Eprint {https://arxiv.org/abs/2004.06419} {arXiv:2004.06419 [astro-ph.HE]}
  \BibitemShut {NoStop}%
\bibitem [{\citenamefont {{Dietrich}}\ \emph {et~al.}(2020)\citenamefont
  {{Dietrich}}, \citenamefont {{Hinderer}},\ and\ \citenamefont
  {{Samajdar}}}]{2020arXiv200402527D}%
  \BibitemOpen
  \bibfield  {author} {\bibinfo {author} {\bibfnamefont {T.}~\bibnamefont
  {{Dietrich}}}, \bibinfo {author} {\bibfnamefont {T.}~\bibnamefont
  {{Hinderer}}},\ and\ \bibinfo {author} {\bibfnamefont {A.}~\bibnamefont
  {{Samajdar}}},\ }\bibfield  {title} {\bibinfo {title} {{Interpreting Binary
  Neutron Star Mergers: Describing the Binary Neutron Star Dynamics, Modelling
  Gravitational Waveforms, and Analyzing Detections}},\ }\href@noop {}
  {\bibfield  {journal} {\bibinfo  {journal} {arXiv e-prints}\ ,\ \bibinfo
  {eid} {arXiv:2004.02527}} (\bibinfo {year} {2020})},\ \Eprint
  {https://arxiv.org/abs/2004.02527} {arXiv:2004.02527 [gr-qc]} \BibitemShut
  {NoStop}%
\bibitem [{\citenamefont {{Sarin}}\ and\ \citenamefont
  {{Lasky}}(2020)}]{2020arXiv201208172S}%
  \BibitemOpen
  \bibfield  {author} {\bibinfo {author} {\bibfnamefont {N.}~\bibnamefont
  {{Sarin}}}\ and\ \bibinfo {author} {\bibfnamefont {P.~D.}\ \bibnamefont
  {{Lasky}}},\ }\bibfield  {title} {\bibinfo {title} {{The evolution of binary
  neutron star post-merger remnants: a review}},\ }\href@noop {} {\bibfield
  {journal} {\bibinfo  {journal} {arXiv e-prints}\ ,\ \bibinfo {eid}
  {arXiv:2012.08172}} (\bibinfo {year} {2020})},\ \Eprint
  {https://arxiv.org/abs/2012.08172} {arXiv:2012.08172 [astro-ph.HE]}
  \BibitemShut {NoStop}%
\bibitem [{\citenamefont {{Abbott}}\ \emph
  {et~al.}(2017{\natexlab{a}})\citenamefont {{Abbott}}, \citenamefont
  {{Abbott}}, \citenamefont {{Abbott}}, \citenamefont {{Acernese}},
  \citenamefont {{Ackley}}, \citenamefont {{Adams}}, \citenamefont {{Adams}},
  \citenamefont {{Addesso}}, \citenamefont {{Adhikari}}, \citenamefont {{Adya}}
  \emph {et~al.}}]{2017PhRvL.119p1101A}%
  \BibitemOpen
  \bibfield  {author} {\bibinfo {author} {\bibfnamefont {B.~P.}\ \bibnamefont
  {{Abbott}}}, \bibinfo {author} {\bibfnamefont {R.}~\bibnamefont {{Abbott}}},
  \bibinfo {author} {\bibfnamefont {T.~D.}\ \bibnamefont {{Abbott}}}, \bibinfo
  {author} {\bibfnamefont {F.}~\bibnamefont {{Acernese}}}, \bibinfo {author}
  {\bibfnamefont {K.}~\bibnamefont {{Ackley}}}, \bibinfo {author}
  {\bibfnamefont {C.}~\bibnamefont {{Adams}}}, \bibinfo {author} {\bibfnamefont
  {T.}~\bibnamefont {{Adams}}}, \bibinfo {author} {\bibfnamefont
  {P.}~\bibnamefont {{Addesso}}}, \bibinfo {author} {\bibfnamefont {R.~X.}\
  \bibnamefont {{Adhikari}}}, \bibinfo {author} {\bibfnamefont {V.~B.}\
  \bibnamefont {{Adya}}}, \emph {et~al.} (\bibinfo {collaboration} {LIGO
  Scientific Collaboration and Virgo Collaboration}),\ }\bibfield  {title}
  {\bibinfo {title} {{GW170817: Observation of Gravitational Waves from a
  Binary Neutron Star Inspiral}},\ }\href
  {https://doi.org/10.1103/PhysRevLett.119.161101} {\bibfield  {journal}
  {\bibinfo  {journal} {\prl}\ }\textbf {\bibinfo {volume} {119}},\ \bibinfo
  {eid} {161101} (\bibinfo {year} {2017}{\natexlab{a}})},\ \Eprint
  {https://arxiv.org/abs/1710.05832} {arXiv:1710.05832 [gr-qc]} \BibitemShut
  {NoStop}%
\bibitem [{\citenamefont {{Abbott}}\ \emph
  {et~al.}(2017{\natexlab{b}})\citenamefont {{Abbott}}, \citenamefont
  {{Abbott}}, \citenamefont {{Abbott}}, \citenamefont {{Acernese}},
  \citenamefont {{Ackley}}, \citenamefont {{Adams}}, \citenamefont {{Adams}},
  \citenamefont {{Addesso}}, \citenamefont {{Adhikari}}, \citenamefont {{Adya}}
  \emph {et~al.}}]{2017ApJ...851L..16A}%
  \BibitemOpen
  \bibfield  {author} {\bibinfo {author} {\bibfnamefont {B.~P.}\ \bibnamefont
  {{Abbott}}}, \bibinfo {author} {\bibfnamefont {R.}~\bibnamefont {{Abbott}}},
  \bibinfo {author} {\bibfnamefont {T.~D.}\ \bibnamefont {{Abbott}}}, \bibinfo
  {author} {\bibfnamefont {F.}~\bibnamefont {{Acernese}}}, \bibinfo {author}
  {\bibfnamefont {K.}~\bibnamefont {{Ackley}}}, \bibinfo {author}
  {\bibfnamefont {C.}~\bibnamefont {{Adams}}}, \bibinfo {author} {\bibfnamefont
  {T.}~\bibnamefont {{Adams}}}, \bibinfo {author} {\bibfnamefont
  {P.}~\bibnamefont {{Addesso}}}, \bibinfo {author} {\bibfnamefont {R.~X.}\
  \bibnamefont {{Adhikari}}}, \bibinfo {author} {\bibfnamefont {V.~B.}\
  \bibnamefont {{Adya}}}, \emph {et~al.} (\bibinfo {collaboration} {LIGO
  Scientific Collaboration and Virgo Collaboration}),\ }\bibfield  {title}
  {\bibinfo {title} {{Search for Post-merger Gravitational Waves from the
  Remnant of the Binary Neutron Star Merger GW170817}},\ }\href
  {https://doi.org/10.3847/2041-8213/aa9a35} {\bibfield  {journal} {\bibinfo
  {journal} {Astrophys. J. Lett.}\ }\textbf {\bibinfo {volume} {851}},\
  \bibinfo {eid} {L16} (\bibinfo {year} {2017}{\natexlab{b}})},\ \Eprint
  {https://arxiv.org/abs/1710.09320} {arXiv:1710.09320 [astro-ph.HE]}
  \BibitemShut {NoStop}%
\bibitem [{\citenamefont {{Abbott}}\ \emph {et~al.}(2020)\citenamefont
  {{Abbott}}, \citenamefont {{Abbott}}, \citenamefont {{Abbott}}, \citenamefont
  {{Abraham}}, \citenamefont {{Acernese}}, \citenamefont {{Ackley}},
  \citenamefont {{Adams}}, \citenamefont {{Adhikari}}, \citenamefont {{Adya}},
  \citenamefont {{Affeldt}} \emph {et~al.}}]{2020ApJ...892L...3A}%
  \BibitemOpen
  \bibfield  {author} {\bibinfo {author} {\bibfnamefont {B.~P.}\ \bibnamefont
  {{Abbott}}}, \bibinfo {author} {\bibfnamefont {R.}~\bibnamefont {{Abbott}}},
  \bibinfo {author} {\bibfnamefont {T.~D.}\ \bibnamefont {{Abbott}}}, \bibinfo
  {author} {\bibfnamefont {S.}~\bibnamefont {{Abraham}}}, \bibinfo {author}
  {\bibfnamefont {F.}~\bibnamefont {{Acernese}}}, \bibinfo {author}
  {\bibfnamefont {K.}~\bibnamefont {{Ackley}}}, \bibinfo {author}
  {\bibfnamefont {C.}~\bibnamefont {{Adams}}}, \bibinfo {author} {\bibfnamefont
  {R.~X.}\ \bibnamefont {{Adhikari}}}, \bibinfo {author} {\bibfnamefont
  {V.~B.}\ \bibnamefont {{Adya}}}, \bibinfo {author} {\bibfnamefont
  {C.}~\bibnamefont {{Affeldt}}}, \emph {et~al.},\ }\bibfield  {title}
  {\bibinfo {title} {{GW190425: Observation of a Compact Binary Coalescence
  with Total Mass {\ensuremath{\sim}} 3.4 M$_{\odot}$}},\ }\href
  {https://doi.org/10.3847/2041-8213/ab75f5} {\bibfield  {journal} {\bibinfo
  {journal} {Astrophys. J. Lett.}\ }\textbf {\bibinfo {volume} {892}},\
  \bibinfo {eid} {L3} (\bibinfo {year} {2020})},\ \Eprint
  {https://arxiv.org/abs/2001.01761} {arXiv:2001.01761 [astro-ph.HE]}
  \BibitemShut {NoStop}%
\bibitem [{\citenamefont {{Clark}}\ \emph {et~al.}(2014)\citenamefont
  {{Clark}}, \citenamefont {{Bauswein}}, \citenamefont {{Cadonati}},
  \citenamefont {{Janka}}, \citenamefont {{Pankow}},\ and\ \citenamefont
  {{Stergioulas}}}]{2014PhRvD..90f2004C}%
  \BibitemOpen
  \bibfield  {author} {\bibinfo {author} {\bibfnamefont {J.}~\bibnamefont
  {{Clark}}}, \bibinfo {author} {\bibfnamefont {A.}~\bibnamefont {{Bauswein}}},
  \bibinfo {author} {\bibfnamefont {L.}~\bibnamefont {{Cadonati}}}, \bibinfo
  {author} {\bibfnamefont {H.~T.}\ \bibnamefont {{Janka}}}, \bibinfo {author}
  {\bibfnamefont {C.}~\bibnamefont {{Pankow}}},\ and\ \bibinfo {author}
  {\bibfnamefont {N.}~\bibnamefont {{Stergioulas}}},\ }\bibfield  {title}
  {\bibinfo {title} {{Prospects for high frequency burst searches following
  binary neutron star coalescence with advanced gravitational wave
  detectors}},\ }\href {https://doi.org/10.1103/PhysRevD.90.062004} {\bibfield
  {journal} {\bibinfo  {journal} {\prd}\ }\textbf {\bibinfo {volume} {90}},\
  \bibinfo {eid} {062004} (\bibinfo {year} {2014})},\ \Eprint
  {https://arxiv.org/abs/1406.5444} {arXiv:1406.5444 [astro-ph.HE]}
  \BibitemShut {NoStop}%
\bibitem [{\citenamefont {{Chatziioannou}}\ \emph {et~al.}(2017)\citenamefont
  {{Chatziioannou}}, \citenamefont {{Clark}}, \citenamefont {{Bauswein}},
  \citenamefont {{Millhouse}}, \citenamefont {{Littenberg}},\ and\
  \citenamefont {{Cornish}}}]{2017PhRvD..96l4035C}%
  \BibitemOpen
  \bibfield  {author} {\bibinfo {author} {\bibfnamefont {K.}~\bibnamefont
  {{Chatziioannou}}}, \bibinfo {author} {\bibfnamefont {J.~A.}\ \bibnamefont
  {{Clark}}}, \bibinfo {author} {\bibfnamefont {A.}~\bibnamefont {{Bauswein}}},
  \bibinfo {author} {\bibfnamefont {M.}~\bibnamefont {{Millhouse}}}, \bibinfo
  {author} {\bibfnamefont {T.~B.}\ \bibnamefont {{Littenberg}}},\ and\ \bibinfo
  {author} {\bibfnamefont {N.}~\bibnamefont {{Cornish}}},\ }\bibfield  {title}
  {\bibinfo {title} {{Inferring the post-merger gravitational wave emission
  from binary neutron star coalescences}},\ }\href
  {https://doi.org/10.1103/PhysRevD.96.124035} {\bibfield  {journal} {\bibinfo
  {journal} {\prd}\ }\textbf {\bibinfo {volume} {96}},\ \bibinfo {eid} {124035}
  (\bibinfo {year} {2017})},\ \Eprint {https://arxiv.org/abs/1711.00040}
  {arXiv:1711.00040 [gr-qc]} \BibitemShut {NoStop}%
\bibitem [{\citenamefont {{Bose}}\ \emph {et~al.}(2018)\citenamefont {{Bose}},
  \citenamefont {{Chakravarti}}, \citenamefont {{Rezzolla}}, \citenamefont
  {{Sathyaprakash}},\ and\ \citenamefont {{Takami}}}]{2018PhRvL.120c1102B}%
  \BibitemOpen
  \bibfield  {author} {\bibinfo {author} {\bibfnamefont {S.}~\bibnamefont
  {{Bose}}}, \bibinfo {author} {\bibfnamefont {K.}~\bibnamefont
  {{Chakravarti}}}, \bibinfo {author} {\bibfnamefont {L.}~\bibnamefont
  {{Rezzolla}}}, \bibinfo {author} {\bibfnamefont {B.~S.}\ \bibnamefont
  {{Sathyaprakash}}},\ and\ \bibinfo {author} {\bibfnamefont {K.}~\bibnamefont
  {{Takami}}},\ }\bibfield  {title} {\bibinfo {title} {{Neutron-Star Radius
  from a Population of Binary Neutron Star Mergers}},\ }\href
  {https://doi.org/10.1103/PhysRevLett.120.031102} {\bibfield  {journal}
  {\bibinfo  {journal} {\prl}\ }\textbf {\bibinfo {volume} {120}},\ \bibinfo
  {eid} {031102} (\bibinfo {year} {2018})},\ \Eprint
  {https://arxiv.org/abs/1705.10850} {arXiv:1705.10850 [gr-qc]} \BibitemShut
  {NoStop}%
\bibitem [{\citenamefont {{Yang}}\ \emph {et~al.}(2018)\citenamefont {{Yang}},
  \citenamefont {{Paschalidis}}, \citenamefont {{Yagi}}, \citenamefont
  {{Lehner}}, \citenamefont {{Pretorius}},\ and\ \citenamefont
  {{Yunes}}}]{2018PhRvD..97b4049Y}%
  \BibitemOpen
  \bibfield  {author} {\bibinfo {author} {\bibfnamefont {H.}~\bibnamefont
  {{Yang}}}, \bibinfo {author} {\bibfnamefont {V.}~\bibnamefont
  {{Paschalidis}}}, \bibinfo {author} {\bibfnamefont {K.}~\bibnamefont
  {{Yagi}}}, \bibinfo {author} {\bibfnamefont {L.}~\bibnamefont {{Lehner}}},
  \bibinfo {author} {\bibfnamefont {F.}~\bibnamefont {{Pretorius}}},\ and\
  \bibinfo {author} {\bibfnamefont {N.}~\bibnamefont {{Yunes}}},\ }\bibfield
  {title} {\bibinfo {title} {{Gravitational wave spectroscopy of binary neutron
  star merger remnants with mode stacking}},\ }\href
  {https://doi.org/10.1103/PhysRevD.97.024049} {\bibfield  {journal} {\bibinfo
  {journal} {\prd}\ }\textbf {\bibinfo {volume} {97}},\ \bibinfo {eid} {024049}
  (\bibinfo {year} {2018})},\ \Eprint {https://arxiv.org/abs/1707.00207}
  {arXiv:1707.00207 [gr-qc]} \BibitemShut {NoStop}%
\bibitem [{\citenamefont {{Torres-Rivas}}\ \emph {et~al.}(2019)\citenamefont
  {{Torres-Rivas}}, \citenamefont {{Chatziioannou}}, \citenamefont
  {{Bauswein}},\ and\ \citenamefont {{Clark}}}]{2019PhRvD..99d4014T}%
  \BibitemOpen
  \bibfield  {author} {\bibinfo {author} {\bibfnamefont {A.}~\bibnamefont
  {{Torres-Rivas}}}, \bibinfo {author} {\bibfnamefont {K.}~\bibnamefont
  {{Chatziioannou}}}, \bibinfo {author} {\bibfnamefont {A.}~\bibnamefont
  {{Bauswein}}},\ and\ \bibinfo {author} {\bibfnamefont {J.~A.}\ \bibnamefont
  {{Clark}}},\ }\bibfield  {title} {\bibinfo {title} {{Observing the
  post-merger signal of GW170817-like events with improved gravitational-wave
  detectors}},\ }\href {https://doi.org/10.1103/PhysRevD.99.044014} {\bibfield
  {journal} {\bibinfo  {journal} {\prd}\ }\textbf {\bibinfo {volume} {99}},\
  \bibinfo {eid} {044014} (\bibinfo {year} {2019})},\ \Eprint
  {https://arxiv.org/abs/1811.08931} {arXiv:1811.08931 [gr-qc]} \BibitemShut
  {NoStop}%
\bibitem [{\citenamefont {{Martynov}}\ \emph {et~al.}(2019)\citenamefont
  {{Martynov}}, \citenamefont {{Miao}}, \citenamefont {{Yang}}, \citenamefont
  {{Vivanco}}, \citenamefont {{Thrane}}, \citenamefont {{Smith}}, \citenamefont
  {{Lasky}}, \citenamefont {{East}}, \citenamefont {{Adhikari}}, \citenamefont
  {{Bauswein}}, \citenamefont {{Brooks}}, \citenamefont {{Chen}}, \citenamefont
  {{Corbitt}}, \citenamefont {{Freise}}, \citenamefont {{Grote}}, \citenamefont
  {{Levin}}, \citenamefont {{Zhao}},\ and\ \citenamefont
  {{Vecchio}}}]{2019PhRvD..99j2004M}%
  \BibitemOpen
  \bibfield  {author} {\bibinfo {author} {\bibfnamefont {D.}~\bibnamefont
  {{Martynov}}}, \bibinfo {author} {\bibfnamefont {H.}~\bibnamefont {{Miao}}},
  \bibinfo {author} {\bibfnamefont {H.}~\bibnamefont {{Yang}}}, \bibinfo
  {author} {\bibfnamefont {F.~H.}\ \bibnamefont {{Vivanco}}}, \bibinfo {author}
  {\bibfnamefont {E.}~\bibnamefont {{Thrane}}}, \bibinfo {author}
  {\bibfnamefont {R.}~\bibnamefont {{Smith}}}, \bibinfo {author} {\bibfnamefont
  {P.}~\bibnamefont {{Lasky}}}, \bibinfo {author} {\bibfnamefont {W.~E.}\
  \bibnamefont {{East}}}, \bibinfo {author} {\bibfnamefont {R.}~\bibnamefont
  {{Adhikari}}}, \bibinfo {author} {\bibfnamefont {A.}~\bibnamefont
  {{Bauswein}}}, \bibinfo {author} {\bibfnamefont {A.}~\bibnamefont
  {{Brooks}}}, \bibinfo {author} {\bibfnamefont {Y.}~\bibnamefont {{Chen}}},
  \bibinfo {author} {\bibfnamefont {T.}~\bibnamefont {{Corbitt}}}, \bibinfo
  {author} {\bibfnamefont {A.}~\bibnamefont {{Freise}}}, \bibinfo {author}
  {\bibfnamefont {H.}~\bibnamefont {{Grote}}}, \bibinfo {author} {\bibfnamefont
  {Y.}~\bibnamefont {{Levin}}}, \bibinfo {author} {\bibfnamefont
  {C.}~\bibnamefont {{Zhao}}},\ and\ \bibinfo {author} {\bibfnamefont
  {A.}~\bibnamefont {{Vecchio}}},\ }\bibfield  {title} {\bibinfo {title}
  {{Exploring the sensitivity of gravitational wave detectors to neutron star
  physics}},\ }\href {https://doi.org/10.1103/PhysRevD.99.102004} {\bibfield
  {journal} {\bibinfo  {journal} {\prd}\ }\textbf {\bibinfo {volume} {99}},\
  \bibinfo {eid} {102004} (\bibinfo {year} {2019})},\ \Eprint
  {https://arxiv.org/abs/1901.03885} {arXiv:1901.03885 [astro-ph.IM]}
  \BibitemShut {NoStop}%
\bibitem [{\citenamefont {{Oliver}}\ \emph {et~al.}(2019)\citenamefont
  {{Oliver}}, \citenamefont {{Keitel}}, \citenamefont {{Miller}}, \citenamefont
  {{Estelles}},\ and\ \citenamefont {{Sintes}}}]{2019MNRAS.485..843O}%
  \BibitemOpen
  \bibfield  {author} {\bibinfo {author} {\bibfnamefont {M.}~\bibnamefont
  {{Oliver}}}, \bibinfo {author} {\bibfnamefont {D.}~\bibnamefont {{Keitel}}},
  \bibinfo {author} {\bibfnamefont {A.~L.}\ \bibnamefont {{Miller}}}, \bibinfo
  {author} {\bibfnamefont {H.}~\bibnamefont {{Estelles}}},\ and\ \bibinfo
  {author} {\bibfnamefont {A.~M.}\ \bibnamefont {{Sintes}}},\ }\bibfield
  {title} {\bibinfo {title} {{Matched-filter study and energy budget suggest no
  detectable gravitational-wave `extended emission' from GW170817}},\ }\href
  {https://doi.org/10.1093/mnras/stz439} {\bibfield  {journal} {\bibinfo
  {journal} {Mon. Not. R. Astron. Soc}\ }\textbf {\bibinfo {volume} {485}},\
  \bibinfo {pages} {843} (\bibinfo {year} {2019})},\ \Eprint
  {https://arxiv.org/abs/1812.06724} {arXiv:1812.06724 [astro-ph.HE]}
  \BibitemShut {NoStop}%
\bibitem [{\citenamefont {{Easter}}\ \emph {et~al.}(2019)\citenamefont
  {{Easter}}, \citenamefont {{Lasky}}, \citenamefont {{Casey}}, \citenamefont
  {{Rezzolla}},\ and\ \citenamefont {{Takami}}}]{2019PhRvD.100d3005E}%
  \BibitemOpen
  \bibfield  {author} {\bibinfo {author} {\bibfnamefont {P.~J.}\ \bibnamefont
  {{Easter}}}, \bibinfo {author} {\bibfnamefont {P.~D.}\ \bibnamefont
  {{Lasky}}}, \bibinfo {author} {\bibfnamefont {A.~R.}\ \bibnamefont
  {{Casey}}}, \bibinfo {author} {\bibfnamefont {L.}~\bibnamefont
  {{Rezzolla}}},\ and\ \bibinfo {author} {\bibfnamefont {K.}~\bibnamefont
  {{Takami}}},\ }\bibfield  {title} {\bibinfo {title} {{Computing fast and
  reliable gravitational waveforms of binary neutron star merger remnants}},\
  }\href {https://doi.org/10.1103/PhysRevD.100.043005} {\bibfield  {journal}
  {\bibinfo  {journal} {\prd}\ }\textbf {\bibinfo {volume} {100}},\ \bibinfo
  {eid} {043005} (\bibinfo {year} {2019})},\ \Eprint
  {https://arxiv.org/abs/1811.11183} {arXiv:1811.11183 [gr-qc]} \BibitemShut
  {NoStop}%
\bibitem [{\citenamefont {{Tsang}}\ \emph {et~al.}(2019)\citenamefont
  {{Tsang}}, \citenamefont {{Dietrich}},\ and\ \citenamefont {{Van Den
  Broeck}}}]{2019PhRvD.100d4047T}%
  \BibitemOpen
  \bibfield  {author} {\bibinfo {author} {\bibfnamefont {K.~W.}\ \bibnamefont
  {{Tsang}}}, \bibinfo {author} {\bibfnamefont {T.}~\bibnamefont
  {{Dietrich}}},\ and\ \bibinfo {author} {\bibfnamefont {C.}~\bibnamefont {{Van
  Den Broeck}}},\ }\bibfield  {title} {\bibinfo {title} {{Modeling the
  postmerger gravitational wave signal and extracting binary properties from
  future binary neutron star detections}},\ }\href
  {https://doi.org/10.1103/PhysRevD.100.044047} {\bibfield  {journal} {\bibinfo
   {journal} {\prd}\ }\textbf {\bibinfo {volume} {100}},\ \bibinfo {eid}
  {044047} (\bibinfo {year} {2019})},\ \Eprint
  {https://arxiv.org/abs/1907.02424} {arXiv:1907.02424 [gr-qc]} \BibitemShut
  {NoStop}%
\bibitem [{\citenamefont {{Breschi}}\ \emph {et~al.}(2019)\citenamefont
  {{Breschi}}, \citenamefont {{Bernuzzi}}, \citenamefont {{Zappa}},
  \citenamefont {{Agathos}}, \citenamefont {{Perego}}, \citenamefont
  {{Radice}},\ and\ \citenamefont {{Nagar}}}]{2019PhRvD.100j4029B}%
  \BibitemOpen
  \bibfield  {author} {\bibinfo {author} {\bibfnamefont {M.}~\bibnamefont
  {{Breschi}}}, \bibinfo {author} {\bibfnamefont {S.}~\bibnamefont
  {{Bernuzzi}}}, \bibinfo {author} {\bibfnamefont {F.}~\bibnamefont {{Zappa}}},
  \bibinfo {author} {\bibfnamefont {M.}~\bibnamefont {{Agathos}}}, \bibinfo
  {author} {\bibfnamefont {A.}~\bibnamefont {{Perego}}}, \bibinfo {author}
  {\bibfnamefont {D.}~\bibnamefont {{Radice}}},\ and\ \bibinfo {author}
  {\bibfnamefont {A.}~\bibnamefont {{Nagar}}},\ }\bibfield  {title} {\bibinfo
  {title} {{Kilohertz gravitational waves from binary neutron star remnants:
  Time-domain model and constraints on extreme matter}},\ }\href
  {https://doi.org/10.1103/PhysRevD.100.104029} {\bibfield  {journal} {\bibinfo
   {journal} {\prd}\ }\textbf {\bibinfo {volume} {100}},\ \bibinfo {eid}
  {104029} (\bibinfo {year} {2019})},\ \Eprint
  {https://arxiv.org/abs/1908.11418} {arXiv:1908.11418 [gr-qc]} \BibitemShut
  {NoStop}%
\bibitem [{\citenamefont {{Hall}}\ and\ \citenamefont
  {{Evans}}(2019)}]{2019CQGra..36v5002H}%
  \BibitemOpen
  \bibfield  {author} {\bibinfo {author} {\bibfnamefont {E.~D.}\ \bibnamefont
  {{Hall}}}\ and\ \bibinfo {author} {\bibfnamefont {M.}~\bibnamefont
  {{Evans}}},\ }\bibfield  {title} {\bibinfo {title} {{Metrics for
  next-generation gravitational-wave detectors}},\ }\href
  {https://doi.org/10.1088/1361-6382/ab41d6} {\bibfield  {journal} {\bibinfo
  {journal} {Classical and Quantum Gravity}\ }\textbf {\bibinfo {volume}
  {36}},\ \bibinfo {eid} {225002} (\bibinfo {year} {2019})},\ \Eprint
  {https://arxiv.org/abs/1902.09485} {arXiv:1902.09485 [astro-ph.IM]}
  \BibitemShut {NoStop}%
\bibitem [{\citenamefont {{Easter}}\ \emph {et~al.}(2020)\citenamefont
  {{Easter}}, \citenamefont {{Ghonge}}, \citenamefont {{Lasky}}, \citenamefont
  {{Casey}}, \citenamefont {{Clark}}, \citenamefont {{Hernandez Vivanco}},\
  and\ \citenamefont {{Chatziioannou}}}]{2020PhRvD.102d3011E}%
  \BibitemOpen
  \bibfield  {author} {\bibinfo {author} {\bibfnamefont {P.~J.}\ \bibnamefont
  {{Easter}}}, \bibinfo {author} {\bibfnamefont {S.}~\bibnamefont {{Ghonge}}},
  \bibinfo {author} {\bibfnamefont {P.~D.}\ \bibnamefont {{Lasky}}}, \bibinfo
  {author} {\bibfnamefont {A.~R.}\ \bibnamefont {{Casey}}}, \bibinfo {author}
  {\bibfnamefont {J.~A.}\ \bibnamefont {{Clark}}}, \bibinfo {author}
  {\bibfnamefont {F.}~\bibnamefont {{Hernandez Vivanco}}},\ and\ \bibinfo
  {author} {\bibfnamefont {K.}~\bibnamefont {{Chatziioannou}}},\ }\bibfield
  {title} {\bibinfo {title} {{Detection and parameter estimation of binary
  neutron star merger remnants}},\ }\href
  {https://doi.org/10.1103/PhysRevD.102.043011} {\bibfield  {journal} {\bibinfo
   {journal} {\prd}\ }\textbf {\bibinfo {volume} {102}},\ \bibinfo {eid}
  {043011} (\bibinfo {year} {2020})},\ \Eprint
  {https://arxiv.org/abs/2006.04396} {arXiv:2006.04396 [astro-ph.HE]}
  \BibitemShut {NoStop}%
\bibitem [{\citenamefont {{Ackley}}\ \emph {et~al.}(2020)\citenamefont
  {{Ackley}}, \citenamefont {{Adya}}, \citenamefont {{Agrawal}}, \citenamefont
  {{Altin}}, \citenamefont {{Ashton}}, \citenamefont {{Bailes}}, \citenamefont
  {{Baltinas}}, \citenamefont {{Barbuio}}, \citenamefont {{Beniwal}},
  \citenamefont {{Blair}} \emph {et~al.}}]{2020PASA...37...47A}%
  \BibitemOpen
  \bibfield  {author} {\bibinfo {author} {\bibfnamefont {K.}~\bibnamefont
  {{Ackley}}}, \bibinfo {author} {\bibfnamefont {V.~B.}\ \bibnamefont
  {{Adya}}}, \bibinfo {author} {\bibfnamefont {P.}~\bibnamefont {{Agrawal}}},
  \bibinfo {author} {\bibfnamefont {P.}~\bibnamefont {{Altin}}}, \bibinfo
  {author} {\bibfnamefont {G.}~\bibnamefont {{Ashton}}}, \bibinfo {author}
  {\bibfnamefont {M.}~\bibnamefont {{Bailes}}}, \bibinfo {author}
  {\bibfnamefont {E.}~\bibnamefont {{Baltinas}}}, \bibinfo {author}
  {\bibfnamefont {A.}~\bibnamefont {{Barbuio}}}, \bibinfo {author}
  {\bibfnamefont {D.}~\bibnamefont {{Beniwal}}}, \bibinfo {author}
  {\bibfnamefont {C.}~\bibnamefont {{Blair}}}, \emph {et~al.},\ }\bibfield
  {title} {\bibinfo {title} {{Neutron Star Extreme Matter Observatory: A
  kilohertz-band gravitational-wave detector in the global network}},\ }\href
  {https://doi.org/10.1017/pasa.2020.39} {\bibfield  {journal} {\bibinfo
  {journal} {Publ. Astron. Soc. Aust.}\ }\textbf {\bibinfo {volume} {37}},\
  \bibinfo {eid} {e047} (\bibinfo {year} {2020})},\ \Eprint
  {https://arxiv.org/abs/2007.03128} {arXiv:2007.03128 [astro-ph.HE]}
  \BibitemShut {NoStop}%
\bibitem [{\citenamefont {{Haster}}\ \emph {et~al.}(2020)\citenamefont
  {{Haster}}, \citenamefont {{Chatziioannou}}, \citenamefont {{Bauswein}},\
  and\ \citenamefont {{Clark}}}]{2020PhRvL.125z1101H}%
  \BibitemOpen
  \bibfield  {author} {\bibinfo {author} {\bibfnamefont {C.-J.}\ \bibnamefont
  {{Haster}}}, \bibinfo {author} {\bibfnamefont {K.}~\bibnamefont
  {{Chatziioannou}}}, \bibinfo {author} {\bibfnamefont {A.}~\bibnamefont
  {{Bauswein}}},\ and\ \bibinfo {author} {\bibfnamefont {J.~A.}\ \bibnamefont
  {{Clark}}},\ }\bibfield  {title} {\bibinfo {title} {{Inference of the Neutron
  Star Equation of State from Cosmological Distances}},\ }\href
  {https://doi.org/10.1103/PhysRevLett.125.261101} {\bibfield  {journal}
  {\bibinfo  {journal} {\prl}\ }\textbf {\bibinfo {volume} {125}},\ \bibinfo
  {eid} {261101} (\bibinfo {year} {2020})},\ \Eprint
  {https://arxiv.org/abs/2004.11334} {arXiv:2004.11334 [gr-qc]} \BibitemShut
  {NoStop}%
\bibitem [{\citenamefont {{Page}}\ \emph {et~al.}(2020)\citenamefont {{Page}},
  \citenamefont {{Goryachev}}, \citenamefont {{Miao}}, \citenamefont {{Chen}},
  \citenamefont {{Ma}}, \citenamefont {{Mason}}, \citenamefont {{Rossi}},
  \citenamefont {{Blair}}, \citenamefont {{Ju}}, \citenamefont {{Blair}},
  \citenamefont {{Schliesser}}, \citenamefont {{Tobar}},\ and\ \citenamefont
  {{Zhao}}}]{2020arXiv200708766P}%
  \BibitemOpen
  \bibfield  {author} {\bibinfo {author} {\bibfnamefont {M.~A.}\ \bibnamefont
  {{Page}}}, \bibinfo {author} {\bibfnamefont {M.}~\bibnamefont {{Goryachev}}},
  \bibinfo {author} {\bibfnamefont {H.}~\bibnamefont {{Miao}}}, \bibinfo
  {author} {\bibfnamefont {Y.}~\bibnamefont {{Chen}}}, \bibinfo {author}
  {\bibfnamefont {Y.}~\bibnamefont {{Ma}}}, \bibinfo {author} {\bibfnamefont
  {D.}~\bibnamefont {{Mason}}}, \bibinfo {author} {\bibfnamefont
  {M.}~\bibnamefont {{Rossi}}}, \bibinfo {author} {\bibfnamefont {C.~D.}\
  \bibnamefont {{Blair}}}, \bibinfo {author} {\bibfnamefont {L.}~\bibnamefont
  {{Ju}}}, \bibinfo {author} {\bibfnamefont {D.~G.}\ \bibnamefont {{Blair}}},
  \bibinfo {author} {\bibfnamefont {A.}~\bibnamefont {{Schliesser}}}, \bibinfo
  {author} {\bibfnamefont {M.~E.}\ \bibnamefont {{Tobar}}},\ and\ \bibinfo
  {author} {\bibfnamefont {C.}~\bibnamefont {{Zhao}}},\ }\bibfield  {title}
  {\bibinfo {title} {{Gravitational wave detectors with broadband high
  frequency sensitivity}},\ }\href@noop {} {\bibfield  {journal} {\bibinfo
  {journal} {arXiv e-prints}\ ,\ \bibinfo {eid} {arXiv:2007.08766}} (\bibinfo
  {year} {2020})},\ \Eprint {https://arxiv.org/abs/2007.08766}
  {arXiv:2007.08766 [physics.optics]} \BibitemShut {NoStop}%
\bibitem [{\citenamefont {{Aggarwal}}\ \emph {et~al.}(2020)\citenamefont
  {{Aggarwal}}, \citenamefont {{Aguiar}}, \citenamefont {{Bauswein}},
  \citenamefont {{Cella}}, \citenamefont {{Clesse}}, \citenamefont {{Cruise}},
  \citenamefont {{Domcke}}, \citenamefont {{Figueroa}}, \citenamefont
  {{Geraci}}, \citenamefont {{Goryachev}} \emph
  {et~al.}}]{2020arXiv201112414A}%
  \BibitemOpen
  \bibfield  {author} {\bibinfo {author} {\bibfnamefont {N.}~\bibnamefont
  {{Aggarwal}}}, \bibinfo {author} {\bibfnamefont {O.~D.}\ \bibnamefont
  {{Aguiar}}}, \bibinfo {author} {\bibfnamefont {A.}~\bibnamefont
  {{Bauswein}}}, \bibinfo {author} {\bibfnamefont {G.}~\bibnamefont {{Cella}}},
  \bibinfo {author} {\bibfnamefont {S.}~\bibnamefont {{Clesse}}}, \bibinfo
  {author} {\bibfnamefont {A.~M.}\ \bibnamefont {{Cruise}}}, \bibinfo {author}
  {\bibfnamefont {V.}~\bibnamefont {{Domcke}}}, \bibinfo {author}
  {\bibfnamefont {D.~G.}\ \bibnamefont {{Figueroa}}}, \bibinfo {author}
  {\bibfnamefont {A.}~\bibnamefont {{Geraci}}}, \bibinfo {author}
  {\bibfnamefont {M.}~\bibnamefont {{Goryachev}}}, \emph {et~al.},\ }\bibfield
  {title} {\bibinfo {title} {{Challenges and Opportunities of Gravitational
  Wave Searches at MHz to GHz Frequencies}},\ }\href@noop {} {\bibfield
  {journal} {\bibinfo  {journal} {arXiv e-prints}\ ,\ \bibinfo {eid}
  {arXiv:2011.12414}} (\bibinfo {year} {2020})},\ \Eprint
  {https://arxiv.org/abs/2011.12414} {arXiv:2011.12414 [gr-qc]} \BibitemShut
  {NoStop}%
\bibitem [{\citenamefont {{Ganapathy}}\ \emph {et~al.}(2021)\citenamefont
  {{Ganapathy}}, \citenamefont {{McCuller}}, \citenamefont {{Rollins}},
  \citenamefont {{Hall}}, \citenamefont {{Barsotti}},\ and\ \citenamefont
  {{Evans}}}]{2021PhRvD.103b2002G}%
  \BibitemOpen
  \bibfield  {author} {\bibinfo {author} {\bibfnamefont {D.}~\bibnamefont
  {{Ganapathy}}}, \bibinfo {author} {\bibfnamefont {L.}~\bibnamefont
  {{McCuller}}}, \bibinfo {author} {\bibfnamefont {J.~G.}\ \bibnamefont
  {{Rollins}}}, \bibinfo {author} {\bibfnamefont {E.~D.}\ \bibnamefont
  {{Hall}}}, \bibinfo {author} {\bibfnamefont {L.}~\bibnamefont {{Barsotti}}},\
  and\ \bibinfo {author} {\bibfnamefont {M.}~\bibnamefont {{Evans}}},\
  }\bibfield  {title} {\bibinfo {title} {{Tuning Advanced LIGO to kilohertz
  signals from neutron-star collisions}},\ }\href
  {https://doi.org/10.1103/PhysRevD.103.022002} {\bibfield  {journal} {\bibinfo
   {journal} {\prd}\ }\textbf {\bibinfo {volume} {103}},\ \bibinfo {eid}
  {022002} (\bibinfo {year} {2021})},\ \Eprint
  {https://arxiv.org/abs/2010.15735} {arXiv:2010.15735 [astro-ph.IM]}
  \BibitemShut {NoStop}%
\bibitem [{\citenamefont {{Lattimer}}\ and\ \citenamefont
  {{Prakash}}(2007)}]{2007PhR...442..109L}%
  \BibitemOpen
  \bibfield  {author} {\bibinfo {author} {\bibfnamefont {J.~M.}\ \bibnamefont
  {{Lattimer}}}\ and\ \bibinfo {author} {\bibfnamefont {M.}~\bibnamefont
  {{Prakash}}},\ }\bibfield  {title} {\bibinfo {title} {{Neutron star
  observations: Prognosis for equation of state constraints}},\ }\href
  {https://doi.org/10.1016/j.physrep.2007.02.003} {\bibfield  {journal}
  {\bibinfo  {journal} {Phys. Rep.}\ }\textbf {\bibinfo {volume} {442}},\
  \bibinfo {pages} {109} (\bibinfo {year} {2007})},\ \Eprint
  {https://arxiv.org/abs/astro-ph/0612440} {arXiv:astro-ph/0612440 [astro-ph]}
  \BibitemShut {NoStop}%
\bibitem [{\citenamefont {{Lattimer}}(2012)}]{2012ARNPS..62..485L}%
  \BibitemOpen
  \bibfield  {author} {\bibinfo {author} {\bibfnamefont {J.~M.}\ \bibnamefont
  {{Lattimer}}},\ }\bibfield  {title} {\bibinfo {title} {{The Nuclear Equation
  of State and Neutron Star Masses}},\ }\href
  {https://doi.org/10.1146/annurev-nucl-102711-095018} {\bibfield  {journal}
  {\bibinfo  {journal} {Annual Review of Nuclear and Particle Science}\
  }\textbf {\bibinfo {volume} {62}},\ \bibinfo {pages} {485} (\bibinfo {year}
  {2012})},\ \Eprint {https://arxiv.org/abs/1305.3510} {arXiv:1305.3510
  [nucl-th]} \BibitemShut {NoStop}%
\bibitem [{\citenamefont {{{\"O}zel}}\ and\ \citenamefont
  {{Freire}}(2016)}]{2016ARA&A..54..401O}%
  \BibitemOpen
  \bibfield  {author} {\bibinfo {author} {\bibfnamefont {F.}~\bibnamefont
  {{{\"O}zel}}}\ and\ \bibinfo {author} {\bibfnamefont {P.}~\bibnamefont
  {{Freire}}},\ }\bibfield  {title} {\bibinfo {title} {{Masses, Radii, and the
  Equation of State of Neutron Stars}},\ }\href
  {https://doi.org/10.1146/annurev-astro-081915-023322} {\bibfield  {journal}
  {\bibinfo  {journal} {Annu. Rev. Astron. Astrophys.}\ }\textbf {\bibinfo
  {volume} {54}},\ \bibinfo {pages} {401} (\bibinfo {year} {2016})},\ \Eprint
  {https://arxiv.org/abs/1603.02698} {arXiv:1603.02698 [astro-ph.HE]}
  \BibitemShut {NoStop}%
\bibitem [{\citenamefont {{Oertel}}\ \emph {et~al.}(2017)\citenamefont
  {{Oertel}}, \citenamefont {{Hempel}}, \citenamefont {{Kl{\"a}hn}},\ and\
  \citenamefont {{Typel}}}]{2017RvMP...89a5007O}%
  \BibitemOpen
  \bibfield  {author} {\bibinfo {author} {\bibfnamefont {M.}~\bibnamefont
  {{Oertel}}}, \bibinfo {author} {\bibfnamefont {M.}~\bibnamefont {{Hempel}}},
  \bibinfo {author} {\bibfnamefont {T.}~\bibnamefont {{Kl{\"a}hn}}},\ and\
  \bibinfo {author} {\bibfnamefont {S.}~\bibnamefont {{Typel}}},\ }\bibfield
  {title} {\bibinfo {title} {{Equations of state for supernovae and compact
  stars}},\ }\href {https://doi.org/10.1103/RevModPhys.89.015007} {\bibfield
  {journal} {\bibinfo  {journal} {Reviews of Modern Physics}\ }\textbf
  {\bibinfo {volume} {89}},\ \bibinfo {eid} {015007} (\bibinfo {year}
  {2017})},\ \Eprint {https://arxiv.org/abs/1610.03361} {arXiv:1610.03361
  [astro-ph.HE]} \BibitemShut {NoStop}%
\bibitem [{\citenamefont {{Lalit}}\ \emph {et~al.}(2019)\citenamefont
  {{Lalit}}, \citenamefont {{Mamun}}, \citenamefont {{Constantinou}},\ and\
  \citenamefont {{Prakash}}}]{2019EPJA...55...10L}%
  \BibitemOpen
  \bibfield  {author} {\bibinfo {author} {\bibfnamefont {S.}~\bibnamefont
  {{Lalit}}}, \bibinfo {author} {\bibfnamefont {M.~A.~A.}\ \bibnamefont
  {{Mamun}}}, \bibinfo {author} {\bibfnamefont {C.}~\bibnamefont
  {{Constantinou}}},\ and\ \bibinfo {author} {\bibfnamefont {M.}~\bibnamefont
  {{Prakash}}},\ }\bibfield  {title} {\bibinfo {title} {{Dense matter equation
  of state for neutron star mergers}},\ }\href
  {https://doi.org/10.1140/epja/i2019-12670-1} {\bibfield  {journal} {\bibinfo
  {journal} {European Physical Journal A}\ }\textbf {\bibinfo {volume} {55}},\
  \bibinfo {eid} {10} (\bibinfo {year} {2019})},\ \Eprint
  {https://arxiv.org/abs/1809.08126} {arXiv:1809.08126 [astro-ph.HE]}
  \BibitemShut {NoStop}%
\bibitem [{\citenamefont {{Hebeler}}(2021)}]{2021PhR...890....1H}%
  \BibitemOpen
  \bibfield  {author} {\bibinfo {author} {\bibfnamefont {K.}~\bibnamefont
  {{Hebeler}}},\ }\bibfield  {title} {\bibinfo {title} {{Three-nucleon forces:
  Implementation and applications to atomic nuclei and dense matter}},\ }\href
  {https://doi.org/10.1016/j.physrep.2020.08.009} {\bibfield  {journal}
  {\bibinfo  {journal} {Phys. Rep}\ }\textbf {\bibinfo {volume} {890}},\
  \bibinfo {pages} {1} (\bibinfo {year} {2021})},\ \Eprint
  {https://arxiv.org/abs/2002.09548} {arXiv:2002.09548 [nucl-th]} \BibitemShut
  {NoStop}%
\bibitem [{\citenamefont {{Andersson}}\ and\ \citenamefont
  {{Kokkotas}}(1998)}]{1998MNRAS.299.1059A}%
  \BibitemOpen
  \bibfield  {author} {\bibinfo {author} {\bibfnamefont {N.}~\bibnamefont
  {{Andersson}}}\ and\ \bibinfo {author} {\bibfnamefont {K.~D.}\ \bibnamefont
  {{Kokkotas}}},\ }\bibfield  {title} {\bibinfo {title} {{Towards gravitational
  wave asteroseismology}},\ }\href
  {https://doi.org/10.1046/j.1365-8711.1998.01840.x} {\bibfield  {journal}
  {\bibinfo  {journal} {Mon. Not. R. Astron. Soc}\ }\textbf {\bibinfo {volume}
  {299}},\ \bibinfo {pages} {1059} (\bibinfo {year} {1998})},\ \Eprint
  {https://arxiv.org/abs/gr-qc/9711088} {arXiv:gr-qc/9711088 [gr-qc]}
  \BibitemShut {NoStop}%
\bibitem [{\citenamefont {{Tsui}}\ and\ \citenamefont
  {{Leung}}(2005)}]{2005MNRAS.357.1029T}%
  \BibitemOpen
  \bibfield  {author} {\bibinfo {author} {\bibfnamefont {L.~K.}\ \bibnamefont
  {{Tsui}}}\ and\ \bibinfo {author} {\bibfnamefont {P.~T.}\ \bibnamefont
  {{Leung}}},\ }\bibfield  {title} {\bibinfo {title} {{Universality in
  quasi-normal modes of neutron stars}},\ }\href
  {https://doi.org/10.1111/j.1365-2966.2005.08710.x} {\bibfield  {journal}
  {\bibinfo  {journal} {Mon. Not. R. Astron. Soc}\ }\textbf {\bibinfo {volume}
  {357}},\ \bibinfo {pages} {1029} (\bibinfo {year} {2005})},\ \Eprint
  {https://arxiv.org/abs/gr-qc/0412024} {arXiv:gr-qc/0412024 [gr-qc]}
  \BibitemShut {NoStop}%
\bibitem [{\citenamefont {{Lau}}\ \emph {et~al.}(2010)\citenamefont {{Lau}},
  \citenamefont {{Leung}},\ and\ \citenamefont {{Lin}}}]{2010ApJ...714.1234L}%
  \BibitemOpen
  \bibfield  {author} {\bibinfo {author} {\bibfnamefont {H.~K.}\ \bibnamefont
  {{Lau}}}, \bibinfo {author} {\bibfnamefont {P.~T.}\ \bibnamefont {{Leung}}},\
  and\ \bibinfo {author} {\bibfnamefont {L.~M.}\ \bibnamefont {{Lin}}},\
  }\bibfield  {title} {\bibinfo {title} {{Inferring Physical Parameters of
  Compact Stars from their f-mode Gravitational Wave Signals}},\ }\href
  {https://doi.org/10.1088/0004-637X/714/2/1234} {\bibfield  {journal}
  {\bibinfo  {journal} {\apj}\ }\textbf {\bibinfo {volume} {714}},\ \bibinfo
  {pages} {1234} (\bibinfo {year} {2010})},\ \Eprint
  {https://arxiv.org/abs/0911.0131} {arXiv:0911.0131 [gr-qc]} \BibitemShut
  {NoStop}%
\bibitem [{\citenamefont {{Chan}}\ \emph {et~al.}(2014)\citenamefont {{Chan}},
  \citenamefont {{Sham}}, \citenamefont {{Leung}},\ and\ \citenamefont
  {{Lin}}}]{2014PhRvD..90l4023C}%
  \BibitemOpen
  \bibfield  {author} {\bibinfo {author} {\bibfnamefont {T.~K.}\ \bibnamefont
  {{Chan}}}, \bibinfo {author} {\bibfnamefont {Y.~H.}\ \bibnamefont {{Sham}}},
  \bibinfo {author} {\bibfnamefont {P.~T.}\ \bibnamefont {{Leung}}},\ and\
  \bibinfo {author} {\bibfnamefont {L.~M.}\ \bibnamefont {{Lin}}},\ }\bibfield
  {title} {\bibinfo {title} {{Multipolar universal relations between f -mode
  frequency and tidal deformability of compact stars}},\ }\href
  {https://doi.org/10.1103/PhysRevD.90.124023} {\bibfield  {journal} {\bibinfo
  {journal} {Phys. Rev. D}\ }\textbf {\bibinfo {volume} {90}},\ \bibinfo {eid}
  {124023} (\bibinfo {year} {2014})},\ \Eprint
  {https://arxiv.org/abs/1408.3789} {arXiv:1408.3789 [gr-qc]} \BibitemShut
  {NoStop}%
\bibitem [{\citenamefont {{Vretinaris}}\ \emph {et~al.}(2020)\citenamefont
  {{Vretinaris}}, \citenamefont {{Stergioulas}},\ and\ \citenamefont
  {{Bauswein}}}]{2020PhRvD.101h4039V}%
  \BibitemOpen
  \bibfield  {author} {\bibinfo {author} {\bibfnamefont {S.}~\bibnamefont
  {{Vretinaris}}}, \bibinfo {author} {\bibfnamefont {N.}~\bibnamefont
  {{Stergioulas}}},\ and\ \bibinfo {author} {\bibfnamefont {A.}~\bibnamefont
  {{Bauswein}}},\ }\bibfield  {title} {\bibinfo {title} {{Empirical relations
  for gravitational-wave asteroseismology of binary neutron star mergers}},\
  }\href {https://doi.org/10.1103/PhysRevD.101.084039} {\bibfield  {journal}
  {\bibinfo  {journal} {\prd}\ }\textbf {\bibinfo {volume} {101}},\ \bibinfo
  {eid} {084039} (\bibinfo {year} {2020})},\ \Eprint
  {https://arxiv.org/abs/1910.10856} {arXiv:1910.10856 [gr-qc]} \BibitemShut
  {NoStop}%
\bibitem [{\citenamefont {{Blacker}}\ \emph {et~al.}(2020)\citenamefont
  {{Blacker}}, \citenamefont {{Bastian}}, \citenamefont {{Bauswein}},
  \citenamefont {{Blaschke}}, \citenamefont {{Fischer}}, \citenamefont
  {{Oertel}}, \citenamefont {{Soultanis}},\ and\ \citenamefont
  {{Typel}}}]{2020PhRvD.102l3023B}%
  \BibitemOpen
  \bibfield  {author} {\bibinfo {author} {\bibfnamefont {S.}~\bibnamefont
  {{Blacker}}}, \bibinfo {author} {\bibfnamefont {N.-U.~F.}\ \bibnamefont
  {{Bastian}}}, \bibinfo {author} {\bibfnamefont {A.}~\bibnamefont
  {{Bauswein}}}, \bibinfo {author} {\bibfnamefont {D.~B.}\ \bibnamefont
  {{Blaschke}}}, \bibinfo {author} {\bibfnamefont {T.}~\bibnamefont
  {{Fischer}}}, \bibinfo {author} {\bibfnamefont {M.}~\bibnamefont {{Oertel}}},
  \bibinfo {author} {\bibfnamefont {T.}~\bibnamefont {{Soultanis}}},\ and\
  \bibinfo {author} {\bibfnamefont {S.}~\bibnamefont {{Typel}}},\ }\bibfield
  {title} {\bibinfo {title} {{Constraining the onset density of the
  hadron-quark phase transition with gravitational-wave observations}},\ }\href
  {https://doi.org/10.1103/PhysRevD.102.123023} {\bibfield  {journal} {\bibinfo
   {journal} {\prd}\ }\textbf {\bibinfo {volume} {102}},\ \bibinfo {eid}
  {123023} (\bibinfo {year} {2020})},\ \Eprint
  {https://arxiv.org/abs/2006.03789} {arXiv:2006.03789 [astro-ph.HE]}
  \BibitemShut {NoStop}%
\bibitem [{\citenamefont {{Chirenti}}\ \emph {et~al.}(2015)\citenamefont
  {{Chirenti}}, \citenamefont {{de Souza}},\ and\ \citenamefont
  {{Kastaun}}}]{2015PhRvD..91d4034C}%
  \BibitemOpen
  \bibfield  {author} {\bibinfo {author} {\bibfnamefont {C.}~\bibnamefont
  {{Chirenti}}}, \bibinfo {author} {\bibfnamefont {G.~H.}\ \bibnamefont {{de
  Souza}}},\ and\ \bibinfo {author} {\bibfnamefont {W.}~\bibnamefont
  {{Kastaun}}},\ }\bibfield  {title} {\bibinfo {title} {{Fundamental
  oscillation modes of neutron stars: Validity of universal relations}},\
  }\href {https://doi.org/10.1103/PhysRevD.91.044034} {\bibfield  {journal}
  {\bibinfo  {journal} {\prd}\ }\textbf {\bibinfo {volume} {91}},\ \bibinfo
  {eid} {044034} (\bibinfo {year} {2015})},\ \Eprint
  {https://arxiv.org/abs/1501.02970} {arXiv:1501.02970 [gr-qc]} \BibitemShut
  {NoStop}%
\bibitem [{\citenamefont {{Bauswein}}\ and\ \citenamefont
  {{Stergioulas}}(2015)}]{Bauswein2015}%
  \BibitemOpen
  \bibfield  {author} {\bibinfo {author} {\bibfnamefont {A.}~\bibnamefont
  {{Bauswein}}}\ and\ \bibinfo {author} {\bibfnamefont {N.}~\bibnamefont
  {{Stergioulas}}},\ }\bibfield  {title} {\bibinfo {title} {{Unified picture of
  the post-merger dynamics and gravitational wave emission in neutron star
  mergers}},\ }\href {https://doi.org/10.1103/PhysRevD.91.124056} {\bibfield
  {journal} {\bibinfo  {journal} {\prd}\ }\textbf {\bibinfo {volume} {91}},\
  \bibinfo {eid} {124056} (\bibinfo {year} {2015})}\BibitemShut {NoStop}%
\bibitem [{\citenamefont {{Chakravarti}}\ and\ \citenamefont
  {{Andersson}}(2020)}]{2020MNRAS.497.5480C}%
  \BibitemOpen
  \bibfield  {author} {\bibinfo {author} {\bibfnamefont {K.}~\bibnamefont
  {{Chakravarti}}}\ and\ \bibinfo {author} {\bibfnamefont {N.}~\bibnamefont
  {{Andersson}}},\ }\bibfield  {title} {\bibinfo {title} {{Exploring
  universality in neutron star mergers}},\ }\href
  {https://doi.org/10.1093/mnras/staa2342} {\bibfield  {journal} {\bibinfo
  {journal} {Mon. Not. R. Astron. Soc}\ }\textbf {\bibinfo {volume} {497}},\
  \bibinfo {pages} {5480} (\bibinfo {year} {2020})},\ \Eprint
  {https://arxiv.org/abs/1906.04546} {arXiv:1906.04546 [gr-qc]} \BibitemShut
  {NoStop}%
\bibitem [{\citenamefont {{Hinderer}}(2008)}]{2008ApJ...677.1216H}%
  \BibitemOpen
  \bibfield  {author} {\bibinfo {author} {\bibfnamefont {T.}~\bibnamefont
  {{Hinderer}}},\ }\bibfield  {title} {\bibinfo {title} {{Tidal Love Numbers of
  Neutron Stars}},\ }\href {https://doi.org/10.1086/533487} {\bibfield
  {journal} {\bibinfo  {journal} {\apj}\ }\textbf {\bibinfo {volume} {677}},\
  \bibinfo {pages} {1216} (\bibinfo {year} {2008})},\ \Eprint
  {https://arxiv.org/abs/0711.2420} {arXiv:0711.2420 [astro-ph]} \BibitemShut
  {NoStop}%
\bibitem [{\citenamefont {{Hinderer}}\ \emph {et~al.}(2010)\citenamefont
  {{Hinderer}}, \citenamefont {{Lackey}}, \citenamefont {{Lang}},\ and\
  \citenamefont {{Read}}}]{2010PhRvD..81l3016H}%
  \BibitemOpen
  \bibfield  {author} {\bibinfo {author} {\bibfnamefont {T.}~\bibnamefont
  {{Hinderer}}}, \bibinfo {author} {\bibfnamefont {B.~D.}\ \bibnamefont
  {{Lackey}}}, \bibinfo {author} {\bibfnamefont {R.~N.}\ \bibnamefont
  {{Lang}}},\ and\ \bibinfo {author} {\bibfnamefont {J.~S.}\ \bibnamefont
  {{Read}}},\ }\bibfield  {title} {\bibinfo {title} {{Tidal deformability of
  neutron stars with realistic equations of state and their gravitational wave
  signatures in binary inspiral}},\ }\href
  {https://doi.org/10.1103/PhysRevD.81.123016} {\bibfield  {journal} {\bibinfo
  {journal} {\prd}\ }\textbf {\bibinfo {volume} {81}},\ \bibinfo {eid} {123016}
  (\bibinfo {year} {2010})},\ \Eprint {https://arxiv.org/abs/0911.3535}
  {arXiv:0911.3535 [astro-ph.HE]} \BibitemShut {NoStop}%
\bibitem [{\citenamefont {{Damour}}\ and\ \citenamefont
  {{Nagar}}(2010)}]{2010PhRvD..81h4016D}%
  \BibitemOpen
  \bibfield  {author} {\bibinfo {author} {\bibfnamefont {T.}~\bibnamefont
  {{Damour}}}\ and\ \bibinfo {author} {\bibfnamefont {A.}~\bibnamefont
  {{Nagar}}},\ }\bibfield  {title} {\bibinfo {title} {{Effective one body
  description of tidal effects in inspiralling compact binaries}},\ }\href
  {https://doi.org/10.1103/PhysRevD.81.084016} {\bibfield  {journal} {\bibinfo
  {journal} {\prd}\ }\textbf {\bibinfo {volume} {81}},\ \bibinfo {eid} {084016}
  (\bibinfo {year} {2010})},\ \Eprint {https://arxiv.org/abs/0911.5041}
  {arXiv:0911.5041 [gr-qc]} \BibitemShut {NoStop}%
\bibitem [{\citenamefont {{Friedman}}\ and\ \citenamefont
  {{Stergioulas}}(2013)}]{2013rrs..book.....F}%
  \BibitemOpen
  \bibfield  {author} {\bibinfo {author} {\bibfnamefont {J.~L.}\ \bibnamefont
  {{Friedman}}}\ and\ \bibinfo {author} {\bibfnamefont {N.}~\bibnamefont
  {{Stergioulas}}},\ }\href@noop {} {\emph {\bibinfo {title} {{Rotating
  Relativistic Stars}}}}\ (\bibinfo {year} {2013})\BibitemShut {NoStop}%
\bibitem [{\citenamefont {{Lioutas}}\ and\ \citenamefont
  {{Stergioulas}}(2018)}]{2018GReGr..50...12L}%
  \BibitemOpen
  \bibfield  {author} {\bibinfo {author} {\bibfnamefont {G.}~\bibnamefont
  {{Lioutas}}}\ and\ \bibinfo {author} {\bibfnamefont {N.}~\bibnamefont
  {{Stergioulas}}},\ }\bibfield  {title} {\bibinfo {title} {{Universal and
  approximate relations for the gravitational-wave damping timescale of f-modes
  in neutron stars}},\ }\href {https://doi.org/10.1007/s10714-017-2331-7}
  {\bibfield  {journal} {\bibinfo  {journal} {General Relativity and
  Gravitation}\ }\textbf {\bibinfo {volume} {50}},\ \bibinfo {eid} {12}
  (\bibinfo {year} {2018})},\ \Eprint {https://arxiv.org/abs/1709.10067}
  {arXiv:1709.10067 [gr-qc]} \BibitemShut {NoStop}%
\bibitem [{\citenamefont {{Oechslin}}\ \emph {et~al.}(2002)\citenamefont
  {{Oechslin}}, \citenamefont {{Rosswog}},\ and\ \citenamefont
  {{Thielemann}}}]{2002PhRvD..65j3005O}%
  \BibitemOpen
  \bibfield  {author} {\bibinfo {author} {\bibfnamefont {R.}~\bibnamefont
  {{Oechslin}}}, \bibinfo {author} {\bibfnamefont {S.}~\bibnamefont
  {{Rosswog}}},\ and\ \bibinfo {author} {\bibfnamefont {F.-K.}\ \bibnamefont
  {{Thielemann}}},\ }\bibfield  {title} {\bibinfo {title} {{Conformally flat
  smoothed particle hydrodynamics application to neutron star mergers}},\
  }\href {https://doi.org/10.1103/PhysRevD.65.103005} {\bibfield  {journal}
  {\bibinfo  {journal} {\prd}\ }\textbf {\bibinfo {volume} {65}},\ \bibinfo
  {eid} {103005} (\bibinfo {year} {2002})},\ \Eprint
  {https://arxiv.org/abs/gr-qc/0111005} {arXiv:gr-qc/0111005 [gr-qc]}
  \BibitemShut {NoStop}%
\bibitem [{\citenamefont {{Oechslin}}\ \emph {et~al.}(2007)\citenamefont
  {{Oechslin}}, \citenamefont {{Janka}},\ and\ \citenamefont
  {{Marek}}}]{2007A&A...467..395O}%
  \BibitemOpen
  \bibfield  {author} {\bibinfo {author} {\bibfnamefont {R.}~\bibnamefont
  {{Oechslin}}}, \bibinfo {author} {\bibfnamefont {H.~T.}\ \bibnamefont
  {{Janka}}},\ and\ \bibinfo {author} {\bibfnamefont {A.}~\bibnamefont
  {{Marek}}},\ }\bibfield  {title} {\bibinfo {title} {{Relativistic neutron
  star merger simulations with non-zero temperature equations of state. I.
  Variation of binary parameters and equation of state}},\ }\href
  {https://doi.org/10.1051/0004-6361:20066682} {\bibfield  {journal} {\bibinfo
  {journal} {Astron. Astrophys.}\ }\textbf {\bibinfo {volume} {467}},\ \bibinfo
  {pages} {395} (\bibinfo {year} {2007})},\ \Eprint
  {https://arxiv.org/abs/astro-ph/0611047} {arXiv:astro-ph/0611047 [astro-ph]}
  \BibitemShut {NoStop}%
\bibitem [{\citenamefont {{Bauswein}}\ \emph
  {et~al.}(2010{\natexlab{a}})\citenamefont {{Bauswein}}, \citenamefont
  {{Oechslin}},\ and\ \citenamefont {{Janka}}}]{2010PhRvD..81b4012B}%
  \BibitemOpen
  \bibfield  {author} {\bibinfo {author} {\bibfnamefont {A.}~\bibnamefont
  {{Bauswein}}}, \bibinfo {author} {\bibfnamefont {R.}~\bibnamefont
  {{Oechslin}}},\ and\ \bibinfo {author} {\bibfnamefont {H.~T.}\ \bibnamefont
  {{Janka}}},\ }\bibfield  {title} {\bibinfo {title} {{Discriminating strange
  star mergers from neutron star mergers by gravitational-wave measurements}},\
  }\href {https://doi.org/10.1103/PhysRevD.81.024012} {\bibfield  {journal}
  {\bibinfo  {journal} {\prd}\ }\textbf {\bibinfo {volume} {81}},\ \bibinfo
  {eid} {024012} (\bibinfo {year} {2010}{\natexlab{a}})},\ \Eprint
  {https://arxiv.org/abs/0910.5169} {arXiv:0910.5169 [astro-ph.SR]}
  \BibitemShut {NoStop}%
\bibitem [{\citenamefont {{Isenberg}}\ and\ \citenamefont
  {{Nester}}(1980)}]{1980grg1.conf...23I}%
  \BibitemOpen
  \bibfield  {author} {\bibinfo {author} {\bibfnamefont {J.}~\bibnamefont
  {{Isenberg}}}\ and\ \bibinfo {author} {\bibfnamefont {J.}~\bibnamefont
  {{Nester}}},\ }\bibfield  {title} {\bibinfo {title} {{Canonical Gravity}},\
  }in\ \href@noop {} {\emph {\bibinfo {booktitle} {General Relativity and
  Gravitation. Vol. 1. One hundred years after the birth of Albert Einstein.
  Edited by A. Held. New York}}},\ Vol.~\bibinfo {volume} {1}\ (\bibinfo {year}
  {1980})\ p.~\bibinfo {pages} {23}\BibitemShut {NoStop}%
\bibitem [{\citenamefont {{Wilson}}\ \emph {et~al.}(1996)\citenamefont
  {{Wilson}}, \citenamefont {{Mathews}},\ and\ \citenamefont
  {{Marronetti}}}]{1996PhRvD..54.1317W}%
  \BibitemOpen
  \bibfield  {author} {\bibinfo {author} {\bibfnamefont {J.~R.}\ \bibnamefont
  {{Wilson}}}, \bibinfo {author} {\bibfnamefont {G.~J.}\ \bibnamefont
  {{Mathews}}},\ and\ \bibinfo {author} {\bibfnamefont {P.}~\bibnamefont
  {{Marronetti}}},\ }\bibfield  {title} {\bibinfo {title} {{Relativistic
  numerical model for close neutron-star binaries}},\ }\href
  {https://doi.org/10.1103/PhysRevD.54.1317} {\bibfield  {journal} {\bibinfo
  {journal} {\prd}\ }\textbf {\bibinfo {volume} {54}},\ \bibinfo {pages} {1317}
  (\bibinfo {year} {1996})},\ \Eprint {https://arxiv.org/abs/gr-qc/9601017}
  {arXiv:gr-qc/9601017 [gr-qc]} \BibitemShut {NoStop}%
\bibitem [{\citenamefont {{Bauswein}}\ \emph
  {et~al.}(2010{\natexlab{b}})\citenamefont {{Bauswein}}, \citenamefont
  {{Janka}},\ and\ \citenamefont {{Oechslin}}}]{2010PhRvD..82h4043B}%
  \BibitemOpen
  \bibfield  {author} {\bibinfo {author} {\bibfnamefont {A.}~\bibnamefont
  {{Bauswein}}}, \bibinfo {author} {\bibfnamefont {H.~T.}\ \bibnamefont
  {{Janka}}},\ and\ \bibinfo {author} {\bibfnamefont {R.}~\bibnamefont
  {{Oechslin}}},\ }\bibfield  {title} {\bibinfo {title} {{Testing
  approximations of thermal effects in neutron star merger simulations}},\
  }\href {https://doi.org/10.1103/PhysRevD.82.084043} {\bibfield  {journal}
  {\bibinfo  {journal} {\prd}\ }\textbf {\bibinfo {volume} {82}},\ \bibinfo
  {eid} {084043} (\bibinfo {year} {2010}{\natexlab{b}})},\ \Eprint
  {https://arxiv.org/abs/1006.3315} {arXiv:1006.3315 [astro-ph.SR]}
  \BibitemShut {NoStop}%
\bibitem [{\citenamefont {{Bauswein}}\ \emph {et~al.}(2020)\citenamefont
  {{Bauswein}}, \citenamefont {{Blacker}}, \citenamefont {{Lioutas}},
  \citenamefont {{Soultanis}}, \citenamefont {{Vijayan}},\ and\ \citenamefont
  {{Stergioulas}}}]{2020arXiv201004461B}%
  \BibitemOpen
  \bibfield  {author} {\bibinfo {author} {\bibfnamefont {A.}~\bibnamefont
  {{Bauswein}}}, \bibinfo {author} {\bibfnamefont {S.}~\bibnamefont
  {{Blacker}}}, \bibinfo {author} {\bibfnamefont {G.}~\bibnamefont
  {{Lioutas}}}, \bibinfo {author} {\bibfnamefont {T.}~\bibnamefont
  {{Soultanis}}}, \bibinfo {author} {\bibfnamefont {V.}~\bibnamefont
  {{Vijayan}}},\ and\ \bibinfo {author} {\bibfnamefont {N.}~\bibnamefont
  {{Stergioulas}}},\ }\bibfield  {title} {\bibinfo {title} {{Systematics of
  prompt black-hole formation in neutron star mergers}},\ }\href@noop {}
  {\bibfield  {journal} {\bibinfo  {journal} {arXiv e-prints}\ ,\ \bibinfo
  {eid} {arXiv:2010.04461}} (\bibinfo {year} {2020})},\ \Eprint
  {https://arxiv.org/abs/2010.04461} {arXiv:2010.04461 [astro-ph.HE]}
  \BibitemShut {NoStop}%
\bibitem [{\citenamefont {Alford}\ \emph {et~al.}(2005)\citenamefont {Alford},
  \citenamefont {Braby}, \citenamefont {Paris},\ and\ \citenamefont
  {Reddy}}]{Alford2005}%
  \BibitemOpen
  \bibfield  {author} {\bibinfo {author} {\bibfnamefont {M.}~\bibnamefont
  {Alford}}, \bibinfo {author} {\bibfnamefont {M.}~\bibnamefont {Braby}},
  \bibinfo {author} {\bibfnamefont {M.}~\bibnamefont {Paris}},\ and\ \bibinfo
  {author} {\bibfnamefont {S.}~\bibnamefont {Reddy}},\ }\bibfield  {title}
  {\bibinfo {title} {Hybrid stars that masquerade as neutron stars},\ }\href
  {https://doi.org/10.1086/430902} {\bibfield  {journal} {\bibinfo  {journal}
  {Astrophys. J}\ }\textbf {\bibinfo {volume} {629}},\ \bibinfo {pages} {969}
  (\bibinfo {year} {2005})}\BibitemShut {NoStop}%
\bibitem [{\citenamefont {Read}\ \emph {et~al.}(2009)\citenamefont {Read},
  \citenamefont {Lackey}, \citenamefont {Owen},\ and\ \citenamefont
  {Friedman}}]{Read2009a}%
  \BibitemOpen
  \bibfield  {author} {\bibinfo {author} {\bibfnamefont {J.~S.}\ \bibnamefont
  {Read}}, \bibinfo {author} {\bibfnamefont {B.~D.}\ \bibnamefont {Lackey}},
  \bibinfo {author} {\bibfnamefont {B.~J.}\ \bibnamefont {Owen}},\ and\
  \bibinfo {author} {\bibfnamefont {J.~L.}\ \bibnamefont {Friedman}},\
  }\bibfield  {title} {\bibinfo {title} {Constraints on a phenomenologically
  parametrized neutron-star equation of state},\ }\href
  {https://doi.org/10.1103/PhysRevD.79.124032} {\bibfield  {journal} {\bibinfo
  {journal} {\prd}\ }\textbf {\bibinfo {volume} {79}},\ \bibinfo {eid} {124032}
  (\bibinfo {year} {2009})}\BibitemShut {NoStop}%
\bibitem [{\citenamefont {Akmal}\ \emph {et~al.}(1998)\citenamefont {Akmal},
  \citenamefont {Pandharipande},\ and\ \citenamefont {Ravenhall}}]{Akmal1998}%
  \BibitemOpen
  \bibfield  {author} {\bibinfo {author} {\bibfnamefont {A.}~\bibnamefont
  {Akmal}}, \bibinfo {author} {\bibfnamefont {V.~R.}\ \bibnamefont
  {Pandharipande}},\ and\ \bibinfo {author} {\bibfnamefont {D.~G.}\
  \bibnamefont {Ravenhall}},\ }\bibfield  {title} {\bibinfo {title} {Equation
  of state of nucleon matter and neutron star structure},\ }\href
  {https://doi.org/10.1103/PhysRevC.58.1804} {\bibfield  {journal} {\bibinfo
  {journal} {\prc}\ }\textbf {\bibinfo {volume} {58}},\ \bibinfo {pages} {1804}
  (\bibinfo {year} {1998})}\BibitemShut {NoStop}%
\bibitem [{\citenamefont {Banik}\ \emph {et~al.}(2014)\citenamefont {Banik},
  \citenamefont {Hempel},\ and\ \citenamefont {Bandyopadhyay}}]{Banik2014}%
  \BibitemOpen
  \bibfield  {author} {\bibinfo {author} {\bibfnamefont {S.}~\bibnamefont
  {Banik}}, \bibinfo {author} {\bibfnamefont {M.}~\bibnamefont {Hempel}},\ and\
  \bibinfo {author} {\bibfnamefont {D.}~\bibnamefont {Bandyopadhyay}},\
  }\bibfield  {title} {\bibinfo {title} {New hyperon equations of state for
  supernovae and neutron stars in density-dependent hadron field theory},\
  }\href {https://doi.org/10.1088/0067-0049/214/2/22} {\bibfield  {journal}
  {\bibinfo  {journal} {Astrophys. J., Suppl. Ser.}\ }\textbf {\bibinfo
  {volume} {214}},\ \bibinfo {eid} {22} (\bibinfo {year} {2014})}\BibitemShut
  {NoStop}%
\bibitem [{\citenamefont {{Goriely}}\ \emph {et~al.}(2010)\citenamefont
  {{Goriely}}, \citenamefont {{Chamel}},\ and\ \citenamefont
  {{Pearson}}}]{Goriely2010}%
  \BibitemOpen
  \bibfield  {author} {\bibinfo {author} {\bibfnamefont {S.}~\bibnamefont
  {{Goriely}}}, \bibinfo {author} {\bibfnamefont {N.}~\bibnamefont
  {{Chamel}}},\ and\ \bibinfo {author} {\bibfnamefont {J.~M.}\ \bibnamefont
  {{Pearson}}},\ }\bibfield  {title} {\bibinfo {title} {{Further explorations
  of Skyrme-Hartree-Fock-Bogoliubov mass formulas. XII. Stiffness and stability
  of neutron-star matter}},\ }\href
  {https://doi.org/10.1103/PhysRevC.82.035804} {\bibfield  {journal} {\bibinfo
  {journal} {\prc}\ }\textbf {\bibinfo {volume} {82}},\ \bibinfo {eid} {035804}
  (\bibinfo {year} {2010})}\BibitemShut {NoStop}%
\bibitem [{\citenamefont {{Hempel}}\ and\ \citenamefont
  {{Schaffner-Bielich}}(2010)}]{Hempel2010}%
  \BibitemOpen
  \bibfield  {author} {\bibinfo {author} {\bibfnamefont {M.}~\bibnamefont
  {{Hempel}}}\ and\ \bibinfo {author} {\bibfnamefont {J.}~\bibnamefont
  {{Schaffner-Bielich}}},\ }\bibfield  {title} {\bibinfo {title} {{A
  statistical model for a complete supernova equation of state}},\ }\href
  {https://doi.org/10.1016/j.nuclphysa.2010.02.010} {\bibfield  {journal}
  {\bibinfo  {journal} {Nucl. Phys. A}\ }\textbf {\bibinfo {volume} {837}},\
  \bibinfo {pages} {210} (\bibinfo {year} {2010})}\BibitemShut {NoStop}%
\bibitem [{\citenamefont {{Typel}}\ \emph {et~al.}(2010)\citenamefont
  {{Typel}}, \citenamefont {{R{\"o}pke}}, \citenamefont {{Kl{\"a}hn}},
  \citenamefont {{Blaschke}},\ and\ \citenamefont {{Wolter}}}]{Typel2010}%
  \BibitemOpen
  \bibfield  {author} {\bibinfo {author} {\bibfnamefont {S.}~\bibnamefont
  {{Typel}}}, \bibinfo {author} {\bibfnamefont {G.}~\bibnamefont
  {{R{\"o}pke}}}, \bibinfo {author} {\bibfnamefont {T.}~\bibnamefont
  {{Kl{\"a}hn}}}, \bibinfo {author} {\bibfnamefont {D.}~\bibnamefont
  {{Blaschke}}},\ and\ \bibinfo {author} {\bibfnamefont {H.~H.}\ \bibnamefont
  {{Wolter}}},\ }\bibfield  {title} {\bibinfo {title} {{Composition and
  thermodynamics of nuclear matter with light clusters}},\ }\href
  {https://doi.org/10.1103/PhysRevC.81.015803} {\bibfield  {journal} {\bibinfo
  {journal} {\prc}\ }\textbf {\bibinfo {volume} {81}},\ \bibinfo {eid} {015803}
  (\bibinfo {year} {2010})}\BibitemShut {NoStop}%
\bibitem [{\citenamefont {{Typel}}(2005)}]{Typel2005}%
  \BibitemOpen
  \bibfield  {author} {\bibinfo {author} {\bibfnamefont {S.}~\bibnamefont
  {{Typel}}},\ }\bibfield  {title} {\bibinfo {title} {{Relativistic model for
  nuclear matter and atomic nuclei with momentum-dependent self-energies}},\
  }\href {https://doi.org/10.1103/PhysRevC.71.064301} {\bibfield  {journal}
  {\bibinfo  {journal} {\prc}\ }\textbf {\bibinfo {volume} {71}},\ \bibinfo
  {eid} {064301} (\bibinfo {year} {2005})}\BibitemShut {NoStop}%
\bibitem [{\citenamefont {Alvarez-Castillo}\ \emph {et~al.}(2016)\citenamefont
  {Alvarez-Castillo}, \citenamefont {Ayriyan}, \citenamefont {Benic},
  \citenamefont {Blaschke}, \citenamefont {Grigorian},\ and\ \citenamefont
  {Typel}}]{Alvarez-Castillo2016}%
  \BibitemOpen
  \bibfield  {author} {\bibinfo {author} {\bibfnamefont {D.}~\bibnamefont
  {Alvarez-Castillo}}, \bibinfo {author} {\bibfnamefont {A.}~\bibnamefont
  {Ayriyan}}, \bibinfo {author} {\bibfnamefont {S.}~\bibnamefont {Benic}},
  \bibinfo {author} {\bibfnamefont {D.}~\bibnamefont {Blaschke}}, \bibinfo
  {author} {\bibfnamefont {H.}~\bibnamefont {Grigorian}},\ and\ \bibinfo
  {author} {\bibfnamefont {S.}~\bibnamefont {Typel}},\ }\bibfield  {title}
  {\bibinfo {title} {New class of hybrid eos and bayesian m - r data
  analysis},\ }\href {https://doi.org/10.1140/epja/i2016-16069-2} {\bibfield
  {journal} {\bibinfo  {journal} {European Physical Journal A}\ }\textbf
  {\bibinfo {volume} {52}},\ \bibinfo {eid} {69} (\bibinfo {year}
  {2016})}\BibitemShut {NoStop}%
\bibitem [{\citenamefont {Fortin}\ \emph {et~al.}(2018)\citenamefont {Fortin},
  \citenamefont {Oertel},\ and\ \citenamefont {Provid{\^e}ncia}}]{Fortin2018}%
  \BibitemOpen
  \bibfield  {author} {\bibinfo {author} {\bibfnamefont {M.}~\bibnamefont
  {Fortin}}, \bibinfo {author} {\bibfnamefont {M.}~\bibnamefont {Oertel}},\
  and\ \bibinfo {author} {\bibfnamefont {C.}~\bibnamefont {Provid{\^e}ncia}},\
  }\bibfield  {title} {\bibinfo {title} {Hyperons in hot dense matter: what do
  the constraints tell us for equation of state?},\ }\bibfield  {journal}
  {\bibinfo  {journal} {Publ. Astron. Soc. Aust.}\ }\textbf {\bibinfo {volume}
  {35}},\ \href {https://doi.org/10.1017/pasa.2018.32} {10.1017/pasa.2018.32}
  (\bibinfo {year} {2018})\BibitemShut {NoStop}%
\bibitem [{\citenamefont {Marques}\ \emph {et~al.}(2017)\citenamefont
  {Marques}, \citenamefont {Oertel}, \citenamefont {Hempel},\ and\
  \citenamefont {Novak}}]{Marques2017}%
  \BibitemOpen
  \bibfield  {author} {\bibinfo {author} {\bibfnamefont {M.}~\bibnamefont
  {Marques}}, \bibinfo {author} {\bibfnamefont {M.}~\bibnamefont {Oertel}},
  \bibinfo {author} {\bibfnamefont {M.}~\bibnamefont {Hempel}},\ and\ \bibinfo
  {author} {\bibfnamefont {J.}~\bibnamefont {Novak}},\ }\bibfield  {title}
  {\bibinfo {title} {{New temperature dependent hyperonic equation of state:
  Application to rotating neutron star models and $I\text{-}Q$ relations}},\
  }\href {https://doi.org/10.1103/PhysRevC.96.045806} {\bibfield  {journal}
  {\bibinfo  {journal} {Phys. Rev. C}\ }\textbf {\bibinfo {volume} {96}},\
  \bibinfo {pages} {045806} (\bibinfo {year} {2017})}\BibitemShut {NoStop}%
\bibitem [{\citenamefont {Wiringa}\ \emph {et~al.}(1988)\citenamefont
  {Wiringa}, \citenamefont {Fiks},\ and\ \citenamefont
  {Fabrocini}}]{Wiringa1988}%
  \BibitemOpen
  \bibfield  {author} {\bibinfo {author} {\bibfnamefont {R.~B.}\ \bibnamefont
  {Wiringa}}, \bibinfo {author} {\bibfnamefont {V.}~\bibnamefont {Fiks}},\ and\
  \bibinfo {author} {\bibfnamefont {A.}~\bibnamefont {Fabrocini}},\ }\bibfield
  {title} {\bibinfo {title} {Equation of state for dense nucleon matter},\
  }\href {https://doi.org/10.1103/PhysRevC.38.1010} {\bibfield  {journal}
  {\bibinfo  {journal} {\prc}\ }\textbf {\bibinfo {volume} {38}},\ \bibinfo
  {pages} {1010} (\bibinfo {year} {1988})}\BibitemShut {NoStop}%
\bibitem [{\citenamefont {Lattimer}\ and\ \citenamefont
  {Douglas~Swesty}(1991)}]{Lattimer1991}%
  \BibitemOpen
  \bibfield  {author} {\bibinfo {author} {\bibfnamefont {J.~M.}\ \bibnamefont
  {Lattimer}}\ and\ \bibinfo {author} {\bibfnamefont {F.}~\bibnamefont
  {Douglas~Swesty}},\ }\bibfield  {title} {\bibinfo {title} {A generalized
  equation of state for hot, dense matter},\ }\href
  {https://doi.org/10.1016/0375-9474(91)90452-C} {\bibfield  {journal}
  {\bibinfo  {journal} {Nuclear Physics A}\ }\textbf {\bibinfo {volume}
  {535}},\ \bibinfo {pages} {331} (\bibinfo {year} {1991})}\BibitemShut
  {NoStop}%
\bibitem [{\citenamefont {Shen}\ \emph {et~al.}(2011)\citenamefont {Shen},
  \citenamefont {Horowitz},\ and\ \citenamefont {Teige}}]{Shen2011}%
  \BibitemOpen
  \bibfield  {author} {\bibinfo {author} {\bibfnamefont {G.}~\bibnamefont
  {Shen}}, \bibinfo {author} {\bibfnamefont {C.~J.}\ \bibnamefont {Horowitz}},\
  and\ \bibinfo {author} {\bibfnamefont {S.}~\bibnamefont {Teige}},\ }\bibfield
   {title} {\bibinfo {title} {New equation of state for astrophysical
  simulations},\ }\href {https://doi.org/10.1103/PhysRevC.83.035802} {\bibfield
   {journal} {\bibinfo  {journal} {\prc}\ }\textbf {\bibinfo {volume} {83}},\
  \bibinfo {eid} {035802} (\bibinfo {year} {2011})}\BibitemShut {NoStop}%
\bibitem [{\citenamefont {Lalazissis}\ \emph {et~al.}(1997)\citenamefont
  {Lalazissis}, \citenamefont {K{\"o}nig},\ and\ \citenamefont
  {Ring}}]{Lalazissis1997a}%
  \BibitemOpen
  \bibfield  {author} {\bibinfo {author} {\bibfnamefont {G.~A.}\ \bibnamefont
  {Lalazissis}}, \bibinfo {author} {\bibfnamefont {J.}~\bibnamefont
  {K{\"o}nig}},\ and\ \bibinfo {author} {\bibfnamefont {P.}~\bibnamefont
  {Ring}},\ }\bibfield  {title} {\bibinfo {title} {New parametrization for the
  lagrangian density of relativistic mean field theory},\ }\href
  {https://doi.org/10.1103/PhysRevC.55.540} {\bibfield  {journal} {\bibinfo
  {journal} {\prc}\ }\textbf {\bibinfo {volume} {55}},\ \bibinfo {pages} {540}
  (\bibinfo {year} {1997})}\BibitemShut {NoStop}%
\bibitem [{\citenamefont {Steiner}\ \emph {et~al.}(2013)\citenamefont
  {Steiner}, \citenamefont {Hempel},\ and\ \citenamefont
  {Fischer}}]{Steiner2013}%
  \BibitemOpen
  \bibfield  {author} {\bibinfo {author} {\bibfnamefont {A.~W.}\ \bibnamefont
  {Steiner}}, \bibinfo {author} {\bibfnamefont {M.}~\bibnamefont {Hempel}},\
  and\ \bibinfo {author} {\bibfnamefont {T.}~\bibnamefont {Fischer}},\
  }\bibfield  {title} {\bibinfo {title} {Core-collapse supernova equations of
  state based on neutron star observations},\ }\href
  {https://doi.org/10.1088/0004-637X/774/1/17} {\bibfield  {journal} {\bibinfo
  {journal} {\apj}\ }\textbf {\bibinfo {volume} {774}},\ \bibinfo {eid} {17}
  (\bibinfo {year} {2013})}\BibitemShut {NoStop}%
\bibitem [{\citenamefont {{Douchin}}\ and\ \citenamefont
  {{Haensel}}(2001)}]{Douchin2001}%
  \BibitemOpen
  \bibfield  {author} {\bibinfo {author} {\bibfnamefont {F.}~\bibnamefont
  {{Douchin}}}\ and\ \bibinfo {author} {\bibfnamefont {P.}~\bibnamefont
  {{Haensel}}},\ }\bibfield  {title} {\bibinfo {title} {{A unified equation of
  state of dense matter and neutron star structure}},\ }\href
  {https://doi.org/10.1051/0004-6361:20011402} {\bibfield  {journal} {\bibinfo
  {journal} {Astron. Astrophys.}\ }\textbf {\bibinfo {volume} {380}},\ \bibinfo
  {pages} {151} (\bibinfo {year} {2001})}\BibitemShut {NoStop}%
\bibitem [{\citenamefont {Sugahara}\ and\ \citenamefont
  {Toki}(1994)}]{Sugahara1994a}%
  \BibitemOpen
  \bibfield  {author} {\bibinfo {author} {\bibfnamefont {Y.}~\bibnamefont
  {Sugahara}}\ and\ \bibinfo {author} {\bibfnamefont {H.}~\bibnamefont
  {Toki}},\ }\bibfield  {title} {\bibinfo {title} {Relativistic mean-field
  theory for unstable nuclei with non-linear {$\sigma$} and {$\omega$} terms},\
  }\href {https://doi.org/10.1016/0375-9474(94)90923-7} {\bibfield  {journal}
  {\bibinfo  {journal} {Nuclear Physics A}\ }\textbf {\bibinfo {volume}
  {579}},\ \bibinfo {pages} {557} (\bibinfo {year} {1994})}\BibitemShut
  {NoStop}%
\bibitem [{\citenamefont {{Hempel}}\ \emph {et~al.}(2012)\citenamefont
  {{Hempel}}, \citenamefont {{Fischer}}, \citenamefont {{Schaffner-Bielich}},\
  and\ \citenamefont {{Liebend{\"o}rfer}}}]{Hempel2012}%
  \BibitemOpen
  \bibfield  {author} {\bibinfo {author} {\bibfnamefont {M.}~\bibnamefont
  {{Hempel}}}, \bibinfo {author} {\bibfnamefont {T.}~\bibnamefont {{Fischer}}},
  \bibinfo {author} {\bibfnamefont {J.}~\bibnamefont {{Schaffner-Bielich}}},\
  and\ \bibinfo {author} {\bibfnamefont {M.}~\bibnamefont
  {{Liebend{\"o}rfer}}},\ }\bibfield  {title} {\bibinfo {title} {{New Equations
  of State in Simulations of Core-collapse Supernovae}},\ }\href
  {https://doi.org/10.1088/0004-637X/748/1/70} {\bibfield  {journal} {\bibinfo
  {journal} {\apj}\ }\textbf {\bibinfo {volume} {748}},\ \bibinfo {eid} {70}
  (\bibinfo {year} {2012})}\BibitemShut {NoStop}%
\bibitem [{\citenamefont {Toki}\ \emph {et~al.}(1995)\citenamefont {Toki},
  \citenamefont {Hirata}, \citenamefont {Sugahara}, \citenamefont {Sumiyoshi},\
  and\ \citenamefont {Tanihata}}]{Toki1995}%
  \BibitemOpen
  \bibfield  {author} {\bibinfo {author} {\bibfnamefont {H.}~\bibnamefont
  {Toki}}, \bibinfo {author} {\bibfnamefont {D.}~\bibnamefont {Hirata}},
  \bibinfo {author} {\bibfnamefont {Y.}~\bibnamefont {Sugahara}}, \bibinfo
  {author} {\bibfnamefont {K.}~\bibnamefont {Sumiyoshi}},\ and\ \bibinfo
  {author} {\bibfnamefont {I.}~\bibnamefont {Tanihata}},\ }\bibfield  {title}
  {\bibinfo {title} {Relativistic many body approach for unstable nuclei and
  supernova},\ }\href {https://doi.org/10.1016/0375-9474(95)00161-S} {\bibfield
   {journal} {\bibinfo  {journal} {Nuclear Physics A}\ }\textbf {\bibinfo
  {volume} {588}},\ \bibinfo {pages} {357} (\bibinfo {year}
  {1995})}\BibitemShut {NoStop}%
\bibitem [{\citenamefont {{Demorest}}\ \emph {et~al.}(2010)\citenamefont
  {{Demorest}}, \citenamefont {{Pennucci}}, \citenamefont {{Ransom}},
  \citenamefont {{Roberts}},\ and\ \citenamefont
  {{Hessels}}}]{2010Natur.467.1081D}%
  \BibitemOpen
  \bibfield  {author} {\bibinfo {author} {\bibfnamefont {P.~B.}\ \bibnamefont
  {{Demorest}}}, \bibinfo {author} {\bibfnamefont {T.}~\bibnamefont
  {{Pennucci}}}, \bibinfo {author} {\bibfnamefont {S.~M.}\ \bibnamefont
  {{Ransom}}}, \bibinfo {author} {\bibfnamefont {M.~S.~E.}\ \bibnamefont
  {{Roberts}}},\ and\ \bibinfo {author} {\bibfnamefont {J.~W.~T.}\ \bibnamefont
  {{Hessels}}},\ }\bibfield  {title} {\bibinfo {title} {{A two-solar-mass
  neutron star measured using Shapiro delay}},\ }\href
  {https://doi.org/10.1038/nature09466} {\bibfield  {journal} {\bibinfo
  {journal} {Nature}\ }\textbf {\bibinfo {volume} {467}},\ \bibinfo {pages}
  {1081} (\bibinfo {year} {2010})},\ \Eprint {https://arxiv.org/abs/1010.5788}
  {arXiv:1010.5788 [astro-ph.HE]} \BibitemShut {NoStop}%
\bibitem [{\citenamefont {{Antoniadis}}\ \emph {et~al.}(2013)\citenamefont
  {{Antoniadis}}, \citenamefont {{Freire}}, \citenamefont {{Wex}},
  \citenamefont {{Tauris}}, \citenamefont {{Lynch}}, \citenamefont {{van
  Kerkwijk}}, \citenamefont {{Kramer}}, \citenamefont {{Bassa}}, \citenamefont
  {{Dhillon}}, \citenamefont {{Driebe}} \emph {et~al.}}]{2013Sci...340..448A}%
  \BibitemOpen
  \bibfield  {author} {\bibinfo {author} {\bibfnamefont {J.}~\bibnamefont
  {{Antoniadis}}}, \bibinfo {author} {\bibfnamefont {P.~C.~C.}\ \bibnamefont
  {{Freire}}}, \bibinfo {author} {\bibfnamefont {N.}~\bibnamefont {{Wex}}},
  \bibinfo {author} {\bibfnamefont {T.~M.}\ \bibnamefont {{Tauris}}}, \bibinfo
  {author} {\bibfnamefont {R.~S.}\ \bibnamefont {{Lynch}}}, \bibinfo {author}
  {\bibfnamefont {M.~H.}\ \bibnamefont {{van Kerkwijk}}}, \bibinfo {author}
  {\bibfnamefont {M.}~\bibnamefont {{Kramer}}}, \bibinfo {author}
  {\bibfnamefont {C.}~\bibnamefont {{Bassa}}}, \bibinfo {author} {\bibfnamefont
  {V.~S.}\ \bibnamefont {{Dhillon}}}, \bibinfo {author} {\bibfnamefont
  {T.}~\bibnamefont {{Driebe}}}, \emph {et~al.},\ }\bibfield  {title} {\bibinfo
  {title} {{A Massive Pulsar in a Compact Relativistic Binary}},\ }\href
  {https://doi.org/10.1126/science.1233232} {\bibfield  {journal} {\bibinfo
  {journal} {Science}\ }\textbf {\bibinfo {volume} {340}},\ \bibinfo {pages}
  {448} (\bibinfo {year} {2013})},\ \Eprint {https://arxiv.org/abs/1304.6875}
  {arXiv:1304.6875 [astro-ph.HE]} \BibitemShut {NoStop}%
\bibitem [{\citenamefont {{Arzoumanian}}\ \emph {et~al.}(2018)\citenamefont
  {{Arzoumanian}}, \citenamefont {{Brazier}}, \citenamefont {{Burke-Spolaor}},
  \citenamefont {{Chamberlin}}, \citenamefont {{Chatterjee}}, \citenamefont
  {{Christy}}, \citenamefont {{Cordes}}, \citenamefont {{Cornish}},
  \citenamefont {{Crawford}}, \citenamefont {{Thankful Cromartie}} \emph
  {et~al.}}]{2018ApJS..235...37A}%
  \BibitemOpen
  \bibfield  {author} {\bibinfo {author} {\bibfnamefont {Z.}~\bibnamefont
  {{Arzoumanian}}}, \bibinfo {author} {\bibfnamefont {A.}~\bibnamefont
  {{Brazier}}}, \bibinfo {author} {\bibfnamefont {S.}~\bibnamefont
  {{Burke-Spolaor}}}, \bibinfo {author} {\bibfnamefont {S.}~\bibnamefont
  {{Chamberlin}}}, \bibinfo {author} {\bibfnamefont {S.}~\bibnamefont
  {{Chatterjee}}}, \bibinfo {author} {\bibfnamefont {B.}~\bibnamefont
  {{Christy}}}, \bibinfo {author} {\bibfnamefont {J.~M.}\ \bibnamefont
  {{Cordes}}}, \bibinfo {author} {\bibfnamefont {N.~J.}\ \bibnamefont
  {{Cornish}}}, \bibinfo {author} {\bibfnamefont {F.}~\bibnamefont
  {{Crawford}}}, \bibinfo {author} {\bibfnamefont {H.}~\bibnamefont {{Thankful
  Cromartie}}}, \emph {et~al.} (\bibinfo {collaboration} {NANOGrav
  Collaboration}),\ }\bibfield  {title} {\bibinfo {title} {{The NANOGrav
  11-year Data Set: High-precision Timing of 45 Millisecond Pulsars}},\ }\href
  {https://doi.org/10.3847/1538-4365/aab5b0} {\bibfield  {journal} {\bibinfo
  {journal} {Astrophys. J., Suppl. Ser.}\ }\textbf {\bibinfo {volume} {235}},\
  \bibinfo {eid} {37} (\bibinfo {year} {2018})},\ \Eprint
  {https://arxiv.org/abs/1801.01837} {arXiv:1801.01837 [astro-ph.HE]}
  \BibitemShut {NoStop}%
\bibitem [{\citenamefont {{Linares}}\ \emph {et~al.}(2018)\citenamefont
  {{Linares}}, \citenamefont {{Shahbaz}},\ and\ \citenamefont
  {{Casares}}}]{2018ApJ...859...54L}%
  \BibitemOpen
  \bibfield  {author} {\bibinfo {author} {\bibfnamefont {M.}~\bibnamefont
  {{Linares}}}, \bibinfo {author} {\bibfnamefont {T.}~\bibnamefont
  {{Shahbaz}}},\ and\ \bibinfo {author} {\bibfnamefont {J.}~\bibnamefont
  {{Casares}}},\ }\bibfield  {title} {\bibinfo {title} {{Peering into the Dark
  Side: Magnesium Lines Establish a Massive Neutron Star in PSR J2215+5135}},\
  }\href {https://doi.org/10.3847/1538-4357/aabde6} {\bibfield  {journal}
  {\bibinfo  {journal} {Astrophys. J}\ }\textbf {\bibinfo {volume} {859}},\
  \bibinfo {eid} {54} (\bibinfo {year} {2018})},\ \Eprint
  {https://arxiv.org/abs/1805.08799} {arXiv:1805.08799 [astro-ph.HE]}
  \BibitemShut {NoStop}%
\bibitem [{\citenamefont {{Cromartie}}\ \emph {et~al.}(2020)\citenamefont
  {{Cromartie}}, \citenamefont {{Fonseca}}, \citenamefont {{Ransom}},
  \citenamefont {{Demorest}}, \citenamefont {{Arzoumanian}}, \citenamefont
  {{Blumer}}, \citenamefont {{Brook}}, \citenamefont {{DeCesar}}, \citenamefont
  {{Dolch}}, \citenamefont {{Ellis}} \emph {et~al.}}]{2020NatAs...4...72C}%
  \BibitemOpen
  \bibfield  {author} {\bibinfo {author} {\bibfnamefont {H.~T.}\ \bibnamefont
  {{Cromartie}}}, \bibinfo {author} {\bibfnamefont {E.}~\bibnamefont
  {{Fonseca}}}, \bibinfo {author} {\bibfnamefont {S.~M.}\ \bibnamefont
  {{Ransom}}}, \bibinfo {author} {\bibfnamefont {P.~B.}\ \bibnamefont
  {{Demorest}}}, \bibinfo {author} {\bibfnamefont {Z.}~\bibnamefont
  {{Arzoumanian}}}, \bibinfo {author} {\bibfnamefont {H.}~\bibnamefont
  {{Blumer}}}, \bibinfo {author} {\bibfnamefont {P.~R.}\ \bibnamefont
  {{Brook}}}, \bibinfo {author} {\bibfnamefont {M.~E.}\ \bibnamefont
  {{DeCesar}}}, \bibinfo {author} {\bibfnamefont {T.}~\bibnamefont {{Dolch}}},
  \bibinfo {author} {\bibfnamefont {J.~A.}\ \bibnamefont {{Ellis}}}, \emph
  {et~al.},\ }\bibfield  {title} {\bibinfo {title} {{Relativistic Shapiro delay
  measurements of an extremely massive millisecond pulsar}},\ }\href
  {https://doi.org/10.1038/s41550-019-0880-2} {\bibfield  {journal} {\bibinfo
  {journal} {Nature Astronomy}\ }\textbf {\bibinfo {volume} {4}},\ \bibinfo
  {pages} {72} (\bibinfo {year} {2020})},\ \Eprint
  {https://arxiv.org/abs/1904.06759} {arXiv:1904.06759 [astro-ph.HE]}
  \BibitemShut {NoStop}%
\bibitem [{\citenamefont {{Abbott}}\ \emph {et~al.}(2019)\citenamefont
  {{Abbott}}, \citenamefont {{Abbott}}, \citenamefont {{Abbott}}, \citenamefont
  {{Acernese}}, \citenamefont {{Ackley}}, \citenamefont {{Adams}},
  \citenamefont {{Adams}}, \citenamefont {{Addesso}}, \citenamefont
  {{Adhikari}}, \citenamefont {{Adya}} \emph {et~al.}}]{2019PhRvX...9a1001A}%
  \BibitemOpen
  \bibfield  {author} {\bibinfo {author} {\bibfnamefont {B.~P.}\ \bibnamefont
  {{Abbott}}}, \bibinfo {author} {\bibfnamefont {R.}~\bibnamefont {{Abbott}}},
  \bibinfo {author} {\bibfnamefont {T.~D.}\ \bibnamefont {{Abbott}}}, \bibinfo
  {author} {\bibfnamefont {F.}~\bibnamefont {{Acernese}}}, \bibinfo {author}
  {\bibfnamefont {K.}~\bibnamefont {{Ackley}}}, \bibinfo {author}
  {\bibfnamefont {C.}~\bibnamefont {{Adams}}}, \bibinfo {author} {\bibfnamefont
  {T.}~\bibnamefont {{Adams}}}, \bibinfo {author} {\bibfnamefont
  {P.}~\bibnamefont {{Addesso}}}, \bibinfo {author} {\bibfnamefont {R.~X.}\
  \bibnamefont {{Adhikari}}}, \bibinfo {author} {\bibfnamefont {V.~B.}\
  \bibnamefont {{Adya}}}, \emph {et~al.} (\bibinfo {collaboration} {LIGO
  Scientific Collaboration and Virgo Collaboration}),\ }\bibfield  {title}
  {\bibinfo {title} {{Properties of the Binary Neutron Star Merger GW170817}},\
  }\href {https://doi.org/10.1103/PhysRevX.9.011001} {\bibfield  {journal}
  {\bibinfo  {journal} {Physical Review X}\ }\textbf {\bibinfo {volume} {9}},\
  \bibinfo {eid} {011001} (\bibinfo {year} {2019})},\ \Eprint
  {https://arxiv.org/abs/1805.11579} {arXiv:1805.11579 [gr-qc]} \BibitemShut
  {NoStop}%
\bibitem [{\citenamefont {{Wen}}\ \emph {et~al.}(2019)\citenamefont {{Wen}},
  \citenamefont {{Li}}, \citenamefont {{Chen}},\ and\ \citenamefont
  {{Zhang}}}]{2019PhRvC..99d5806W}%
  \BibitemOpen
  \bibfield  {author} {\bibinfo {author} {\bibfnamefont {D.-H.}\ \bibnamefont
  {{Wen}}}, \bibinfo {author} {\bibfnamefont {B.-A.}\ \bibnamefont {{Li}}},
  \bibinfo {author} {\bibfnamefont {H.-Y.}\ \bibnamefont {{Chen}}},\ and\
  \bibinfo {author} {\bibfnamefont {N.-B.}\ \bibnamefont {{Zhang}}},\
  }\bibfield  {title} {\bibinfo {title} {{GW170817 implications on the
  frequency and damping time of f -mode oscillations of neutron stars}},\
  }\href {https://doi.org/10.1103/PhysRevC.99.045806} {\bibfield  {journal}
  {\bibinfo  {journal} {\prc}\ }\textbf {\bibinfo {volume} {99}},\ \bibinfo
  {eid} {045806} (\bibinfo {year} {2019})},\ \Eprint
  {https://arxiv.org/abs/1901.03779} {arXiv:1901.03779 [gr-qc]} \BibitemShut
  {NoStop}%
\bibitem [{\citenamefont {{Yagi}}\ and\ \citenamefont
  {{Yunes}}(2013)}]{2013Sci...341..365Y}%
  \BibitemOpen
  \bibfield  {author} {\bibinfo {author} {\bibfnamefont {K.}~\bibnamefont
  {{Yagi}}}\ and\ \bibinfo {author} {\bibfnamefont {N.}~\bibnamefont
  {{Yunes}}},\ }\bibfield  {title} {\bibinfo {title} {{I-Love-Q: Unexpected
  Universal Relations for Neutron Stars and Quark Stars}},\ }\href
  {https://doi.org/10.1126/science.1236462} {\bibfield  {journal} {\bibinfo
  {journal} {Science}\ }\textbf {\bibinfo {volume} {341}},\ \bibinfo {pages}
  {365} (\bibinfo {year} {2013})},\ \Eprint {https://arxiv.org/abs/1302.4499}
  {arXiv:1302.4499 [gr-qc]} \BibitemShut {NoStop}%
\bibitem [{\citenamefont {{Piekarewicz}}\ and\ \citenamefont
  {{Fattoyev}}(2019)}]{2019PhRvC..99d5802P}%
  \BibitemOpen
  \bibfield  {author} {\bibinfo {author} {\bibfnamefont {J.}~\bibnamefont
  {{Piekarewicz}}}\ and\ \bibinfo {author} {\bibfnamefont {F.~J.}\ \bibnamefont
  {{Fattoyev}}},\ }\bibfield  {title} {\bibinfo {title} {{Impact of the neutron
  star crust on the tidal polarizability}},\ }\href
  {https://doi.org/10.1103/PhysRevC.99.045802} {\bibfield  {journal} {\bibinfo
  {journal} {Phys. Rev. C}\ }\textbf {\bibinfo {volume} {99}},\ \bibinfo {eid}
  {045802} (\bibinfo {year} {2019})},\ \Eprint
  {https://arxiv.org/abs/1812.09974} {arXiv:1812.09974 [nucl-th]} \BibitemShut
  {NoStop}%
\bibitem [{\citenamefont {{Gamba}}\ \emph {et~al.}(2020)\citenamefont
  {{Gamba}}, \citenamefont {{Read}},\ and\ \citenamefont
  {{Wade}}}]{2020CQGra..37b5008G}%
  \BibitemOpen
  \bibfield  {author} {\bibinfo {author} {\bibfnamefont {R.}~\bibnamefont
  {{Gamba}}}, \bibinfo {author} {\bibfnamefont {J.~S.}\ \bibnamefont
  {{Read}}},\ and\ \bibinfo {author} {\bibfnamefont {L.~E.}\ \bibnamefont
  {{Wade}}},\ }\bibfield  {title} {\bibinfo {title} {{The impact of the crust
  equation of state on the analysis of GW170817}},\ }\href
  {https://doi.org/10.1088/1361-6382/ab5ba4} {\bibfield  {journal} {\bibinfo
  {journal} {Classical and Quantum Gravity}\ }\textbf {\bibinfo {volume}
  {37}},\ \bibinfo {eid} {025008} (\bibinfo {year} {2020})},\ \Eprint
  {https://arxiv.org/abs/1902.04616} {arXiv:1902.04616 [gr-qc]} \BibitemShut
  {NoStop}%
\bibitem [{\citenamefont {{Ducoin}}\ \emph {et~al.}(2011)\citenamefont
  {{Ducoin}}, \citenamefont {{Margueron}}, \citenamefont {{Provid{\^e}ncia}},\
  and\ \citenamefont {{Vida{\~n}a}}}]{2011PhRvC..83d5810D}%
  \BibitemOpen
  \bibfield  {author} {\bibinfo {author} {\bibfnamefont {C.}~\bibnamefont
  {{Ducoin}}}, \bibinfo {author} {\bibfnamefont {J.}~\bibnamefont
  {{Margueron}}}, \bibinfo {author} {\bibfnamefont {C.}~\bibnamefont
  {{Provid{\^e}ncia}}},\ and\ \bibinfo {author} {\bibfnamefont
  {I.}~\bibnamefont {{Vida{\~n}a}}},\ }\bibfield  {title} {\bibinfo {title}
  {{Core-crust transition in neutron stars: Predictivity of density
  developments}},\ }\href {https://doi.org/10.1103/PhysRevC.83.045810}
  {\bibfield  {journal} {\bibinfo  {journal} {\prc}\ }\textbf {\bibinfo
  {volume} {83}},\ \bibinfo {eid} {045810} (\bibinfo {year} {2011})},\ \Eprint
  {https://arxiv.org/abs/1102.1283} {arXiv:1102.1283 [nucl-th]} \BibitemShut
  {NoStop}%
\bibitem [{\citenamefont {{Bauswein}}\ \emph {et~al.}(2014)\citenamefont
  {{Bauswein}}, \citenamefont {{Stergioulas}},\ and\ \citenamefont
  {{Janka}}}]{2014PhRvD..90b3002B}%
  \BibitemOpen
  \bibfield  {author} {\bibinfo {author} {\bibfnamefont {A.}~\bibnamefont
  {{Bauswein}}}, \bibinfo {author} {\bibfnamefont {N.}~\bibnamefont
  {{Stergioulas}}},\ and\ \bibinfo {author} {\bibfnamefont {H.~T.}\
  \bibnamefont {{Janka}}},\ }\bibfield  {title} {\bibinfo {title} {{Revealing
  the high-density equation of state through binary neutron star mergers}},\
  }\href {https://doi.org/10.1103/PhysRevD.90.023002} {\bibfield  {journal}
  {\bibinfo  {journal} {\prd}\ }\textbf {\bibinfo {volume} {90}},\ \bibinfo
  {eid} {023002} (\bibinfo {year} {2014})},\ \Eprint
  {https://arxiv.org/abs/1403.5301} {arXiv:1403.5301 [astro-ph.SR]}
  \BibitemShut {NoStop}%
\bibitem [{\citenamefont {Wendland}(1995)}]{Wendland1995}%
  \BibitemOpen
  \bibfield  {author} {\bibinfo {author} {\bibfnamefont {H.}~\bibnamefont
  {Wendland}},\ }\bibfield  {title} {\bibinfo {title} {Piecewise polynomial,
  positive definite and compactly supported radial functions of minimal
  degree},\ }\href {https://doi.org/10.1007/BF02123482} {\bibfield  {journal}
  {\bibinfo  {journal} {Adv Comput Math}\ }\textbf {\bibinfo {volume} {4}},\
  \bibinfo {pages} {389} (\bibinfo {year} {1995})}\BibitemShut {NoStop}%
\bibitem [{\citenamefont {{Dehnen}}\ and\ \citenamefont
  {{Aly}}(2012)}]{Dehnen2012}%
  \BibitemOpen
  \bibfield  {author} {\bibinfo {author} {\bibfnamefont {W.}~\bibnamefont
  {{Dehnen}}}\ and\ \bibinfo {author} {\bibfnamefont {H.}~\bibnamefont
  {{Aly}}},\ }\bibfield  {title} {\bibinfo {title} {{Improving convergence in
  smoothed particle hydrodynamics simulations without pairing instability}},\
  }\href {https://doi.org/10.1111/j.1365-2966.2012.21439.x} {\bibfield
  {journal} {\bibinfo  {journal} {Mon. Not. R. Astron. Soc}\ }\textbf {\bibinfo
  {volume} {425}},\ \bibinfo {pages} {1068} (\bibinfo {year}
  {2012})}\BibitemShut {NoStop}%
\bibitem [{\citenamefont {{De}}\ \emph {et~al.}(2018)\citenamefont {{De}},
  \citenamefont {{Finstad}}, \citenamefont {{Lattimer}}, \citenamefont
  {{Brown}}, \citenamefont {{Berger}},\ and\ \citenamefont
  {{Biwer}}}]{2018PhRvL.121i1102D}%
  \BibitemOpen
  \bibfield  {author} {\bibinfo {author} {\bibfnamefont {S.}~\bibnamefont
  {{De}}}, \bibinfo {author} {\bibfnamefont {D.}~\bibnamefont {{Finstad}}},
  \bibinfo {author} {\bibfnamefont {J.~M.}\ \bibnamefont {{Lattimer}}},
  \bibinfo {author} {\bibfnamefont {D.~A.}\ \bibnamefont {{Brown}}}, \bibinfo
  {author} {\bibfnamefont {E.}~\bibnamefont {{Berger}}},\ and\ \bibinfo
  {author} {\bibfnamefont {C.~M.}\ \bibnamefont {{Biwer}}},\ }\bibfield
  {title} {\bibinfo {title} {{Tidal Deformabilities and Radii of Neutron Stars
  from the Observation of GW170817}},\ }\href
  {https://doi.org/10.1103/PhysRevLett.121.091102} {\bibfield  {journal}
  {\bibinfo  {journal} {\prl}\ }\textbf {\bibinfo {volume} {121}},\ \bibinfo
  {eid} {091102} (\bibinfo {year} {2018})},\ \Eprint
  {https://arxiv.org/abs/1804.08583} {arXiv:1804.08583 [astro-ph.HE]}
  \BibitemShut {NoStop}%
\bibitem [{\citenamefont {{Schmidt}}\ and\ \citenamefont
  {{Hinderer}}(2019)}]{2019PhRvD.100b1501S}%
  \BibitemOpen
  \bibfield  {author} {\bibinfo {author} {\bibfnamefont {P.}~\bibnamefont
  {{Schmidt}}}\ and\ \bibinfo {author} {\bibfnamefont {T.}~\bibnamefont
  {{Hinderer}}},\ }\bibfield  {title} {\bibinfo {title} {{Frequency domain
  model of f -mode dynamic tides in gravitational waveforms from compact binary
  inspirals}},\ }\href {https://doi.org/10.1103/PhysRevD.100.021501} {\bibfield
   {journal} {\bibinfo  {journal} {\prd}\ }\textbf {\bibinfo {volume} {100}},\
  \bibinfo {eid} {021501} (\bibinfo {year} {2019})},\ \Eprint
  {https://arxiv.org/abs/1905.00818} {arXiv:1905.00818 [gr-qc]} \BibitemShut
  {NoStop}%
\bibitem [{\citenamefont {{Guerra Chaves}}\ and\ \citenamefont
  {{Hinderer}}(2019)}]{2019JPhG...46l3002G}%
  \BibitemOpen
  \bibfield  {author} {\bibinfo {author} {\bibfnamefont {A.}~\bibnamefont
  {{Guerra Chaves}}}\ and\ \bibinfo {author} {\bibfnamefont {T.}~\bibnamefont
  {{Hinderer}}},\ }\bibfield  {title} {\bibinfo {title} {{Probing the equation
  of state of neutron star matter with gravitational waves from binary
  inspirals in light of GW170817: a brief review}},\ }\href
  {https://doi.org/10.1088/1361-6471/ab45be} {\bibfield  {journal} {\bibinfo
  {journal} {Journal of Physics G Nuclear Physics}\ }\textbf {\bibinfo {volume}
  {46}},\ \bibinfo {eid} {123002} (\bibinfo {year} {2019})},\ \Eprint
  {https://arxiv.org/abs/1912.01461} {arXiv:1912.01461 [nucl-th]} \BibitemShut
  {NoStop}%
\bibitem [{\citenamefont {{Pratten}}\ \emph {et~al.}(2020)\citenamefont
  {{Pratten}}, \citenamefont {{Schmidt}},\ and\ \citenamefont
  {{Hinderer}}}]{2020NatCo..11.2553P}%
  \BibitemOpen
  \bibfield  {author} {\bibinfo {author} {\bibfnamefont {G.}~\bibnamefont
  {{Pratten}}}, \bibinfo {author} {\bibfnamefont {P.}~\bibnamefont
  {{Schmidt}}},\ and\ \bibinfo {author} {\bibfnamefont {T.}~\bibnamefont
  {{Hinderer}}},\ }\bibfield  {title} {\bibinfo {title} {{Gravitational-wave
  asteroseismology with fundamental modes from compact binary inspirals}},\
  }\href {https://doi.org/10.1038/s41467-020-15984-5} {\bibfield  {journal}
  {\bibinfo  {journal} {Nature Communications}\ }\textbf {\bibinfo {volume}
  {11}},\ \bibinfo {eid} {2553} (\bibinfo {year} {2020})},\ \Eprint
  {https://arxiv.org/abs/1905.00817} {arXiv:1905.00817 [gr-qc]} \BibitemShut
  {NoStop}%
\bibitem [{\citenamefont {{Chatziioannou}}(2020)}]{2020GReGr..52..109C}%
  \BibitemOpen
  \bibfield  {author} {\bibinfo {author} {\bibfnamefont {K.}~\bibnamefont
  {{Chatziioannou}}},\ }\bibfield  {title} {\bibinfo {title} {{Neutron-star
  tidal deformability and equation-of-state constraints}},\ }\href
  {https://doi.org/10.1007/s10714-020-02754-3} {\bibfield  {journal} {\bibinfo
  {journal} {General Relativity and Gravitation}\ }\textbf {\bibinfo {volume}
  {52}},\ \bibinfo {eid} {109} (\bibinfo {year} {2020})},\ \Eprint
  {https://arxiv.org/abs/2006.03168} {arXiv:2006.03168 [gr-qc]} \BibitemShut
  {NoStop}%
\bibitem [{\citenamefont {{Doneva}}\ \emph {et~al.}(2013)\citenamefont
  {{Doneva}}, \citenamefont {{Gaertig}}, \citenamefont {{Kokkotas}},\ and\
  \citenamefont {{Kr{\"u}ger}}}]{2013PhRvD..88d4052D}%
  \BibitemOpen
  \bibfield  {author} {\bibinfo {author} {\bibfnamefont {D.~D.}\ \bibnamefont
  {{Doneva}}}, \bibinfo {author} {\bibfnamefont {E.}~\bibnamefont {{Gaertig}}},
  \bibinfo {author} {\bibfnamefont {K.~D.}\ \bibnamefont {{Kokkotas}}},\ and\
  \bibinfo {author} {\bibfnamefont {C.}~\bibnamefont {{Kr{\"u}ger}}},\
  }\bibfield  {title} {\bibinfo {title} {{Gravitational wave asteroseismology
  of fast rotating neutron stars with realistic equations of state}},\ }\href
  {https://doi.org/10.1103/PhysRevD.88.044052} {\bibfield  {journal} {\bibinfo
  {journal} {\prd}\ }\textbf {\bibinfo {volume} {88}},\ \bibinfo {eid} {044052}
  (\bibinfo {year} {2013})},\ \Eprint {https://arxiv.org/abs/1305.7197}
  {arXiv:1305.7197 [astro-ph.SR]} \BibitemShut {NoStop}%
\bibitem [{\citenamefont {{Kr{\"u}ger}}\ and\ \citenamefont
  {{Kokkotas}}(2020)}]{2020PhRvL.125k1106K}%
  \BibitemOpen
  \bibfield  {author} {\bibinfo {author} {\bibfnamefont {C.~J.}\ \bibnamefont
  {{Kr{\"u}ger}}}\ and\ \bibinfo {author} {\bibfnamefont {K.~D.}\ \bibnamefont
  {{Kokkotas}}},\ }\bibfield  {title} {\bibinfo {title} {{Fast Rotating
  Relativistic Stars: Spectra and Stability without Approximation}},\ }\href
  {https://doi.org/10.1103/PhysRevLett.125.111106} {\bibfield  {journal}
  {\bibinfo  {journal} {\prl}\ }\textbf {\bibinfo {volume} {125}},\ \bibinfo
  {eid} {111106} (\bibinfo {year} {2020})},\ \Eprint
  {https://arxiv.org/abs/1910.08370} {arXiv:1910.08370 [gr-qc]} \BibitemShut
  {NoStop}%
\bibitem [{\citenamefont {{Bauswein}}\ \emph {et~al.}(2019)\citenamefont
  {{Bauswein}}, \citenamefont {{Bastian}}, \citenamefont {{Blaschke}},
  \citenamefont {{Chatziioannou}}, \citenamefont {{Clark}}, \citenamefont
  {{Fischer}},\ and\ \citenamefont {{Oertel}}}]{2019PhRvL.122f1102B}%
  \BibitemOpen
  \bibfield  {author} {\bibinfo {author} {\bibfnamefont {A.}~\bibnamefont
  {{Bauswein}}}, \bibinfo {author} {\bibfnamefont {N.-U.~F.}\ \bibnamefont
  {{Bastian}}}, \bibinfo {author} {\bibfnamefont {D.~B.}\ \bibnamefont
  {{Blaschke}}}, \bibinfo {author} {\bibfnamefont {K.}~\bibnamefont
  {{Chatziioannou}}}, \bibinfo {author} {\bibfnamefont {J.~A.}\ \bibnamefont
  {{Clark}}}, \bibinfo {author} {\bibfnamefont {T.}~\bibnamefont {{Fischer}}},\
  and\ \bibinfo {author} {\bibfnamefont {M.}~\bibnamefont {{Oertel}}},\
  }\bibfield  {title} {\bibinfo {title} {{Identifying a First-Order Phase
  Transition in Neutron-Star Mergers through Gravitational Waves}},\ }\href
  {https://doi.org/10.1103/PhysRevLett.122.061102} {\bibfield  {journal}
  {\bibinfo  {journal} {\prl}\ }\textbf {\bibinfo {volume} {122}},\ \bibinfo
  {eid} {061102} (\bibinfo {year} {2019})},\ \Eprint
  {https://arxiv.org/abs/1809.01116} {arXiv:1809.01116 [astro-ph.HE]}
  \BibitemShut {NoStop}%
\bibitem [{\citenamefont {{Weih}}\ \emph {et~al.}(2020)\citenamefont {{Weih}},
  \citenamefont {{Hanauske}},\ and\ \citenamefont
  {{Rezzolla}}}]{2020PhRvL.124q1103W}%
  \BibitemOpen
  \bibfield  {author} {\bibinfo {author} {\bibfnamefont {L.~R.}\ \bibnamefont
  {{Weih}}}, \bibinfo {author} {\bibfnamefont {M.}~\bibnamefont {{Hanauske}}},\
  and\ \bibinfo {author} {\bibfnamefont {L.}~\bibnamefont {{Rezzolla}}},\
  }\bibfield  {title} {\bibinfo {title} {{Postmerger Gravitational-Wave
  Signatures of Phase Transitions in Binary Mergers}},\ }\href
  {https://doi.org/10.1103/PhysRevLett.124.171103} {\bibfield  {journal}
  {\bibinfo  {journal} {\prl}\ }\textbf {\bibinfo {volume} {124}},\ \bibinfo
  {eid} {171103} (\bibinfo {year} {2020})},\ \Eprint
  {https://arxiv.org/abs/1912.09340} {arXiv:1912.09340 [gr-qc]} \BibitemShut
  {NoStop}%
\bibitem [{\citenamefont {{Bauswein}}\ and\ \citenamefont
  {{Blacker}}(2020)}]{2020EPJST.229.3595B}%
  \BibitemOpen
  \bibfield  {author} {\bibinfo {author} {\bibfnamefont {A.}~\bibnamefont
  {{Bauswein}}}\ and\ \bibinfo {author} {\bibfnamefont {S.}~\bibnamefont
  {{Blacker}}},\ }\bibfield  {title} {\bibinfo {title} {{Impact of quark
  deconfinement in neutron star mergers and hybrid star mergers}},\ }\href
  {https://doi.org/10.1140/epjst/e2020-000138-7} {\bibfield  {journal}
  {\bibinfo  {journal} {European Physical Journal Special Topics}\ }\textbf
  {\bibinfo {volume} {229}},\ \bibinfo {pages} {3595} (\bibinfo {year}
  {2020})},\ \Eprint {https://arxiv.org/abs/2006.16183} {arXiv:2006.16183
  [astro-ph.HE]} \BibitemShut {NoStop}%
\end{thebibliography}%

\end{document}